\documentclass[sn-mathphys,Numbered]{sn-jnl}

\RequirePackage{multirow,booktabs,subfigure,color,array,hhline,makecell}
\usepackage{indentfirst}
\usepackage{graphicx}
\usepackage{amsmath,amssymb,amsfonts}%
\usepackage{amsthm}%
\usepackage{mathrsfs}%
\usepackage[title]{appendix}%
\usepackage{xcolor}%
\usepackage{textcomp}%
\usepackage{manyfoot}%
\usepackage{booktabs}%
\usepackage{algpseudocode}%
\usepackage{listings}%
\usepackage{changes}
\RequirePackage{algorithm,algpseudocode}

\newtheorem{thm}{Theorem}
\newtheorem{defin}{Definition}
\newtheorem{lem}{Lemma}
\newtheorem{assum}{Assumption}
\newtheorem{rem}{Remark}

\newtheorem{con}{Condition}

\newtheorem{Ex}{Example}

\begin{document}

\title[Overlapping community detection in weighted networks]{Overlapping community detection in weighted networks}

\author*[1]{\fnm{Huan} \sur{Qing}}\email{qinghuan@u.nus.edu}
\affil[1]{\orgdiv{School of Economics and Finance}, \orgname{Chongqing University of Technology}, \city{Chongqing}, \postcode{400054}, \state{Chongqing}, \country{China}}


\abstract{Over the past decade, community detection in overlapping un-weighted networks, where nodes can belong to multiple communities, has been one of the most popular topics in modern network science. However, community detection in overlapping weighted networks, where edge weights can be any real value, remains challenging. In this article, we propose a generative model called the weighted degree-corrected mixed membership (WDCMM) model to model such weighted networks. This model adopts the same factorization for the expectation of the adjacency matrix as the previous degree-corrected mixed membership (DCMM) model. Our WDCMM extends the DCMM from un-weighted networks to weighted networks by allowing the elements of the adjacency matrix to be generated from distributions beyond Bernoulli. We first address the community membership estimation of the model by applying a spectral algorithm and establishing a theoretical guarantee of consistency. Then, we propose overlapping weighted modularity to measure the quality of overlapping community detection for both assortative and dis-assortative weighted networks. To determine the number of communities, we incorporate the algorithm into the proposed modularity. We demonstrate the advantages of the model and the modularity through applications to simulated data and real-world networks.}
\keywords{Overlapping community detection, overlapping weighted modularity, overlapping weighted networks}



\maketitle
\section{Introduction}\label{sec1}
Community detection is a popular data mining problem in network science and it is to partition a network into several components, where each component is called a community (also referred to as a module or a cluster, or a block). Generally, nodes within the same community have more interactions than nodes across communities. Community detection has many real-world applications in various domains, like sociology, biology, marketing, and computer science \cite{girvan2002community,newman2003the,barabasi2004network,guimera2005functional,palla2007quantifying,andersen2012overlapping,garcia2018applications}. For example, community detection in biological networks helps researchers to analyze the interaction of brain regions and have a better understanding of the influence of the interaction on brain functions \cite{garcia2018applications,huang2019ocomm,tooley2022functional}. Community detection in social networks can help to identify crime organizations \cite{magalingam2015using,legramanti2022extended}. Community detection in online social networks and food social networks can facilitate recommender systems \cite{sahebi2011community,gasparetti2021community,rostami2023novel}. Community detection in coauthorship and citation networks can be used to identify meaningful communities for scientists \cite{Newman2004,ji2022co}.

For its wide applications, in past decades, substantial methods have been developed to learn community structure for non-overlapping un-weighted networks in which each node belongs to a single community and the edges or links between nodes are either present or not \cite{fortunato2010community,fortunato2016community,papadopoulos2012community}, where un-weighted networks are also referred to as binary networks \cite{lu2014algorithms}. For community detection in non-overlapping un-weighted networks, perhaps the most popular and well-studied generative model is the Stochastic Blockmodel (SBM) \cite{SBM}. The SBM assumes that the probability of generating an edge between two nodes depends on their belonging communities. Under SBM, nodes within the same community have the same expected degree while node degrees always vary in real-world networks because there are ``hub'' nodes with many more edges than other nodes in the same community. To solve this limitation, the Degree-Corrected Stochastic Blockmodel (DCSBM) \cite{DCSBM} was proposed by introducing node heterogeneity parameters. Many works have been developed to study SBM and DCSBM, to name a few, \cite{rohe2011spectral,lei2015consistency,SCORE,joseph2016impact,abbe2016exact,chen2018convexified}. One limitation of SBM and DCSBM is, they assume that the network has non-overlapping properties while nodes can belong to multiple communities simultaneously in real-world social as well as other types of networks (for example, a movie can be classified into multiple genres, individuals belong to multiple social groups, scientists may belong to more than one
research group, and proteins have multiple functions) \cite{palla2005uncovering,xie2013overlapping,javed2018community}. To address this limitation, the Mixed Membership Stochastic Blockmodel (MMSB) \cite{MMSB} and the Degree-Corrected Mixed Membership (DCMM) model \cite{MixedSCORE} extend SBM and DCSBM to overlapping un-weighted networks, where DCMM can also be viewed as an extension of MMSB by considering node heterogeneity and a reparametrization of the overlapping continuous community assignment model (OCCAM) considered in \citep{OCCAM}. In recent years, some spectral algorithms with theoretical guarantees on consistency have been proposed to study MMSB and DCMM, to name a few, \cite{MaoSVM,mao2020estimating, qing2023regularized,MixedSCORE}. A thorough survey of the community detection approaches for un-weighted networks has been provided by \cite{jin2021survey}.

\begin{figure}
\centering
\subfigure[Slovene Parliamentary Party \cite{ferligoj1996analysis}]{\includegraphics[width=0.32\textwidth]{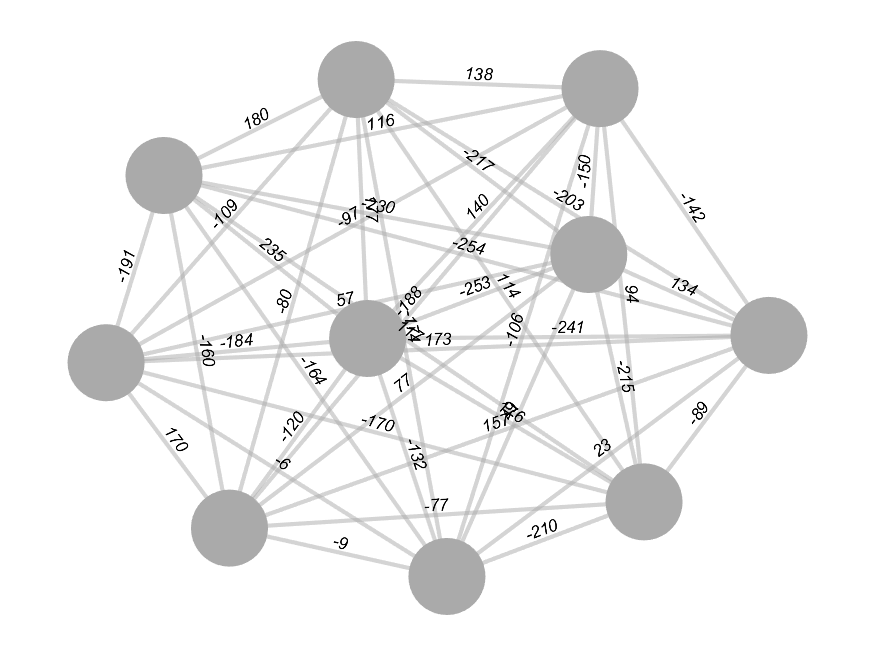}}
\subfigure[Gahuku-Gama subtribes \cite{read1954cultures}]{\includegraphics[width=0.32\textwidth]{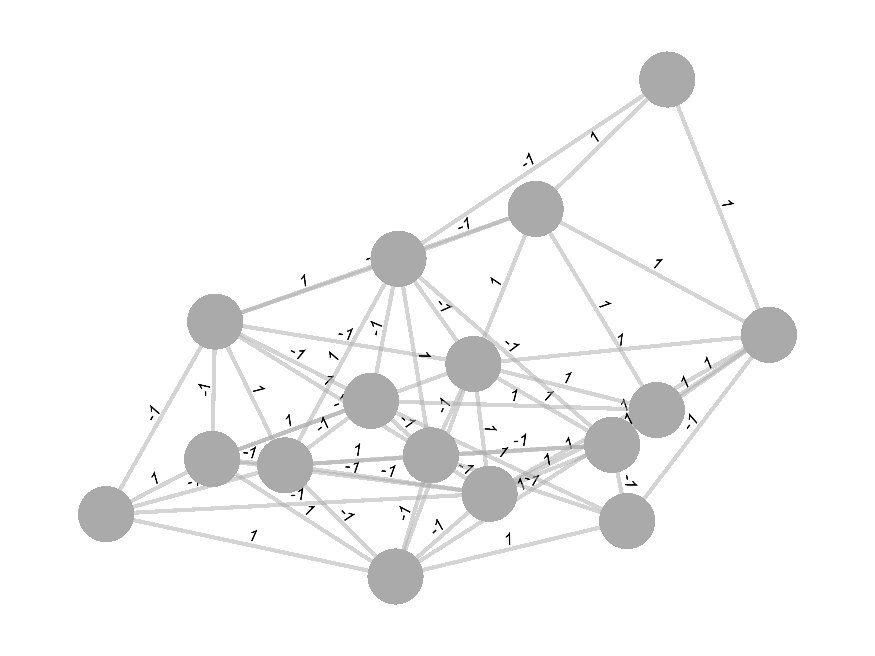}}
\subfigure[Karate-club-weighted \cite{zachary1977information}]{\includegraphics[width=0.32\textwidth]{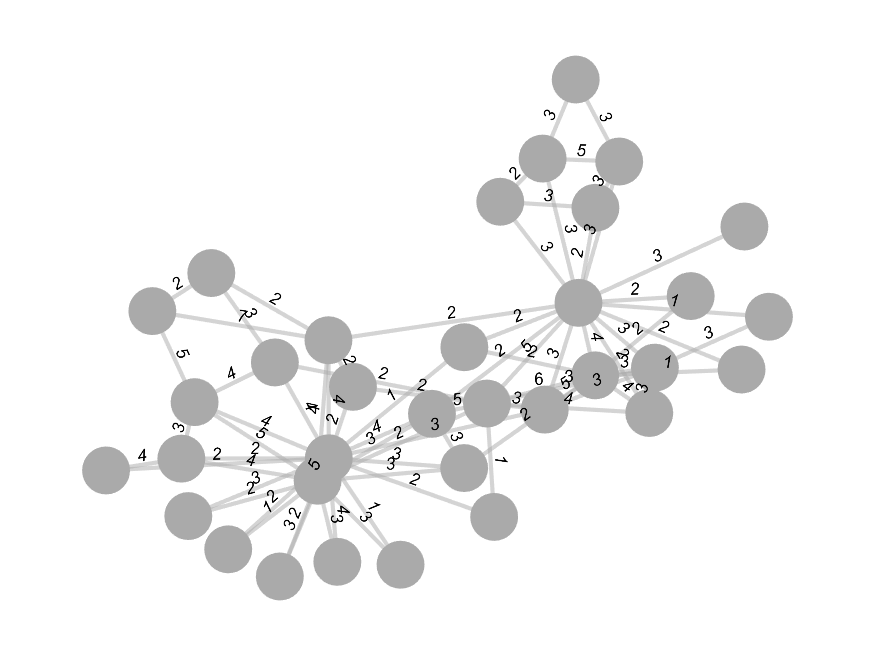}}
\caption{Illustrative examples of three real-world weighted networks. For the Slovene Parliamentary Party network, edge weight means political space distance between parties. For the Gahuku-Gama subtribes network, edge weight represents friendship. For the Karate-club-weighted network, edge weight indicates the relative strength of the associations. For visualization, we do not show node labels.}
\label{NetReal3}
\end{figure}

Edge weights are commonplace in complex networks and they provide more information for community detection power than edges \cite{newman2004analysis}. Edge weights usually tell us how strong or weak a relationship is. For example, in a coauthorship network \cite{Newman2004}, the edge weight between two authors is the number of papers they co-authored. In a signed network \cite{ferligoj1996analysis,tang2016survey,kirkley2019balance,shi2019dynamics}, positive edges represent friendship/trust and negative edges mean animosity/mistrust. Three examples of real-world weighted networks in which edge weights indicate the strength of the relationship are shown in Figure \ref{NetReal3}. However, the aforementioned models ignore edge weights and they only model un-weighted networks. To address this limitation, some statistical models have been developed in recent years. For non-overlapping weighted networks, several Weighted Stochastic Blockmodels (WSBMs) \citep{aicher2015learning, jog2015information,peixoto2018nonparametric,palowitch2018significance,xu2020optimal,ng2021weighted,guo2024efficacy,gallagher2024spectral}, the Distribution-Free model (DFM) \citep{qing2021DFM}, and the Degree-Corrected Distribution-Free model (DCDFM) \citep{DCDFM}, and the bipartite versions of DFM and DCDFM \citep{BiDFMs}
extend SBM and DCSBM to weighted networks by allowing edge weights to follow distributions more than the Bernoulli distribution. Recently, \citep{gallagher2024spectral} demonstrated that appropriate transformations of the adjacency matrix elements for a weighted network can be highly beneficial under WSBMs. Additionally, \citep{cape2024robust} also considered WSBMs that permit the adjacency matrix elements to follow any distribution and established the estimation consistency of a robust spectral clustering method for weighted networks. However, the above models for weighted networks assume that each node belongs to a single community and they can not capture the community structure of overlapping weighted networks. Though the weighted version of MMSB (WMMSB) \cite{dulac2020mixed} and the multi-way blockmodels (MWB) \cite{airoldi2013multi} can model overlapping weighted networks, they both face some limitations. WMMSB requires edge weights to be generated from Poisson distribution, so it fails to model un-weighted networks, signed networks, networks with positive and negative edge weights, and networks with decimal edge weights. MWB requires edge weights to follow Normal or Bernoulli distributions, thus it fails to describe the community structure of signed networks and networks in which edge weights are nonnegative integers. Meanwhile, both WMMSB and MWB ignore the variation of node degrees. To close these limitations of WMMSB and MWB, we aim to provide a model for overlapping weighted networks. The main contributions of this article are summarized as follows.

\begin{itemize}
  \item We provide a generative model called the Weighted Degree-Corrected Mixed Membership (WDCMM) model for overlapping networks with weighted edges. WDCMM shares the same factorization for the expectation of the adjacency matrix as DCMM and extends its applicability. This extension allows the elements of the adjacency matrix to be generated from various distributions, such as Normal, Poisson, and Uniform, in addition to the Bernoulli distribution. Moreover, signed networks can also be modeled by WDCMM. Further examples of distributions that WDCMM can handle are listed in Table \ref{table-bounds}.
  \item We use a spectral algorithm with a theoretical guarantee of consistency to fit WDCMM. The theoretical error rate for the algorithm when edge weights follow a specific distribution can be obtained immediately from the main result and we also provide several examples to show this.
  \item We provide overlapping weighted modularity in Equation (\ref{Modularity}) to measure the quality of overlapping community detection for both assortative and dis-assortative weighted networks with positive and negative edge weights.
  \item We provide a method to infer the number of communities for weighted networks by combining the algorithm with the overlapping weighted modularity.
  \item We implement experiments on computer-generated overlapping weighted networks and real-world networks to demonstrate the advantages of the proposed model and the overlapping weighted modularity.
\end{itemize}

The rest of the paper is organized as follows. Section \ref{SecWDCMM} introduces the model. Section \ref{SecAlgorithm} introduces the algorithm fitting the model. Section \ref{SecMain} presents the consistency of the algorithm under the model and provides some examples for different distributions. Section \ref{SecK} introduces the overlapping weighted modularity and the method to determine the number of communities. Section \ref{SecExperiments} conducts extensive experiments. Section \ref{SecConclusion} concludes. All proofs are in the Appendix.
\section{The Weighted Degree-Corrected Mixed Membership model}\label{SecWDCMM}
\begin{table}[ht]
\centering
\scriptsize
\resizebox{\columnwidth}{!}{
\begin{tabular}{cc|cc}
\hline
Symbol& Description&Symbol& Description\\
\hline
$\mathbb{R}$&Real number set&$\mathbb{R}_{+}$&Set of nonnegative real numbers\\
$\mathcal{N}$&Weighted network&$P\in\mathbb{R}^{K\times K}$& Block matrix ($P=P'$)\\
$K$&Number of communities&$\|x\|_{q}$&$\ell_{q}$-norm for vector $x$\\
$n$&Number of nodes&$X'$& $X$'s transpose\\
$[m]$&$\{1,2,\ldots,m\}$ for positive integer $m$&$\mathbb{E}$&Expectation\\
$A\in\mathbb{R}^{n\times n}$&Adjacency matrix&$\|X\|_{2\rightarrow\infty}$&$X$'s maximum $\ell_{2}$-norm\\
$\mathcal{C}_{(k)}$&$k$-th community&$\|X\|_{F}$&Frobenius norm of $X$\\
$\Pi\in[0,1]^{n\times K}$&Membership matrix&$X(i,:)$&$X$'s $i$-th row\\
$\mathcal{I}$&Index set of pure nodes&$X(:,j)$&$X$'s $j$-th column\\
$\rho$&Sparsity parameter&$X(\mathcal{I},:)$&Rows in the index set $\mathcal{I}$ of $X$\\
$\mathcal{F}$&Distribution&$\mathrm{rank}(X)$&$X$'s rank\\
$\theta$&$n\times 1$ node heterogeneity vector&$\Theta$&Diagonal matrix with $\Theta(i,i)=\theta(i)$ for $i\in[n]$\\
$\theta_{\mathrm{max}}$&Maximum element of $\theta$&$\theta_{\mathrm{min}}$&Minimum element of $\theta$\\
$\sigma_{k}(X)$&$X$'s $k$-th largest singular value&$\lambda_{k}(X)$&$X$'s $k$-th largest eigenvalue in magnitude\\
$\Omega$&$\Theta\Pi P\Pi'\Theta$&$\kappa(X)$&$X$'s condition number\\
$P_{\mathrm{max}}$&Maximum entry of $P$&$\Omega_{\mathrm{max}}$&Maximum entry of $\Omega$\\
$U\in\mathbb{R}^{n\times K}$&Top $K$ eigenvectors of $\Omega$&$\mathcal{P}\in\{0,1\}^{K\times K}$&Permutation matrix\\
$\hat{U}\in\mathbb{R}^{n\times K}$&Top $K$ eigenvectors of $A$&$|a|$&Absolute value for real value $a$\\
$\Lambda$&Diagonal matrix with $\Lambda(k,k)=\lambda_{k}(\Omega)$ for $k\in[K]$&$\hat{\Lambda}$&Diagonal matrix with $\hat{\Lambda}(k,k)=\lambda_{k}(A)$ for $k\in[K]$\\
$U_{*}$&$U$'s row-normalization version&$\hat{U}_{*}$&$\hat{U}$'s row-normalization version\\
$\mathbb{N}$&Set of nonnegative integers&$e_{i}$&$e_{i}(j)=1(i=j)$\\
$\hat{\Pi}\in[0,1]^{n\times K}$&Estimated memberships&$\mathbb{P}$&Probability\\
$\tau$&$\mathrm{max}_{i\in[n],j\in[n]}|A(i,j)-\Omega(i,j)|$&$\gamma$&$\mathrm{max}_{i\in[n],j\in[n]}\frac{\mathbb{E}[(A(i,j)-\Omega(i,j))^{2}]}{\theta(i)\theta(j)}$\\
$\mathrm{diag}(M)$&Diagonal matrix with $(i,i)$-th entry $M(i,i)$&$\mathrm{max}(0,M)$&Matrix with $(i,j)$-th entry $\mathrm{max}(0,M(i,j))$\\
$M^{-1}$&Inverse of matrix $M$&$I_{K}$&$K\times K$ identity matrix\\
$Q_{ovw}$&Overlapping weighted modularity&$n_{0}$&Number of pure nodes in each community\\
$\zeta_{\mathrm{mixed}}$&Proportion of highly mixed nodes&$\zeta_{\mathrm{pure}}$&Proportion of highly pure nodes\\
$\varsigma$&Balance parameter&$\hat{\mathcal{I}}$&Estimated index set\\
$\eta_{\mathcal{N}}$&Index reflecting $\mathcal{N}$'s assortativity&$N_{M}$&Diagonal matrix with positive diagonal elements\\
$\mathbb{N}_{+}$&Set of positive integers&$\mathbb{R}_{+}$&Set of positive real numbers\\
\hline
\end{tabular}
}
\caption{Table of main symbols.}
\label{table-symbol}
\end{table}

Table \ref{table-symbol} summarizes the main symbols used in this paper. Given a weighted network $\mathcal{N}$ with $n$ nodes (network is also known as graph and node is also known as a vertex in literature), let $A\in \mathbb{R}^{n\times n}$ be its adjacency matrix such that $A(i,j)$ denotes the edge weight between nodes $i$ and $j$ for $i,j\in[n]$. Since we only consider un-directed networks in this article, $A$ is symmetric. All entries of $A$ are allowed to be any finite real values in this article. We assume there are $K$ communities
\begin{align}\label{DefinCommunity}
\mathcal{C}_{(1)},\mathcal{C}_{(2)},\ldots,\mathcal{C}_{(K)}.
\end{align}
We work with the mixed membership (also known as overlapping) weighted network in which a node can belong to multiple communities. Let $\Pi$ be an $n\times K$ matrix such that $\Pi(i,k)$ denotes the weight of node $i$ belonging to community $k$ for $i\in[n],k\in[K]$. We assume that $\Pi$ should satisfy the following conditions to make our model identifiable:
\begin{align}
&\mathrm{rank}(\Pi)=K, \Pi(i,k)\geq 0, \sum_{k=1}^{K}\Pi(i,k)=1 \mathrm{~for~}i\in[n],k\in[K],\label{DefinePI}\\
&\mathrm{Each~of~the ~}K\mathrm{~communities~has~at~least~one~pure~node}\label{pure},
\end{align}
where node $i$ is a pure node if one element of $\Pi(i,:)$ is 1 and it is a mixed node otherwise. Let $\mathcal{I}$ be an index set of nodes corresponding to $K$ pure nodes, one from each community, i.e., $\mathcal{I}=\{s_{1},s_{2},\ldots,s_{K}\}$, where $s_{k}$ is a pure node in the $k$-th community $\mathcal{C}_{(k)}$ for $k\in[K]$. Similar to \cite{mao2020estimating}, we let $\Pi(\mathcal{I},:)$ be the $K\times K$ identity matrix for convenience. Let $P$ be a $K\times K$ block matrix (we also call $P$ connectivity matrix occasionally in this article) and satisfy the following conditions:
\begin{align}\label{definP}
 P=P', P\in\mathbb{R}^{K\times K},~\mathrm{rank}(P)=K,\mathrm{and~}P\mathrm{~has~unit~diagonals},
\end{align}
where the requirement for $P$ to have diagonal elements is imposed to ensure the identifiability of the proposed model. For further details, please refer to Remark \ref{IDWDCMM} provided later. Note that since we consider weighted networks and $A$'s entries can by any finite real values, $P$ can have negative elements and $P$ is not a probability matrix like that of traditional models SBM \cite{SBM}, DCSBM \cite{DCSBM}, MMSB \cite{MMSB}, and DCMM \cite{MixedSCORE} for un-weighted networks. For convenience, let $P_{\mathrm{max}}=\mathrm{max}_{k,l\in[K]}|P(k,l)|$. Note that $P_{\mathrm{max}}$ can be larger than 1 because we will consider node heterogeneity later. Unless specified, throughout this article, $K$ is assumed to be a known integer.

Introduce the $n\times 1$ vector $\theta$ such that $\theta(i)$ is the node heterogeneity of node $i$ for $i\in[n]$, where $\theta$ controls the degree variations for all nodes. Let $\Theta$ be an $n\times n$ diagonal matrix whose $i$-th diagonal element is $\theta(i)$. Set $\theta_{\mathrm{max}}=\mathrm{max}_{i\in[n]}\theta(i)$ and $\theta_{\mathrm{min}}=\mathrm{min}_{i\in[n]}\theta(i)$. Node heterogeneity should be positive, so we have
\begin{align}\label{thetapositive}
\theta_{\mathrm{min}}>0.
\end{align}
For \emph{arbitrary distribution} $\mathcal{F}$, our model generates $A$ for the weighted network $\mathcal{N}$ by assuming that $A(i,j)$ are independent random variables generated from $\mathcal{F}$ with expectation
\begin{align}\label{DefinOmega}
\mathbb{E}[A(i,j)]=\Omega(i,j), \mathrm{where~}\Omega:=\Theta\Pi P\Pi'\Theta \mathrm{~for~}i,j\in[n].
\end{align}
Equation (\ref{DefinOmega}) means that $A$'s expectation matrix $\Omega$ should have a block structure $\Theta \Pi P\Pi'\Theta$ reflecting community membership information, and this is the reason we call $P$ a block matrix. Meanwhile, Equation (\ref{DefinOmega}) also means that $A$ is a random matrix generated from any distribution $\mathcal{F}$ with expectation matrix $\Omega$ and there is no limitation on $\mathcal{F}$ as long as Equation (\ref{DefinOmega}) holds. Furthermore, the expectation adjacency matrix $\Omega$ shares the same factorization as the $\Omega$ in Equation (2.3) of \citep{MixedSCORE} under the DCMM model.
\begin{defin}\label{DefinWDCMM}
Call Equations (\ref{DefinCommunity})-(\ref{DefinOmega}) the Weighted Degree-Corrected Mixed Membership (WDCMM) model and denote it by $WDCMM_{n}(K, P, \Pi, \Theta,\mathcal{F})$.
\end{defin}

In Definition \ref{DefinWDCMM}, we do not constrain any special type of the distribution $\mathcal{F}$ that can be handled by the WDCMM model. This is because our WDCMM allows $A$'s elements to be generated from any distribution $\mathcal{F}$ as long as $A$'s expectation under $\mathcal{F}$ is $\Omega$, which enjoys a block structure related to nodes' mixed memberships. In other words, the only limitation of our WDCMM model on the distribution $\mathcal{F}$ is that $\mathbb{E}[A(i,j)]=\Omega(i,j)$ for $i,j\in[n]$ under the distribution $\mathcal{F}$, i.e., $\mathcal{F}$'s first moment exists and has a block structure $\Omega$. For example, $\mathcal{F}$ can be distributions like Normal, Bernoulli, Poisson, Uniform, Binomial, Negative binomial, Exponential, Gamma, Beta, Geometric, Laplace, Lognormal, Pareto, Logistic, and Weibull distributions as described in \citep{rohatgi2015introduction}, as well as a discrete distribution for signed networks. The details of these distributions are provided below, for $i,j\in[n]$:
\begin{itemize}
  \item $\mathcal{F}$ can be \textbf{Normal distribution} by letting $A(i,j)\sim \text{Normal}(\Omega(i,j), \sigma^2_A)$. For this distribution, $A(i,j)\in\mathbb{R}$, $\mathbb{E}[A(i,j)]=\Omega(i,j)$ holds, and there is no limitation on $\Omega(i,j)$.
  \item $\mathcal{F}$ can be \textbf{Bernoulli distribution} by letting $A(i,j)\sim \text{Bernoulli}(\Omega(i,j))$. For this distribution, $A(i,j)\in\{0,1\}$, $\mathbb{E}[A(i,j)]=\Omega(i,j)$ holds, and $\Omega(i,j)$ should be a probability in $[0,1]$.
  \item $\mathcal{F}$ can be \textbf{Poisson distribution} by letting $A(i,j)\sim \text{Poisson}(\Omega(i,j))$. For this distribution, $A(i,j)$ is a nonnegative integer, $\mathbb{E}[A(i,j)]=\Omega(i,j)$ holds, and $\Omega(i,j)$ should be a nonnegative value.
  \item $\mathcal{F}$ can be \textbf{Uniform distribution} by letting $A(i,j)\sim \text{Uniform}(a, 2\Omega(i,j)-a)$ (or $A(i,j)\sim \text{Uniform}(2\Omega(i,j)-a, a)$) for any finite value $a$. For this distribution, $A(i,j)\in(a, 2\Omega(i,j)-a)$ (or $A(i,j)\in(2\Omega(i,j)-a, a)$), $\mathbb{E}[A(i,j)]=\Omega(i,j)$ holds, and $\Omega(i,j)$ should be larger (or smaller) than $a$.
  \item $\mathcal{F}$ can also be a \textbf{discrete distribution} such that $P(A(i,j)=1)=\frac{1+\Omega(i,j)}{2}$ and $P(A(i,j)=-1)=\frac{1-\Omega(i,j)}{2}$. For this case, $A$ is the adjacency matrix of a signed network, $\mathbb{E}[A(i,j)]=\Omega(i,j)$ holds, and $\Omega(i,j)$ should range in $[-1,1]$.
  \item $\mathcal{F}$ can be \textbf{Binomial distribution} by letting $A(i,j)\sim \text{Binomial}(m, \frac{\Omega(i,j)}{m})$, where $m$ is a positive integer. For this distribution, $A(i,j)$ is a nonnegative integer that is no larger than $m$, $\mathbb{E}[A(i,j)]=\Omega(i,j)$ holds, and $\Omega(i,j)$ should be a value ranging in $[0, m]$ since $\frac{\Omega(i,j)}{m}$ is a probability.
  \item $\mathcal{F}$ can be \textbf{Negative binomial distribution} by letting $A(i,j)\sim \text{Negative binomial}(m, \frac{m}{m+\Omega(i,j)})$, where $m$ is a positive integer. For this distribution, $A(i,j)$ is a nonnegative integer, $\mathbb{E}[A(i,j)]=\Omega(i,j)$ holds, and $\Omega(i,j)$ should be nonnegative.
  \item $\mathcal{F}$ can be \textbf{Exponential distribution} by letting $A(i,j)\sim \text{Exponential}(\frac{1}{\Omega(i,j)})$. For this distribution, $A(i,j)$ is nonnegative, $\mathbb{E}[A(i,j)]=\Omega(i,j)$ holds, and $\Omega(i,j)$ should be positive.
  \item $\mathcal{F}$ can be \textbf{Gamma distribution} by letting $A(i,j)\sim \text{Gamma}(a, \frac{\Omega(i,j)}{a})$ (or $A(i,j)\sim \text{Gamma}(\frac{\Omega(i,j)}{a}, a)$) for any $a>0$. For this distribution, $A(i,j)$ is positive, $\mathbb{E}[A(i,j)]=\Omega(i,j)$ holds, and $\Omega(i,j)$ should be positive.
  \item $\mathcal{F}$ can be \textbf{Beta distribution} by letting $A(i,j)\sim \text{Beta}(a\frac{\Omega(i,j)}{1-\Omega(i,j)}, a)$ (or $A(i,j)\sim \text{Beta}(a, a\frac{1-\Omega(i,j)}{\Omega(i,j)})$) for any $a>0$. For this distribution, $A(i,j)\in(0,1)$, $\mathbb{E}[A(i,j)]=\Omega(i,j)$ holds, and $\Omega(i,j)$ should range in $(0,1)$.
  \item $\mathcal{F}$ can be \textbf{Geometric distribution} by letting $A(i,j)\sim \text{Geometric}(\frac{1}{\Omega(i,j)})$. For this distribution, $A(i,j)$ is a positive integer, $\mathbb{E}[A(i,j)]=\Omega(i,j)$ holds, and $\Omega(i,j)$ should be no smaller than 1 given that $\frac{1}{\Omega(i,j)}$ is a probability.
  \item $\mathcal{F}$ can be \textbf{Laplace distribution} (also known as Double exponential distribution) by letting $A(i,j)\sim \text{Laplace}(\Omega(i,j), a)$ for any $a>0$. For this distribution, $A(i,j)\in\mathbb{R}$, $\mathbb{E}[A(i,j)]=\Omega(i,j)$ holds, and $\Omega(i,j)$ can be any finite value.
  \item $\mathcal{F}$ can be \textbf{Lognormal distribution} by letting $A(i,j)\sim \text{Lognormal }(\mathrm{log}(\Omega(i,j))-\frac{a^{2}}{2},a^{2})$ for $a>0$. For this distribution, $A(i,j)>0$, $\mathbb{E}[A(i,j)]=\Omega(i,j)$ holds, and $\Omega(i,j)$ should be positive.
  \item $\mathcal{F}$ can be \textbf{Pareto distribution} by letting $A(i,j)\sim \text{Pareto}(\frac{a-1}{a}\Omega(i,j),a)$ for $a>2$. For this distribution, $A(i,j)>\frac{a-1}{a}\Omega(i,j)$, $\mathbb{E}[A(i,j)]=\Omega(i,j)$ holds, and $\Omega(i,j)$ should be positive.
  \item $\mathcal{F}$ can be \textbf{Logistic distribution} by letting $A(i,j)\sim \text{Logistic}(\Omega(i,j), a)$ for any $a>0$. For this distribution, $A(i,j)\in\mathbb{R}$, $\mathbb{E}[A(i,j)]=\Omega(i,j)$ holds, and $\Omega(i,j)$ can be any finite value.
  \item $\mathcal{F}$ can be \textbf{Weibull distribution} by letting $A(i,j)\sim \text{Weibull}(a, (\frac{\Omega(i,j)}{\Gamma(1+\frac{1}{a})})^a)$ for $a>0$, where $\Gamma(\cdot)$ is the gamma function. For this distribution, $A(i,j)$ is nonnegative, $\mathbb{E}[A(i,j)]=\Omega(i,j)$ holds, and $\Omega(i,j)$ should be positive.
\end{itemize}

Certainly, not all distributions satisfy Equation (\ref{DefinOmega}). Below are some examples:

\begin{itemize}
  \item $\mathcal{F}$ cannot be a degenerate distribution with parameter $a$, where the random variable is constant and equals $a$ with probability 1.
  \item $\mathcal{F}$ cannot be a t-distribution whose mean is 0.
  \item $\mathcal{F}$ cannot be a Cauchy distribution, as its mean does not exist.
  \item $\mathcal{F}$ cannot be a discrete uniform distribution on the first $m$ natural numbers because its expectation, $\frac{m+1}{2}$, is a fixed value that cannot capture the block structure in $\Omega$.
  \item Given that the Hypergeometric distribution describes the probability of obtaining exactly $m$ successes in a sample of $N_2$ items drawn without replacement from a population of $N$ items, where $N_1$ items are considered successes, its mean, $\frac{N_1N_2}{N}$, cannot capture the block structure in $\Omega$. Therefore, $\mathcal{F}$ cannot be a Hypergeometric distribution. Similarly, $\mathcal{F}$ cannot be a Negative-Hypergeometric distribution.
  \item Given that the mean of a Chi-square distribution ($\chi^2$-distribution) with $m$ degrees of freedom is $m$, a positive integer than cannot capture the block structure in $\Omega$. Thus, $\mathcal{F}$ cannot be a $\chi^2$-distribution.
\end{itemize}

Under WDCMM, node $i$'s expectation degree is $\sum_{j=1}^{n}\Omega(i,j)=\theta(i)\Pi(i,:)\sum_{j=1}^{n}P\Pi'(j,:)$. We see that if there is a node $\bar{i}$ such that $\Pi(\bar{i},:)=\Pi(i,:)$ (i.e., node $\bar{i}$ has the same community membership as node $i$) for $\bar{i}\neq i$, the expectation degree of node $\bar{i}$ may still differ from that of node $i$ if $\theta(\bar{i})\neq\theta(i)$, and this is the reason we call $\theta(i)$ heterogeneity parameter.
\begin{rem}\label{IDWDCMM}
By Proposition A.1 of \cite{MixedSCORE}, WDCMM is identifiable when Equations (\ref{DefinePI})-(\ref{definP}) hold, where Proposition A.1 of \cite{MixedSCORE} has no constraint on $A$'s distribution and it allows $P$ to have negative elements as long as $P$ has unit-diagonals. However, if $P$ has non-unit diagonals, WDCMM is unidentifiable unless all entries of $\theta$ are the same.
\end{rem}
\begin{rem}
In this remark, we compare our WDCMM with some previous models.
\begin{itemize}
  \item When $\mathcal{F}$ is Bernoulli distribution, WDCMM reduces to DCMM \cite{MixedSCORE}, i.e., WDCMM extends DCMM from un-weighted networks to weighted networks. Given that DCMM includes several well-known models, including MMSB, DCSBM, SBM, and the Erd\"os-R\'enyi random graph \cite{erdos1960evolution} as special cases, our WDCMM similarly regards these models as special cases. Specifically, by setting $\theta(i) = \sqrt{\rho}$ for all $i \in [n]$, we observe that $\Omega = \rho \Pi P \Pi'$, which aligns with the expectation of $A$ as presented in Equation (1) of \citep{mao2020estimating}. Thus, in this particular case, WDCMM degenerates into MMSB. Here, $\rho > 0$ is known as the sparsity parameter \citep{lei2015consistency,mao2020estimating}, which governs the network's sparsity. In this paper, we will theoretically investigate the influence of $\rho$ on the performance of a spectral method by incorporating $\rho$ into our final theoretical bound.
  \item When all nodes are pure, WDCMM reduces to DCDFM \cite{DCDFM}, i.e., WDCMM extends DCDFM from non-overlapping weighted networks to overlapping weighted networks. If we further assume that all entries of $\theta$ are the same, WDCMM reduces to DFM \cite{qing2021DFM}.
\end{itemize}
\end{rem}
After designing our model WDCMM, a random adjacency matrix $A$ with community membership $\Pi$ can be generated from WDCMM by the following steps.
\begin{itemize}
  \item[Step (a)] Fix $n, K, \Pi, P$, and $\Theta$ satisfying conditions in Equations (\ref{DefinePI})-(\ref{thetapositive}).
  \item[Step (b)] Compute $\Omega=\Theta \Pi P\Pi'\Theta$ by Equation (\ref{DefinOmega}).
  \item[Step (c)] For $1\leq i\leq j\leq n$, let $A(i,j)$ be a random variable generated from distribution $\mathcal{F}$ with expectation $\Omega(i,j)$. Set $A(j,i)=A(i,j)$ since we only consider un-directed weighted networks. If we do not consider self-connected nodes, let $A$'s diagonal entries be zeros.
\end{itemize}

After generating the random adjacency matrix $A$ by Steps (a)-(c), in this article, we aim to answer the following questions:
\begin{itemize}
  \item[Q (1)] Can we design an algorithm to estimate $\Pi$ with known $A$ and $K$ when $A$ is generated from arbitrary distribution $\mathcal{F}$ satisfying Equation (\ref{DefinOmega}) under our WDCMM?
  \item[Q (2)] When there exists an algorithm to estimate $\Pi$ such that Q (1) is solved, let $\hat{\Pi}$ be the estimated membership matrix. For any distribution $\mathcal{F}$ satisfying Equation (\ref{DefinOmega}), does $\hat{\Pi}$ concentrate around $\Pi$ or can we obtain a theoretical upper bound of the difference between $\hat{\Pi}$ and $\Pi$? Does the upper bound enjoy consistency such that it goes to zero as the network size $n$ goes to infinity?
  \item[Q (3)] Can we design a method to determine the number of communities $K$ for weighted networks generated from WDCMM?
\end{itemize}

In the next few sections, we will answer the above questions by applying an efficient spectral algorithm to fit WDCMM, building a theoretical guarantee of consistency for the algorithm under WDCMM, and proposing overlapping weighted modularity to infer $K$.
\section{Algorithm}\label{SecAlgorithm}
In this section, we aim to answer Q (1) by designing an efficient spectral algorithm to fit WDCMM. Because $\Omega$ has a low dimensional structure with rank $K$ when $\mathrm{rank}(\Pi)=K, \mathrm{rank}(P)=K$, and all diagonal entries of $\Theta$ are positive, to answer Q1, one natural way to design an algorithm to fit WDCMM is from the oracle case with known $\Omega$. A good algorithm fitting WDCMM should satisfy the following nice property: it returns $\Pi$ exactly when using $\Omega$ to replace $A$. Otherwise, if the algorithm can not return $\Pi$ when the input matrix is $\Omega$, it is hard to say that it fits the model well.

\begin{figure}
\centering
{\includegraphics[width=0.666666\textwidth]{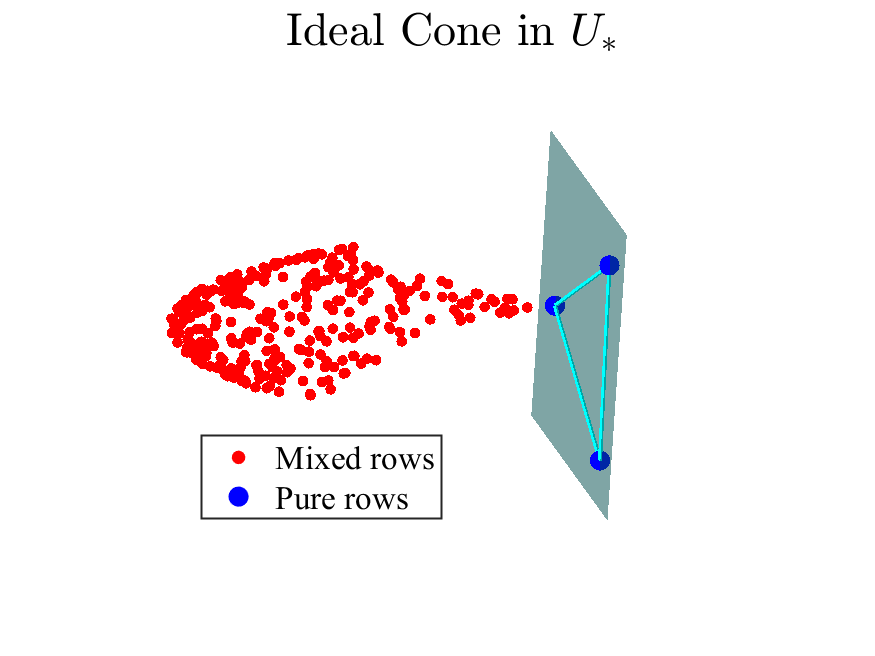}}
\caption{Illustration of the Ideal Cone structure embedded within $U_{*}$ for the case when $K=3$. Here, dots represent the rows of $U_{*}$, while the hyperplane is constituted by the three rows of $U_{*}(\mathcal{I},:)$. Notably, all mixed rows of $U_{*}$ are located on one side of this hyperplane, where we call $U_{*}(i,:)$ a mixed row if the corresponding node $i$ is a mixed node and a pure row otherwise. In this graphical depiction, for every mixed node $i$, we assign its mixed membership as $\Pi(i,1)=r_{1}$, $\Pi(i,2)=r_{2}$, and $\Pi(i,3)=1-r_{1}-r_{2}$, where $r_{1}$ and $r_{2}$ are determined by $\frac{\mathrm{rand}(1)}{2}$, with $\mathrm{rand}(1)$ being a random value drawn from the $\mathrm{Uniform}(0,1)$ distribution. For visualization, these points have been projected and rotated from $\mathbb{R}^{3}$ into $\mathbb{R}^{2}$.}
\label{ICUstar}
\end{figure}

To design an algorithm from the oracle case, because $\Omega$ has a low dimensional structure with $K$ nonzero eigenvalues, we let $\Omega=U\Lambda U'$ be the eigendecomposition of $\Omega$ such that $U\in\mathbb{R}^{n\times K}, \Lambda\in \mathbb{R}^{K\times K}$, and $U'U=I_{K}$, where $\Lambda$'s $k$-th diagonal entry is the leading $k$ eigenvalue of $\Omega$ for $k\in[K]$. From now on, we aim at recovering the mixed membership matrix $\Pi$ exactly from $\Omega$'s eigendecomposition $U\Lambda U'$. To achieve this goal, we let $U_{*}$ be an $n\times K$ matrix such that its $i$-th row is $U_{*}(i,:)=\frac{U(i,:)}{\|U(i,:)\|_{F}}$ for $i\in[n]$. Let $N_{U}$ be an $n\times n$ diagonal matrix such that its $i$-th diagonal entry is $N_{U}(i,i)=\frac{1}{\|U(i,:)\|_{F}}$ for $i\in[n]$, i.e., we have $U_{*}=N_{U}U$. Lemma 3 of \cite{Criterion} is a distribution-free theoretical result and it shows that there exists a matrix $Y\in\mathbb{R}^{n\times K}_{+}$ such that
\begin{align}\label{eq1}
U_{*}=YU_{*}(\mathcal{I},:),
\end{align}
where $Y=N_{M}\Pi\Theta^{-1}(\mathcal{I},\mathcal{I})N^{-1}_{U}(\mathcal{I},\mathcal{I})$ and $N_{M}$ is a diagonal matrix with all diagonal elements being positive. Given that no row of $Y$ is 0, $U_{*}(\mathcal{I},:)$ corresponds to $U_{*}$'s $K$ rows, and each row of $U_{*}$ has a unit $l_{2}$ norm, as stated in Problem 1 in \citep{MaoSVM}, the structure $U_{*}=YU_{*}(\mathcal{I},:)$ is known as Ideal Cone \cite{MaoSVM}. Figure \ref{ICUstar} illustrates the Ideal Cone structure inherent in $U_{*}$ when there are three communities. We observe that the mixed rows of $U_{*}$ corresponding to mixed nodes are located at one side of the hyperplane formed by the $K$ pure rows of $U_{*}$ corresponding to pure nodes.

By applying Equation (\ref{eq1}), we derive
\begin{align}\label{eq2}
U_{*}U^{-1}_{*}(\mathcal{I},:)N_{U}(\mathcal{I},\mathcal{I})\Theta(\mathcal{I},\mathcal{I})=N_{M}\Pi.
\end{align}

Recalling that $U_{*}=N_{U}U$, and combing it with Equation (\ref{eq2}), we obtain
\begin{align}\label{eq3}
UU^{-1}_{*}(\mathcal{I},:)N_{U}(\mathcal{I},\mathcal{I})\Theta(\mathcal{I},\mathcal{I})=N^{-1}_{U}N_{M}\Pi.
\end{align}

In Equation (\ref{eq3}), given that both $N_{U}$ and $N_{M}$ are diagonal matrices, $\Pi$ can be immediately recovered from $U$ and $\mathcal{I}$ once $\Theta(\mathcal{I},\mathcal{I})$ is known. To determine this term, we note that under the WDCMM model, $\Omega=\Theta \Pi P\Pi'\Theta=U\Lambda U'$. Consequently, $\Omega(\mathcal{I},\mathcal{I})=\Theta(\mathcal{I},\mathcal{I})\Pi(\mathcal{I},:)P\Pi'(\mathcal{I},:)\Theta(\mathcal{I},\mathcal{I})=U(\mathcal{I},:)\Lambda U'(\mathcal{I},:)$. Recall that $\Pi(\mathcal{I},:)$ is a $K$-by-$K$ identity matrix, it follows that $\Theta(\mathcal{I},\mathcal{I})P\Theta(\mathcal{I},\mathcal{I})=U(\mathcal{I},:)\Lambda U'(\mathcal{I},:)$. Recall that the connectivity matrix $P$ is required to have unit diagonals in Equation (\ref{definP}) for the model's identifiability, combing this unit-diagonal condition with $\Theta(\mathcal{I},\mathcal{I})P\Theta(\mathcal{I},\mathcal{I})=U(\mathcal{I},:)\Lambda U'(\mathcal{I},:)$ immediately yields $\Theta(\mathcal{I},\mathcal{I})=\sqrt{\mathrm{diag}(U(\mathcal{I},:)\Lambda U'(\mathcal{I},:))}$. This indicates that a portion of the degree heterogeneity matrix $\Theta$ can be expressed using a portion of the eigendecomposition of $\Omega$. Now, substituting $\sqrt{\mathrm{diag}(U(\mathcal{I},:)\Lambda U'(\mathcal{I},:))}$ for $\Theta(\mathcal{I},\mathcal{I})$ in Equation (\ref{eq3}), we obtain
\begin{align}\label{eq4}
N^{-1}_{U}N_{M}\Pi=UU^{-1}_{*}(\mathcal{I},:)N_{U}(\mathcal{I},\mathcal{I})\sqrt{\mathrm{diag}(U(\mathcal{I},:)\Lambda U'(\mathcal{I},:))}.
\end{align}

Given that $N_{U}(\mathcal{I},\mathcal{I})$ is a diagonal matrix and $U_{*}=N_{U}U$, it follows that
\begin{align*}
N_{U}(\mathcal{I},\mathcal{I})\sqrt{\mathrm{diag}(U(\mathcal{I},:)\Lambda U'(\mathcal{I},:))}=\sqrt{\mathrm{diag}(U_{*}(\mathcal{I},:)\Lambda U'_{*}(\mathcal{I},:))},
\end{align*}
which simplifies Equation (\ref{eq4}) to:
\begin{align}\label{eq5}
N^{-1}_{U}N_{M}\Pi=UU^{-1}_{*}(\mathcal{I},:)\sqrt{\mathrm{diag}(U_{*}(\mathcal{I},:)\Lambda U'_{*}(\mathcal{I},:))}.
\end{align}

Define $Z=UU^{-1}_{*}(\mathcal{I},:)\sqrt{\mathrm{diag}(U_{*}(\mathcal{I},:)\Lambda U'_{*}(\mathcal{I},:))}$. By Equation (\ref{eq5}), and given that $N^{-1}_{U}N_{M}$ is a diagonal matrix, we immediately deduce
\begin{align}\label{eq6}
\Pi(i,:)=\frac{Z(i,:)}{\|Z(i,:)\|_{1}}\mathrm{~for~}i\in[n].
\end{align}

Thus, we can recover $\Pi$ from $Z$ using Equation (\ref{eq6}) as long as we know the index set $\mathcal{I}$ when $\Omega$ is given, where $Z$ is directly related to the eigendecomposition of $\Omega$ based on Equation (\ref{eq5}). In the aforementioned analysis, we observe that:
\begin{itemize}
  \item Equations (\ref{eq4})-(\ref{eq6}) are valid only when the equality $\Theta(\mathcal{I},\mathcal{I})=\sqrt{\mathrm{diag}(U(\mathcal{I},:)\Lambda U'(\mathcal{I},:))}$ holds. This equality, in turn, depends on the unit-diagonal condition imposed on $P$. This further underscores the significance of the unit-diagonal condition on $P$, as it guarantees the identifiability of the WDCMM model. Indeed, if the unit-diagonal condition fails to hold, it would be impossible to accurately recover $\Pi$ from Equation (\ref{eq6}), rendering the model unidentifiable.
  \item If we disregard $U_{*}$, the row-normalized counterpart of $U$, it becomes impossible to exactly recover $\Pi$ from $U$ due to the degree heterogeneity parameter $\theta$. To mitigate the influence of $\theta$, akin to the methodologies employed in \citep{lei2015consistency, qin2013regularized, MaoSVM}, we proceed to normalize each row of $U$ to have unit $l_{2}$ norms.
  \item Upon analyzing Equations (\ref{eq1})-(\ref{eq6}), readers may observe that we impose no constraints on the elements of the population adjacency matrix $\Omega$, as our focus lies solely on analyzing the structure of $\Omega$'s eigendecomposition. Similarly, readers may notice that apart from the unit-diagonal condition on the connectivity matrix $P$ and the requirement that all elements of $\theta$ must be positive, there are no additional restrictions on $P$ and $\theta$. For instance, the off-diagonal elements of $P$ can be either negative or greater than 1, and the elements of $\theta$ can exceed or fall below 1 in this paper. In essence, we do not impose any constraints on the elements of $\Omega$. These observations underscore the fact that there are no restrictions on the distribution of $A$'s elements, which enhances the generality of our WDCMM model and highlights the advantages of our algorithm, presented later, for weighted networks. In contrast, Mao et al.\citep{MaoSVM} and Jin et al. \citep{MixedSCORE}, who consider the mixed membership estimation task for un-weighted networks, must impose the conditions that all elements of $P$ are non-negative and $0\leq\mathrm{max}_{i,j\in[n]}\Omega(i,j)\leq1$ in their theoretical analysis.
\end{itemize}

Since $U_{*}=YU_{*}(\mathcal{I},:)$ is an Ideal Cone structure, the SVM-cone algorithm designed in \cite{MaoSVM} can be applied to exactly recover $U_{*}(\mathcal{I},:)$ when the inputs are $U_{*}$ and $K$. From the Ideal Cone structure depicted in Figure \ref{ICUstar}, it is evident that the mixed rows in $U_{*}$, which correspond to mixed nodes, are clearly distinguishable from the pure rows in $U_{*}$, which correspond to pure nodes. The SVM-cone algorithm proposed in \citep{MaoSVM} is capable of efficiently identifying the pure nodes within the Ideal Cone structure of $U_{*}$.

\begin{rem}
Let $\mathcal{I}$ and $\bar{\mathcal{I}}$ be two index sets, where $\mathcal{I}=\{s_1,s_2,\ldots,s_K\}$ and $\bar{\mathcal{I}}=\{\bar{s}_1,\bar{s}_2,\ldots,\bar{s}_K\}$ with $s_k$ and $\bar{s}_k$ being pure nodes in the $k$-th community for $k\in[K]$. We now prove the statement $U_{*}(\mathcal{I},:)=U_{*}(\bar{\mathcal{I}},:)$.

First, we define the matrix $M=\Pi\Theta^{-1}(\mathcal{I},\mathcal{I})U(\mathcal{I},:)$. By Lemma 3 of \cite{Criterion}, we know that for any $i\in[n]$, the $(i,i)$-th entry of $N_M$ is given by $N_M(i,i)=\frac{1}{\|M(i,:)\|_F}$. Since $U_*=YU_*(\mathcal{I},:)=N_M\Pi\Theta^{-1}(\mathcal{I},\mathcal{I})N^{-1}_U(\mathcal{I},\mathcal{I})U_*(\mathcal{I},:)$, we see that for any two distinct nodes $i,\bar{i}$ with the same membership (i.e., $\Pi(i,:)=\Pi(\bar{i},:)$), we have $U_*(i,:)=U_*(\bar{i},:)$, which leads to $U_*(\mathcal{I},:)=U_*(\bar{\mathcal{I}},:)$.

This result means that even if the SVM-cone algorithm finds the index set $\bar{\mathcal{I}}$ instead of $\mathcal{I}$, we still obtain the $K\times K$ corner matrix $U_*(\mathcal{I},:)$ since $U_*(\mathcal{I},:)=U_*(\bar{\mathcal{I}},:)$. In other words, for each community, the choice of which pure node $s_k$ is picked into the index set $\mathcal{I}$ is not important as long as it is a pure node for $k\in[K]$ for the algorithm to correctly recover the underlying communities.
\end{rem}

Given $\Omega$ and $K$, we provide the Ideal ScD algorithm below, where S denotes SVM, c denotes cone, and D denotes distribution-free. Input $\Omega, K$. Output: $\Pi$.
\begin{itemize}
  \item Compute the leading $K$ eigendecomposition of $\Omega$ to get $U$ and $\Lambda$. Get $U_{*}$ from $U$.
  \item Apply SVM-cone algorithm \cite{MaoSVM} to $U_{*}$ with $K$ communities to get the index set $\mathcal{I}$.
  \item Set $Z=UU^{-1}_{*}(\mathcal{I},:)\sqrt{\mathrm{diag}(U_{*}(\mathcal{I},:)\Lambda U'_{*}(\mathcal{I},:))}$.
  \item Set $\Pi(i,:)=\frac{Z(i,:)}{\|Z(i,:)\|_{1}}$ for $i\in[n]$.
\end{itemize}

Since the SVM-cone algorithm exactly returns $U_{*}(\mathcal{I},:)$, the Ideal ScD exactly recovers $\Pi$, and this also guarantees the identifiability of our WDCMM.

For the real case when we only know $A$ and $K$, we aim at estimating $\Pi$ when $A$ is generated from any distribution $\mathcal{F}$ satisfying Equation (\ref{DefinOmega}) under our WDCMM. Algorithm \ref{alg:ScD}, which we call ScD, is a natural extension of the Ideal ScD.
\begin{algorithm}
\caption{ScD}
\label{alg:ScD}
\begin{algorithmic}[1]
\Require The adjacency matrix $A\in \mathbb{R}^{n\times n}$ and the number of communities $K$.
\Ensure The estimated $n\times K$ membership matrix $\hat{\Pi}$.
\State Let $\tilde{A}=\hat{U}\hat{\Lambda}\hat{U}'$ be the leading $K$ eigendecomposition of $A$ such that $\hat{\Lambda}$ contains $A$'s leading $K$ eigenvalues. Let $S_{+}=\{i\in[n]: \|\hat{U}(i,:)\|_{F}>0\}$ and $\hat{U}^{+}=\hat{U}(S_{+},:)$. Let $\hat{U}_{*}$ be the row-normalization of $\hat{U}^{+}$ such that $\hat{U}_{*}(i,:)=\frac{\hat{U}^{+}(i,:)}{\|\hat{U}^{+}(i,:)\|_{F}}$ for $i\in S_{+}$.
\State Apply SVM-cone algorithm \cite{MaoSVM} to $\hat{U}_{*}$ with $K$ communities to get the estimated index set $\hat{\mathcal{I}}$.
\State Set $\hat{Z}=\mathrm{max}(0,\hat{U}^{+}\hat{U}^{-1}_{*}(\hat{\mathcal{I}},:)\sqrt{\mathrm{diag}(\hat{U}_{*}(\hat{\mathcal{I}},:)\hat{\Lambda}\hat{U}'_{*}(\hat{\mathcal{I}},:))})$.
\State Estimate $\Pi(i,:)$ by setting $\hat{\Pi}(i,:)=\hat{Z}(i,:)/\|\hat{Z}(i,:)\|_{1}$ for $i\in S_{+}$ and $\hat{\Pi}(i,:)=(\frac{1}{K},\frac{1}{K},\ldots, \frac{1}{K})$ for $i\notin S_{+}$.
\end{algorithmic}
\end{algorithm}
\begin{rem}
In Algorithm \ref{alg:ScD}, we notice that when the network is extremely sparse, leading to a significant number of zeros in the adjacency matrix $A$, some rows of $\hat{U}$ may turn into zero vectors. To tackle this problem, we employ the strategy presented in Algorithm 2 of \citep{lei2015consistency}. Specifically, we estimate mixed memberships exclusively for the non-zero rows and allocate the mixed memberships of the zero rows equally across all communities.
\end{rem}
\begin{rem}\label{BinaryPi}
To obtain binary rather fuzzy community membership, we can threshold each entry of $\hat{\Pi}$ to obtain a binary membership matrix $\hat{\Pi}^{0}\in\{0,1\}^{n\times K}$ such that
\begin{align*}
\hat{\Pi}^{0}(i,k)=\begin{cases}
1& \mbox{when~} \hat{\Pi}(i,k)\geq \delta_{K},\\
0, & \mbox{otherwise},
\end{cases}
\end{align*}
where $\delta_{K}$ is a threshold and we set it as $\frac{1}{K}$ in this paper. Follow a similar definition of $\hat{\Pi}^{0}$, we can define $\Pi^{0}\in\{0,1\}^{n\times K}$ from $\Pi$. The two binary membership matrices will be used in defining our overlapping weighted modularity in Equation (\ref{Modularity}).
\end{rem}
Because $A$'s expectation is $\Omega$ under the WDCMM, $\hat{U}, \hat{\Lambda}, \hat{Z}$, and $\hat{\Pi}$ are good estimations of $U, \Lambda, Z$, and $\Pi$, respectively. Again, ScD is the SVM-cone-DCMMSB algorithm of \cite{MaoSVM} and we re-name the SVM-cone-DCMMSB as ScD to emphasize its distribution-free property. Note that in \cite{MaoSVM}, the ScD algorithm is only used to estimate $\Pi$ when $A$ is generated from a Bernoulli distribution under DCMM while the ScD used in this article can estimate $\Pi$ when $A$ is generated from any distribution $\mathcal{F}$ as long as Equation (\ref{DefinOmega}) holds under our WDCMM. Furthermore, building spectral algorithms via applications of the eigendecomposition of the adjacency matrix or the Laplacian matrix to fit statistical models for un-weighted networks is a popular way in community detection areas, see \cite{rohe2011spectral,lei2015consistency,joseph2016impact,SCORE,MixedSCORE,MaoSVM,mao2020estimating}.

The computationally intensive steps of ScD are the leading $K$ eigendecomposition (i.e., SVD) and SVM-cone. The computational cost of SVD is $O(n^3)$. Because $A$ for real-world networks is usually sparse, applying the power method \cite{newman2006modularity}, the computation complexity for SVD is $O(Kn^{2})$ \cite{SCORE,MixedSCORE}. The complexity of the SVM-cone algorithm is $O(nK^{2})$ \cite{chang2011libsvm,MaoSVM}. Since $K\ll n$ in this article, as a result, the total complexity of ScD is $O(Kn^{2})$. In this paper, the notation $a_{1} = O(a_{2})$ indicates that $a_{1}$ and $a_{2}$ are of the same order, meaning $a_{1}$ grows (or decreases) at the same rate as $a_{2}$ increases (or decreases). For instance, $O(Kn^{2})$ signifies that the running time of the ScD algorithm is dominated by terms that scale as the product of $n^{2}$ and $K$, while $a = O(1)$ implies that $a$ is considered to be a constant.
\begin{rem}
(Comparison to existing approaches) Here, we compare our ScD algorithm with the SVM-cone-DCMMSB algorithm presented in \citep{MaoSVM}, the Mixed-SCORE algorithm developed in \citep{MixedSCORE}, and the OCCAM algorithm introduced in \citep{OCCAM}.
\begin{itemize}
    \item \textbf{SVM-cone-DCMMSB:} As mentioned earlier, our ScD algorithm is essentially the same as the SVM-cone-DCMMSB algorithm due to the following reasons. Our WDCMM model generalizes the DCMM model studied in \citep{MaoSVM} from un-weighted networks to weighted networks. Both algorithms estimate the mixed membership matrix $\Pi$ using the Ideal Cone structure $U_{*} = YU_{*}(\mathcal{I},:)$, where such structure is initially observed in \citep{MaoSVM}. However, a key difference lies in the constraints on the adjacency matrix $A$. Our ScD algorithm has no constraints on $A$ since it focuses on weighted networks, whereas the SVM-cone-DCMMSB algorithm considers binary adjacency matrices as it is designed for un-weighted networks.

    \item \textbf{Mixed-SCORE:} Recall that our WDCMM model extends the DCMM model introduced in \citep{MixedSCORE} to weighted networks, utilizing the row-wise ratios strategy presented in \citep{SCORE, MixedSCORE} to eliminate the effect of $\theta$, one can also develop an efficient spectral method to estimate mixed memberships under the WDCMM model. This method is indeed the Mixed-SCORE algorithm developed in \citep{MixedSCORE}. According to Theorem 1.1 in the latest arXiv version of \citep{MixedSCORE}, Mixed-SCORE can exactly recover $\Pi$ when the input matrix is the population adjacency matrix $\Omega$. Therefore, the Mixed-SCORE algorithm can be applied to fit our WDCMM model. Developing theoretical guarantees for Mixed-SCORE under WDCMM remains an interesting direction for future work.

    \item \textbf{OCCAM:} The OCCAM model introduced in \citep{OCCAM} can be viewed as equivalent to the DCMM model \citep{MixedSCORE}. One might naturally wonder whether the OCCAM algorithm designed in \citep{OCCAM} could also be applied to fit our WDCMM model. However, this is not the case. The OCCAM algorithm employs the K-medians clustering algorithm to identify cluster centers, which prevents it from exactly recovering the mixed membership matrix $\Pi$ when using the population adjacency matrix $\Omega$ to replace the adjacency matrix $A$ in OCCAM. Consequently, this algorithm is not suitable for fitting our WDCMM model.
\end{itemize}
\end{rem}
\section{Asymptotic Consistency}\label{SecMain}
In this section, we aim to answer Q (2) by building a theoretical guarantee on ScD's consistency for any distribution $\mathcal{F}$ satisfying Equation (\ref{DefinOmega}) under the WDCMM model. Set $\tau=\mathrm{max}_{i,j\in[n]}|A(i,j)-\Omega(i,j)|$ and $\gamma=\mathrm{max}_{i,j\in[n]}\frac{\mathrm{Var}(A(i,j))}{\theta(i)\theta(j)}$, where $\mathrm{Var}(A(i,j))=\mathbb{E}[(A(i,j)-\Omega(i,j))^{2}]$ denotes the variance of $A(i,j)$. $\tau$ and $\gamma$ are two parameters related to distribution $\mathcal{F}$. For different distributions $\mathcal{F}$, the upper bounds of $\tau$ and $\gamma$ can be different, see Examples \ref{NormalF}-\ref{SignedF} for detail. We need the following assumption for our theoretical analysis.
\begin{assum}\label{assumeTau}
Under $WDCMM_{n}(K, P, \Pi, \Theta,\mathcal{F})$, assume that $\tau$ is finite or at least finite with high probability.
\end{assum}
In this paper, we do not require \(\tau\) to be a constant; rather, we assume that \(\tau\) is bounded above by some quantity (which may depend on \(n\)) with high probability. This is sufficient for applying the row-wise eigenspace error bound from Theorem 4.2.1 of \citep{chen2021spectral}, which only requires that the entries of \(A - \Omega\) are uniformly bounded in absolute value by some \(\tau\) (not necessarily constant). Although \(\tau\) may grow with \(n\) for certain distributions (e.g., Normal and Poisson), Assumption \ref{assumeTau} ensures that it is finite with high probability. Indeed, Lemmas \ref{NormalTau} and \ref{PoissonTau} in Appendix \ref{ExtraSec} establish that \(\tau\) is bounded with high probability for Normal and Poisson distributions, respectively, while Lemma \ref{TauRoughLemma} provides a general finite-with-high-probability bound of \(\tau\) for any distribution with finite variance.
\begin{assum}\label{assumesparsity}
Under $WDCMM_{n}(K, P, \Pi, \Theta,\mathcal{F})$, assume $\gamma\theta_{\mathrm{max}}\|\theta\|_{1}\geq\tau^{2}\mathrm{log}(n)$.
\end{assum}
Assumption \ref{assumesparsity} provides a lower bound requirement on $\gamma\theta_{\mathrm{max}}\|\theta\|_{1}$ for our theoretical analysis and it functions similar to Assumption 3.1 of \cite{mao2020estimating} and Assumption 1 of \cite{MixedSCORE}. Note that Assumption \ref{assumesparsity} presumes the fulfillment of Assumption \ref{assumeTau}, implying that $\tau$ can be regarded as a constant. We also need the following condition which simplifies our theoretical result.
\begin{con}\label{condition}
Under $WDCMM_{n}(K, P, \Pi, \Theta,\mathcal{F})$, assume $\kappa(P)=O(1), K=O(1), \lambda_{K}(\Pi'\Pi)=O(\frac{n}{K}), \theta_{\mathrm{max}}=O(\sqrt{\rho})$, and $\theta_{\mathrm{min}}=O(\sqrt{\rho})$ for some $\rho>0$.
\end{con}
In Condition \ref{condition}, $\kappa(P)=O(1)$ is mild because it means the connectivity matrix $P$ is well-conditioned, $K=O(1)$ means we only consider a constant number of communities for our theoretical analysis, $\lambda_{K}(\Pi'\Pi)=O(\frac{n}{K})$ means the size of each community is in the same order, and $\theta_{\mathrm{max}}=O(\sqrt{\rho}), \theta_{\mathrm{min}}=O(\sqrt{\rho})$ mean that all entries of $\theta$ are in the same order. So, we see that Condition \ref{condition} is mild. Recall that when $\mathcal{F}$ is Bernoulli distribution ( and all nodes are pure), and $\theta(i)=\sqrt{\rho}$ for $i\in[n]$, WDCMM reduces to the classical MMSB (and SBM) model. For this case, the sparsity parameter $\rho$ governs network's sparsity \cite{lei2015consistency,mao2020estimating}. Meanwhile, Corollary 3.1 in \citep{mao2020estimating} and Assumptions 1-3 in \citep{MixedSCORE} also adopt similar conditions to streamline their theoretical analysis. We are ready to present the main theorem:
\begin{thm}\label{Main}
Under $WDCMM_{n}(K,P,\Pi,\Theta,\mathcal{F})$, let $\hat{\Pi}$ be obtained from Algorithm \ref{alg:ScD}, when Assumption \ref{assumeTau}, Assumption \ref{assumesparsity}, and Condition \ref{condition} hold, there exists a permutation matrix $\mathcal{P}\in\mathbb{R}^{K\times K}$ such that with probability at least $1-o(n^{-5})$, we have
\begin{align*}	\mathrm{max}_{i\in[n]}\|e'_{i}(\hat{\Pi}-\Pi\mathcal{P})\|_{1}=O(\frac{\sqrt{\gamma\mathrm{log}(n)}}{\sigma_{K}(P)\sqrt{\rho n}}).
\end{align*}
\end{thm}
Proof of Theorem \ref{Main} is provided in Appendix \ref{AppendixMain}. In Theorem \ref{Main}, when $P, K, \theta_{\mathrm{min}}$, and distribution $\mathcal{F}$ are fixed, increasing $n$ decreases ScD's error rate. In particular, when $n$ goes to infinity, the theoretical upper bound of the error rate in Theorem \ref{Main} goes to zero, and this guarantees ScD's consistency under our WDCMM. In Theorem \ref{Main}, only parameters $\gamma$ and $\tau$ are directly related to distribution $\mathcal{F}$, so our main theoretical result in Theorem \ref{Main} is general since we do not constrain $A$'s distribution $\mathcal{F}$. In the following examples, we will show that Assumption \ref{assumeTau} is satisfied,  $\gamma$ is finite, and we can obtain a more accurate theoretical upper bound of ScD's error rate by analyzing $\gamma$'s upper bound for some specific distributions.
\begin{Ex}\label{NormalF}
When $\mathcal{F}$ is \textbf{Normal distribution} such that $A(i,j)\sim \mathrm{Normal}(\Omega(i,j),\sigma^{2}_{A})$ for $\sigma_{A}>0$, i.e., $A(i,j)\in\mathbb{R}$, we have $\mathbb{E}[A(i,j)]=\Omega(i,j)$ satisfying Equation (\ref{DefinOmega}), $P$'s non-diagonal elements can be any finite real values because the mean of Normal distribution can be negative, $\Omega(i,j)$'s range is $(-\infty, \infty)$, $\tau$ is finite with high probability by Lemma \ref{NormalTau} in Appendix \ref{ExtraSec}, and $\gamma=\mathrm{max}_{i,j\in[n]}\frac{\mathrm{Var}(A(i,j))}{\theta(i)\theta(j)}\leq\frac{\sigma^{2}_{A}}{\theta^{2}_{\mathrm{min}}}=O(\frac{\sigma^{2}_{A}}{\rho})$ is finite. Setting $\gamma=O(\frac{\sigma^{2}_{A}}{\rho})$ in Theorem \ref{Main} obtains $O(\frac{\sigma_{A}\sqrt{\mathrm{log}(n)}}{\sigma_{K}(P)\rho\sqrt{n}})$, the theoretical upper bound of ScD's error rate when $\mathcal{F}$ is Normal distribution. We see that increasing $\rho$ (or decreasing $\sigma^{2}_{A}$) decreases ScD's error rate for Normal distribution.
\end{Ex}
\begin{Ex}\label{BernoulliF}
When $\mathcal{F}$ is \textbf{Bernoulli distribution} such that $A(i,j)\sim\mathrm{Bernoulli}(\Omega(i,j))$ and our WDCMM degenerates to the DCMM model, i.e., $A(i,j)\in\{0,1\}$, we have $\mathbb{E}[A(i,j)]=\Omega(i,j)$ satisfying Equation (\ref{DefinOmega}), all entries of $P$ should be nonnegative, $\Omega(i,j)$ is a probability in $[0,1]$, and $\tau=1$. For this case, we have $\mathrm{Var}(A(i,j))=\Omega(i,j)(1-\Omega(i,j))\leq \Omega(i,j)=\theta(i)\theta(j)\Pi(i,:)P\Pi'(j,:)\leq\theta(i)\theta(j)P_{\mathrm{max}}$, which suggests that $\gamma=P_{\mathrm{max}}$, a finite number. Setting $\gamma=P_{\mathrm{max}}$ in Theorem \ref{Main} gets $O(\frac{\sqrt{\mathrm{log}(n)}}{\sigma_{K}(P)\sqrt{\rho n}})$, the theoretical upper bound of ScD's error rate. We find that increasing $\rho$ decreases ScD's error rate. Setting $\gamma=P_{\mathrm{max}}$ and $\tau=1$, Assumption \ref{assumesparsity} is $\rho n\geq O(\mathrm{log}(n))$ by Condition \ref{condition}, and it means a lower bound requirement on network sparsity. So $\rho$ controls network sparsity for Bernoulli distribution.
\end{Ex}
\begin{rem}
When $\mathcal{F}$ is Bernoulli distribution, we observe that our sparsity requirement, $\rho n \geq O(\mathrm{log}(n))$, aligns with the sparsity conditions stated in Theorem 3.1 of \citep{lei2015consistency}, Equation (2.9) of \citep{SCORE}, Equation (2.8) in Assumption 2 of \citep{wang2020spectral}, and Assumption 1 of \citep{MixedSCORE}. Notably, such sparsity condition is mild as it merely necessitate that the average node degree increases at a rate faster than $\mathrm{log}(n)$. Meanwhile, our error bound $O(\frac{\sqrt{\mathrm{log}(n)}}{\sigma_{K}(P)\sqrt{\rho n}})$ also matches with that in Equation (F.76) of \citep{MixedSCORE} or the one in Theorem 2.2 in the latest arXiv version of \citep{MixedSCORE} when Condition \ref{condition} holds. This indicates that our proposed method achieves a comparable theoretical guarantee in terms of error rate. Additionally, by the analysis in Remarks 5, 11, and 13 of \citep{Criterion}, we know that the error bounds in Theorem 3.2 of \citep{mao2020estimating} and Theorem 3.2 of \citep{MaoSVM} should multiply a $\log^{\xi}(n)$ term for some constant $\xi > 1$. After multiplying this term, their error bounds become $O\left(\frac{\log^{\xi}(n)}{\sigma_{K}(P)\sqrt{\rho n}}\right)$, which is larger than ours.
\end{rem}
\begin{Ex}\label{PoissonF}
When $\mathcal{F}$ is \textbf{Poisson distribution} such that $A(i,j)\sim \mathrm{Poisson}(\Omega(i,j))$, i.e., $A(i,j)\in\mathbb{N}$, we have $\mathbb{E}[A(i,j)]=\Omega(i,j)$ satisfying Equation (\ref{DefinOmega}), $P$'s elements should be nonnegative, $\Omega(i,j)$'s range is $[0,\infty)$, $\tau$ is finite with high probability by Lemma \ref{PoissonTau} in Appendix \ref{ExtraSec}, and $\gamma=\mathrm{max}_{i,j\in[n]}\frac{\mathrm{Var}(A(i,j))}{\theta(i)\theta(j)}=\mathrm{max}_{i,j\in[n]}\frac{\Omega(i,j)}{\theta(i)\theta(j)}\leq P_{\mathrm{max}}$, i.e., $\gamma$ is finite. Setting $\gamma=P_{\mathrm{max}}$ in Theorem \ref{Main} gets $O(\frac{\sqrt{\mathrm{log}(n)}}{\sigma_{K}(P)\sqrt{\rho n}})$, the theoretical result for Poisson distribution. We find that increasing $\rho$ decreases ScD's error rate. 
\end{Ex}
\begin{Ex}\label{UniformF}
When $\mathcal{F}$ is \textbf{Uniform distribution} such that $A(i,j)\sim\mathrm{Uniform}(0,2\Omega(i,j))$ for $i,j\in[n]$, i.e., $A(i,j)\in(0,2\Omega(i,j))$, we have $\mathbb{E}[A(i,j)]=\frac{0+2\Omega(i,j)}{2}=\Omega(i,j)$ satisfying Equation (\ref{DefinOmega}), $P$'s entries should be nonnegative, $\Omega(i,j)$'s range is $[0,+\infty)$, $\tau$'s upper bound is $2\theta^{2}_{\mathrm{max}}P_{\mathrm{max}}$, and $\gamma=\mathrm{max}_{i,j\in[n]}\frac{\mathbb{E}[(A(i,j)-\Omega(i,j))^{2}]}{\theta(i)\theta(j)}=\mathrm{max}_{i,j\in[n]}\frac{4\Omega^{2}(i,j)}{12\theta(i)\theta(j)}\leq\mathrm{max}_{i,j\in[n]}\frac{(\theta(i)\theta(j)P_{\mathrm{max}})^{2}}{3\theta(i)\theta(j)}=O(\frac{\rho P^{2}_{\mathrm{max}}}{3})$. Setting $\gamma=O(\frac{\rho P^{2}_{\mathrm{max}}}{3})$ in Theorem \ref{Main}, we find that $\rho$ disappears in $O(\frac{\sqrt{\mathrm{log}(n)}}{\sigma_{K}(P)\sqrt{n}})$ (the theoretical upper bound of ScD's error rate), which indicates that increasing $\rho$ has almost no significant impact on the behavior of ScD.
\end{Ex}
\begin{Ex}\label{SignedF}
WDCMM can also model overlapping \textbf{signed networks} by setting $\mathbb{P}(A(i,j)=1)=\frac{1+\Omega(i,j)}{2}$ and $\mathbb{P}(A(i,j)=-1)=\frac{1-\Omega(i,j)}{2}$, i.e., $A(i,j)\in\{1,-1\}$ for $i\neq j$. For signed networks, $\mathbb{E}[A(i,j)]=\Omega(i,j)$ holds by basic computation, $P$'s non-diagonal elements can be negative, $\Omega(i,j)$'s range is $[-1,1]$ because $\frac{1+\Omega(i,j)}{2}$ and $\frac{1-\Omega(i,j)}{2}$ are probabilities, $\tau$ is $2$, and $\mathrm{Var}(A(i,j))=1-\Omega^{2}(i,j)\leq1$, i.e., $\gamma\leq\frac{1}{\theta^{2}_{\mathrm{min}}}=O(\frac{1}{\rho})$ is finite. Setting $\gamma=O(\frac{1}{\rho})$ in Theorem \ref{Main} obtains $O(\frac{\sqrt{\mathrm{log}(n)}}{\sigma_{K}(P)\rho\sqrt{n}})$, the theoretical results for signed networks. We see that increasing $\rho$ decreases ScD's error rate.
\end{Ex}
Generally speaking, benefiting from the distribution-free property of our WDCMM, $\mathcal{F}$ can be any distribution as long as Equation (\ref{DefinOmega}) holds, as previously discussed in the analysis following Definition \ref{DefinWDCMM}. Furthermore, by adopting a similar approach (i.e., analyzing $\tau$'s finiteness and $\gamma$'s upper bound under different distributions) to the analyses presented in Examples \ref{NormalF}-\ref{SignedF}, we can delve deeper into the theoretical upper bound of ScD's error rate for various other distributions listed after Definition \ref{DefinWDCMM}. For brevity, we list these theoretical results for different distributions in Table \ref{table-bounds}. Following a similar analysis to Examples \ref{NormalF}-\ref{SignedF}, we can analyze ScD's performance under different distributions. To save space, we omit the detailed analysis here.
\begin{table}[ht]
\centering
\scriptsize
\resizebox{\columnwidth}{!}{
\begin{tabular}{cccccccc}
\hline
Distribution $\mathcal{F}$&$A(i,j)$'s range&$\Omega(i,j)$'s range&$\tau$'s upper bound& $\gamma$'s upper bound&Theoretical upper bound of ScD' error rate\\
\hline
$\mathrm{Normal}(\Omega(i,j),\sigma^{2}_{A})$ for $\sigma_{A}>0$&$\mathbb{R}$&$\mathbb{R}$&FWHP&$O(\frac{\sigma^{2}_{A}}{\rho})$&$O(\frac{\sigma_{A}\sqrt{\mathrm{log}(n)}}{\sigma_{K}(P)\rho\sqrt{n}})$\\
$\mathrm{Bernoulli}(\Omega(i,j))$&$\{0,1\}$&$[0,1]$&1&$P_{\mathrm{max}}$&$O(\frac{\sqrt{\mathrm{log}(n)}}{\sigma_{K}(P)\sqrt{\rho n}})$\\
$\mathrm{Poisson}(\Omega(i,j))$&$\mathbb{N}$&$\mathbb{R}_{+}$&FWHP&$P_{\mathrm{max}}$&$O(\frac{\sqrt{\mathrm{log}(n)}}{\sigma_{K}(P)\sqrt{\rho n}})$\\
$\mathrm{Uniform}(0,2\Omega(i,j))$&$(0,2\Omega(i,j))$&$\mathbb{R}_{+}$&$2\theta^{2}_{\mathrm{max}}P_{\mathrm{max}}$&$O(\frac{\rho P^{2}_{\mathrm{max}}}{3})$&$O(\frac{\sqrt{\mathrm{log}(n)}}{\sigma_{K}(P)\sqrt{n}})$\\
Signed network&$\{-1,1\}$&$[-1,1]$&2&$O(\frac{1}{\rho})$&$O(\frac{\sqrt{\mathrm{log}(n)}}{\sigma_{K}(P)\rho\sqrt{n}})$\\
$\mathrm{Binomial}(m,\frac{\Omega(i,j)}{m})$&$\{0,1,2\ldots,m\}$&$[0,m]$&$m$&$P_{\mathrm{max}}$&$O(\frac{\sqrt{\mathrm{log}(n)}}{\sigma_{K}(P)\sqrt{\rho n}})$\\
$\mathrm{Negative~binomial}(m,\frac{m}{m+\Omega(i,j)})$&$\mathbb{N}$&$\mathbb{R}_{+}$&FWHP&$O(\frac{\rho P^{2}_{\mathrm{max}}}{m})$&$O(\frac{\sqrt{\mathrm{log}(n)}}{\sigma_{K}(P)\sqrt{n}})$\\
$\mathrm{Exponential}(\frac{1}{\Omega(i,j)})$&$\mathbb{R}_{+}$&$\mathbb{R}_{>0}$&FWHP&$O(\rho P^{2}_{\mathrm{max}})$&$O(\frac{\sqrt{\mathrm{log}(n)}}{\sigma_{K}(P)\sqrt{n}})$\\
$\mathrm{Gamma}(a,\frac{\Omega(i,j)}{a})$ for $a>0$&$\mathbb{R}_{>0}$&$\mathbb{R}_{>0}$&FWHP&$O(\frac{\rho P^{2}_{\mathrm{max}}}{a})$&$O(\frac{\sqrt{\mathrm{log}(n)}}{\sigma_{K}(P)\sqrt{n}})$\\
$\mathrm{Gamma}(\frac{\Omega(i,j)}{a},a)$ for $a>0$&$\mathbb{R}_{>0}$&$\mathbb{R}_{>0}$&FWHP&$aP_{\mathrm{max}}$&$O(\frac{\sqrt{\mathrm{log}(n)}}{\sigma_{K}(P)\sqrt{\rho n}})$\\
$\mathrm{Beta}(\frac{a\Omega(i,j)}{1-\Omega(i,j)},a)$ for $a>0$&$(0,1)$&$(0,1)$&1&$\frac{P_{\mathrm{max}}}{a}$&$O(\frac{\sqrt{\mathrm{log}(n)}}{\sigma_{K}(P)\sqrt{\rho n}})$\\
$\mathrm{Beta}(a,\frac{a(1-\Omega(i,j))}{\Omega(i,j)})$ for $a>0$&$(0,1)$&$(0,1)$&1&$O(\frac{\rho P^{2}_{\mathrm{max}}}{a})$&$O(\frac{\sqrt{\mathrm{log}(n)}}{\sigma_{K}(P)\sqrt{n}})$\\
$\mathrm{Geometric}(\frac{1}{\Omega(i,j)})$&$\mathbb{N}_{+}$&$[1,+\infty)$&FWHP&$O(\rho P^{2}_{\mathrm{max}})$&$O(\frac{\sqrt{\mathrm{log}(n)}}{\sigma_{K}(P)\sqrt{n}})$\\
$\mathrm{Laplace}(\Omega(i,j),a)$ for $a>0$&$\mathbb{R}$&$\mathbb{R}$&FWHP&$O(\frac{a^{2}}{\rho})$&$O(\frac{\sqrt{\mathrm{log}(n)}}{\sigma_{K}(P)\rho\sqrt{n}})$\\
$\mathrm{Lognormal}(\mathrm{log}(\Omega(i,j))-\frac{a^{2}}{2},a^{2})$ for $a>0$&$\mathbb{R}_{>0}$&$\mathbb{R}_{>0}$&FWHP&$O(\rho P^{2}_{\mathrm{max}}(\mathrm{exp}(a^{2})-1))$&$O(\frac{\sqrt{\mathrm{log}(n)}}{\sigma_{K}(P)\sqrt{n}})$\\
$\mathrm{Pareto}(\frac{a-1}{a}\Omega(i,j),a)$ for $a>2$&$(\frac{a-1}{a}\Omega(i,j),+\infty)$&$\mathbb{R}_{>0}$&FWHP&$O(\frac{\rho P^{2}_{\mathrm{max}}}{a(a-2)})$&$O(\frac{\sqrt{\mathrm{log}(n)}}{\sigma_{K}(P)\sqrt{n}})$\\
$\mathrm{Logistic}(\Omega(i,j),a)$ for $a>0$&$\mathbb{R}$&$\mathbb{R}$&FWHP&$O(\frac{a^{2}}{\rho})$&$O(\frac{\sqrt{\mathrm{log}(n)}}{\sigma_{K}(P)\rho\sqrt{n}})$\\
$\mathrm{Weibull}(a,(\frac{\Omega(i,j)}{\Gamma(1+\frac{1}{a})})^{a})$ for $a>0$&$\mathbb{R}_{+}$&$\mathbb{R}_{>0}$&FWHP&$O(\frac{\rho P^{2}_{\mathrm{max}}\Gamma(1+\frac{2}{a})}{\Gamma^{2}(1+\frac{1}{a})})$&$O(\frac{\sqrt{\mathrm{log}(n)}}{\sigma_{K}(P)\sqrt{n}})$\\
\hline
\end{tabular}
}
\caption{Theoretical results for different distributions under the WDCMM model, where FWHP stands for ``finite with high probability". Here, $i$ and $j$ range in $[n]$. The FWHP property of $\tau$ is guaranteed by Lemma \ref{NormalTau} for the Normal distribution, Lemma \ref{PoissonTau} for the Poisson distribution, and Lemma \ref{TauRoughLemma} for other distributions, as detailed in Appendix \ref{ExtraSec}.}
\label{table-bounds}
\end{table}

By Examples \ref{NormalF}, \ref{UniformF}, and \ref{SignedF}, we find that the non-diagonal elements of $A$ are always nonzero, which suggests that there is always a link between any two nodes. However, this is impractical for real-world large-scale networks in which most node pairs have no connections \cite{lei2015consistency}. Treat an edge with weight 0 as a missing edge. We can generate missing edges for networks generated from WDCMM such that $A(i,j)$ is multiplied by $\mathcal{A}(i,j)$ for $i,j\in[n]$, where $\mathcal{A}\in\{0,1\}^{n\times n}$ is a symmetric and connected adjacency matrix for an un-weighted network and it can be generated from any model for un-directed and un-weighted networks such as SBM, DCSBM, MMSB, DCMM, and  Erd\"os-R\'enyi random graph.
\section{Estimation of the Number of Communities}\label{SecK}
In this section, we aim to answer Q (3). Algorithm \ref{alg:ScD} requires a known number of communities $K$.  However, for real-world networks, $K$ is usually unknown, which diminishes the usefulness of a community detection approach. Here, to estimate $K$, we introduce overlapping weighted modularity.

Because $A$ considered in this article may have negative elements, $A$ can be written as $A=A^{+}-A^{-}$, where $A^{+}=\mathrm{max}(0,A)$ and $A^{-}=\mathrm{max}(0,-A)$. Let $d^{+}$ be an $n\times 1$ vector such that  $d^{+}(i)=\sum_{j=1}^{n}A^{+}(i,j)$ for $i\in[n]$. Let $m^{+}=\sum_{i=1}^{n}d^{+}_{i}$. Similarly, we can define $d^{-}$ and $m^{-}$ from $A^{-}$. Let $O$ be an $n\times 1$ vector such that $O(i)$ denotes the number of communities that node $i$ belongs to, i.e., $O(i)=\sum_{k\in[K]}\hat{\Pi}^{0}(i,k)$ for $i\in[n]$, where $\hat{\Pi}^{0}$ is a binary membership matrix computed by Remark \ref{BinaryPi}. Let $\hat{\mathcal{C}}_{k}=\{i:\hat{\Pi}^{0}(i,k)=1,i\in[n]\}$ be the set of nodes belonging to community $k$ for $k\in[K]$. For convenience, let $\hat{\mathcal{C}}$ be the set of the $K$ estimated communities $\hat{\mathcal{C}}_{k}$ for $k\in[K]$. Based on the binary membership matrix $\hat{\Pi}^{0}$, overlapping modularity $Q^{+}$ for $A$'s positive entries and overlapping modularity $Q^{-}$ for $A$'s negative entries are defined as
\begin{align*}
&Q^{+}=\frac{1_{m^{+}>0}}{m^{+}}\sum_{k=1}^{K}\sum_{i\in\hat{\mathcal{C}}_{k},j\in\hat{\mathcal{C}}_{k}}\frac{1}{O(i)O(j)}(A^{+}(i,j)-\frac{d^{+}(i)d^{+}(j)}{m^{+}}),\\
& Q^{-}=\frac{1_{m^{-}>0}}{m^{-}}\sum_{k=1}^{K}\sum_{i\in\hat{\mathcal{C}}_{k},j\in\hat{\mathcal{C}}_{k}}\frac{1}{O(i)O(j)}(A^{-}(i,j)-\frac{d^{-}(i)d^{-}(j)}{m^{-}}),
\end{align*}
where $1_{m^{+}>0}$ is indicator function such that $1_{m^{+}>0}=1$ if $m^{+}>0$ and 0 otherwise. Similar for  $1_{m^{-}>0}$. \cite{newman2002assortative} discovered that social networks often exhibit assortativity, where nodes preferentially connect to others similar to them. Specifically, nodes within the same group tend to have more connections than nodes from different groups \citep{newman2002assortative,newman2003mixing,radicchi2004defining,catanzaro2004assortative,noldus2015assortativity}. Conversely, technological and biological networks often display dis-assortativity, where nodes in the same group have fewer connections than nodes in different groups. Following \cite{amini2018semidefinite}, our WDCMM model can generate assortative weighted networks by setting the maximum off-diagonal element of $P$ smaller than its minimum diagonal element. Conversely, dis-assortative weighted networks can be modeled by setting the minimum off-diagonal element of $P$ larger than its maximum diagonal element. In this paper, we introduce $\eta_{\mathcal{N}}$ as an index to represent the assortativity of a weighted network $\mathcal{N}$, where $\eta_{\mathcal{N}}=1$ indicates an assortative network, and $\eta_{\mathcal{N}}=-1$ indicates a dis-assortative network. Inspired by the concept of weighted modularity for signed networks \cite{gomez2009analysis}, we define our overlapping weighted modularity as follows:
\begin{align}\label{Modularity}
Q_{ovw}=(\frac{m^{+}}{m^{+}+m^{-}}Q^{+}-\frac{m^{-}}{m^{+}+m^{-}}Q^{-})\eta_{\mathcal{N}},
\end{align}
where $ovw$ means ``overlapping weighted". This measure evaluates the quality of overlapping community partition in both assortative and dis-assortative weighted networks. When $\eta_{\mathcal{N}}=1$ (assortative network), $Q_{ovw}=\frac{m^{+}}{m^{+}+m^{-}}Q^{+}-\frac{m^{-}}{m^{+}+m^{-}}Q^{-}$, and it quantifies the quality of overlapping community partitions in assortative networks. When $\eta_{\mathcal{N}}=-1$ (dis-assortative network), $Q_{ovw}=\frac{m^{-}}{m^{+}+m^{-}}Q^{-}-\frac{m^{+}}{m^{+}+m^{-}}Q^{+}$, and it assesses the quality of overlapping community detection in dis-assortative networks. Similar to the Newman-Girvan modularity \cite{newman2004finding,newman2006modularity} and weighted modularity \cite{gomez2009analysis}, a larger overlapping weighted modularity indicates a better community partition. Our overlapping weighted modularity includes several popular modularity as special cases when $\eta_{N}=1$ (assortative network).
\begin{itemize}
  \item When $A$ has both positive and negative entries and $\hat{\Pi}^{0}$  degenerates such that each node only belongs to one community, our overlapping weighted modularity reduces to the modularity for signed networks provided in Equation (17) of \cite{gomez2009analysis}.
  \item When all entries of $A$ are nonnegative (i.e., $m^{-}=0$), our overlapping weighted modularity degenerates to the overlapping modularity provided in Equation (2) of \cite{shen2009detect}.
  \item When $\hat{\Pi}^{0}$ degenerates such that each node only belongs to one community and all entries of $A$ are nonnegative, our overlapping weighted modularity degenerates to the famous Newman-Girvan modularity \cite{newman2004finding,newman2006modularity}.
\end{itemize}

To infer the number of communities, we adopt the suggestion proposed in \cite{nepusz2008fuzzy,qing2024applications}, that is, we iteratively increase $k$ and choose the one maximizing our overlapping weighted modularity in Equation (\ref{Modularity}). We let $\hat{\Pi}^{0}$ be obtained from our ScD algorithm and call our method for inferring $K$ as KScD. Given our newly defined overlapping weighted modularity $Q_{ovw}$ for measuring the quality of overlapping community partitions in assortative and dis-assortative weighted networks, it is natural to question its effectiveness. To address this, we will examine the Accuracy rate (defined later) of our KScD method in estimating $K$ for both assortative and dis-assortative weighted networks in the next section. If our KScD (designed by maximizing $Q_{ovw}$) estimates $K$ with a high Accuracy rate, it indicates that our modularity $Q_{ovw}$ is a reliable measure. Numerical results in the subsequent section confirm that our overlapping weighted modularity $Q_{ovw}$ is indeed a valid measure.
\section{Experiments}\label{SecExperiments}
We conduct extensive experiments to validate our theoretical results, demonstrate the effectiveness of ScD in estimating membership matrix $\Pi$, and demonstrate the effectiveness of our overlapping weighted modularity by showing that KScD can infer $K$ for both assortative and dis-assortative networks generated from WDCMM. The experimental part is implemented with Matlab R2021b on a standard personal computer (Thinkpad X1 Carbon Gen 8).
\subsection{ Baseline Methods}
For the task of estimating memberships, we select three mixed membership community detection approaches as baseline methods of our WDCMM model.
\begin{itemize}
  \item \textbf{GeoNMF} \cite{GeoNMF} adapts nonnegative matrix factorization to estimate mixed memberships for networks generated from MMSB.
  \item \textbf{SPACL} \cite{mao2020estimating} is a spectral algorithm for estimating mixed memberships for networks generated from MMSB.
  \item  \textbf{Mixed-SCORE} \cite{MixedSCORE} and \textbf{OCCAM} \cite{OCCAM} are two spectral methods for estimating mixed memberships for networks generated from DCMM.
\end{itemize}
\begin{rem}\label{GeoNMFMixedSCORE}
GeoNMF, Mixed-SCORE, and OCCAM may fail to output for the case that some elements in the adjacency matrix $A$ are negative. To make them work when $A$ has negative elements, we make all entries of $A$ nonnegative by adding a sufficiently large constant for these three methods.
\end{rem}

For the task of inferring $K$, we compare KScD with the following four methods.
\begin{itemize}
  \item \textbf{KSPACL} is the method to determine $K$ by finding the number of clusters that maximizes our overlapping weighted modularity when $\hat{\Pi}^{0}$ is obtained from SPACL.
  \item \textbf{KMixed-SCORE} is the method to infer $K$ by maximizing our overlapping weighted modularity when $\hat{\Pi}^{0}$ is obtained from Mixed-SCORE.
  \item \textbf{NB} and \textbf{BHac} \cite{le2022estimating} are two methods for inferring the number of communities and they determine $K$ via non-backtracking matrix and Bethe Hessian matrix, respectively.
\end{itemize}
\subsection{Evaluation metrics}
For the task of estimating memberships, we consider the following metrics depending on the fact that whether the ground truth membership matrix $\Pi$ is known.

\textbf{Metrics with ground truth.} For this case, we introduce three widely used evaluation metrics: Hamming error \cite{MixedSCORE} and Relative error \cite{mao2020estimating}.
\begin{itemize}
  \item \textbf{Hamming error} measures the $l_{1}$ difference between $\Pi$ and $\hat{\Pi}$:
      \begin{align*}
\mathrm{Hamming~error=}\mathrm{min}_{\mathcal{P}\in\{ K\times K\mathrm{~permutation~matrix}\}}\frac{1}{n}\|\hat{\Pi}-\Pi\mathcal{P}\|_{1}.
\end{align*}
Hamming error is nonnegative with an upper bound 1, and it is the smaller the better. Note that Theorem \ref{Main} provides a theoretical upper bound of ScD's Hamming error for each node.
  \item \textbf{Relative error} evaluates the relative $l_{2}$ distance between $\Pi$ and $\hat{\Pi}$:
      \begin{align*}
\mathrm{Relative~error}=\mathrm{min}_{\mathcal{P}\in\{ K\times K\mathrm{~permutation~matrix}\}}\frac{\|\hat{\Pi}-\Pi\mathcal{P}\|_{F}}{\|\Pi\|_{F}}.
\end{align*}
A smaller Relative error indicates that $\hat{\Pi}$ is a better estimation of membership matrix $\Pi$.
\item \textbf{Overlapping normalized mutual information (ONMI)} defined in \citep{lancichinetti2009detecting} can also be applied to measure the effectiveness of the proposed methods while comparing with ground truth. Since ONMI works for binary overlapping memberships \citep{OCCAM,MaoSVM} while $\Pi\in[0,1]^{n\times K}$ and $\hat{\Pi}\in[0,1]^{n\times K}$ are continuous memberships in this paper, to compute ONMI, we use $\Pi^{0}$ computed from $\Pi$ and $\hat{\Pi}^{0}$ computed from $\hat{\Pi}$. ONMI is no larger than 1 and a larger ONMI means a better estimation of the membership matrix $\Pi$.
\end{itemize}

\textbf{Metrics without ground truth.} For this case, our overlapping weighted modularity in Equation (\ref{Modularity}) is capable of evaluating the quality of detected communities, and a higher value of overlapping weighted modularity indicates more accurately detected communities.

\textbf{Accuracy rate.} For the task of estimating $K$, we use the Accuracy rate to evaluate the precision of KScD and its competitors, where Accuracy rate is defined as the proportion of times an approach exactly determines $K$ \cite{le2022estimating}.
\subsection{Simulations}
\begin{figure}
\centering
\resizebox{\columnwidth}{!}{
\subfigure[$\eta_{\mathcal{N}}=1$]{\includegraphics[width=0.2\textwidth]{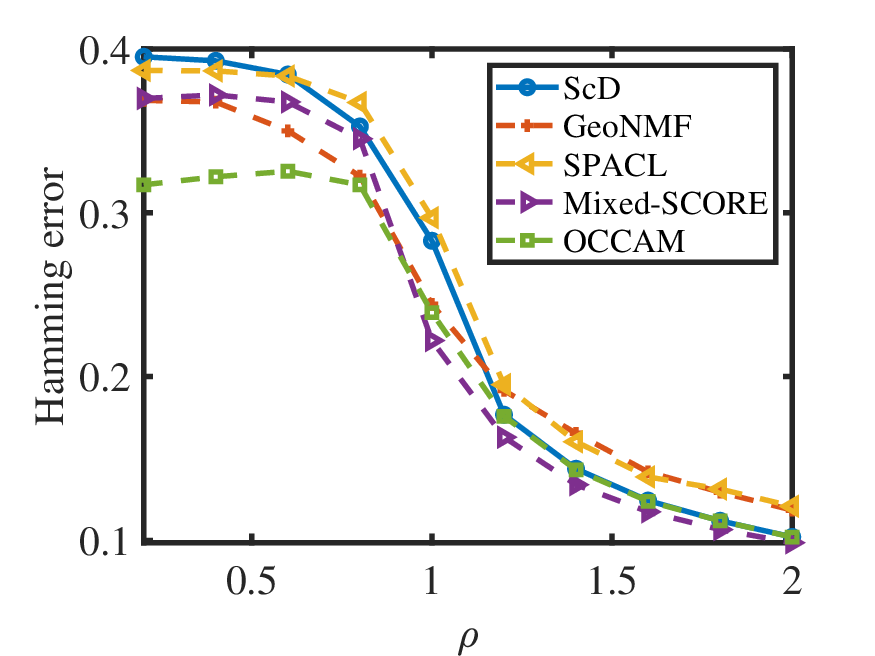}}
\subfigure[$\eta_{\mathcal{N}}=1$]{\includegraphics[width=0.2\textwidth]{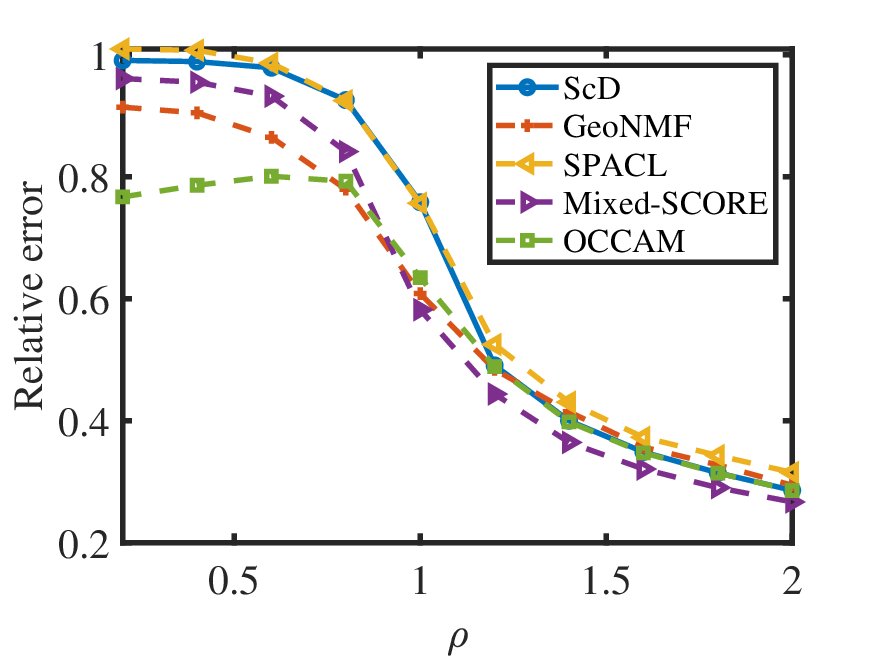}}
\subfigure[$\eta_{\mathcal{N}}=1$]{\includegraphics[width=0.2\textwidth]{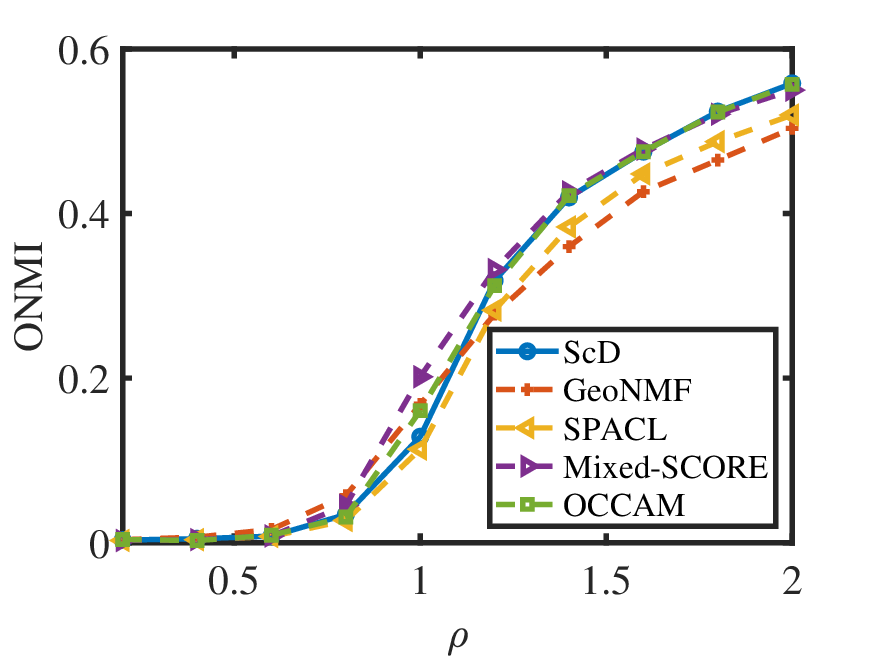}}
\subfigure[$\eta_{\mathcal{N}}=1$]{\includegraphics[width=0.2\textwidth]{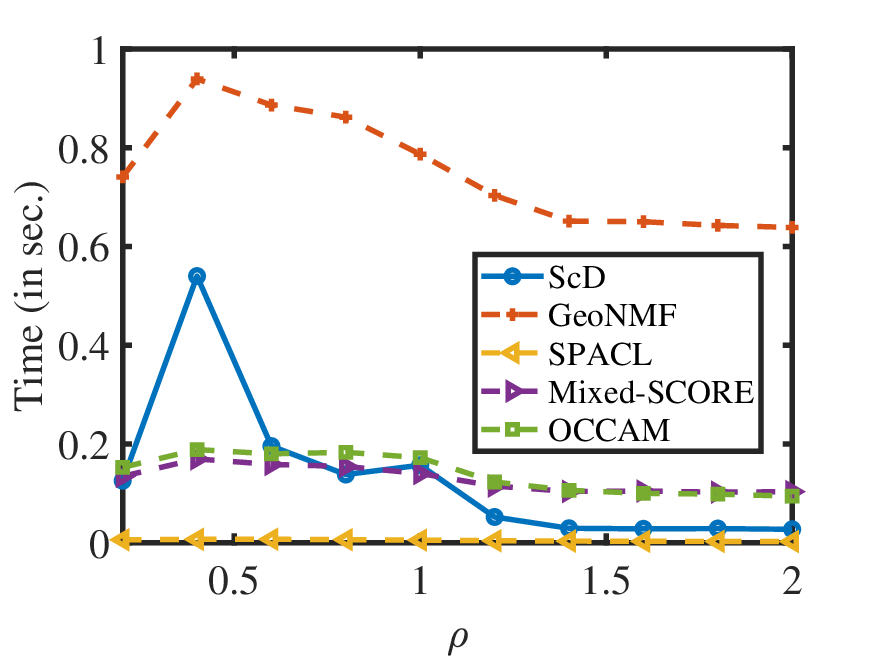}}
\subfigure[$\eta_{\mathcal{N}}=1$]{\includegraphics[width=0.2\textwidth]{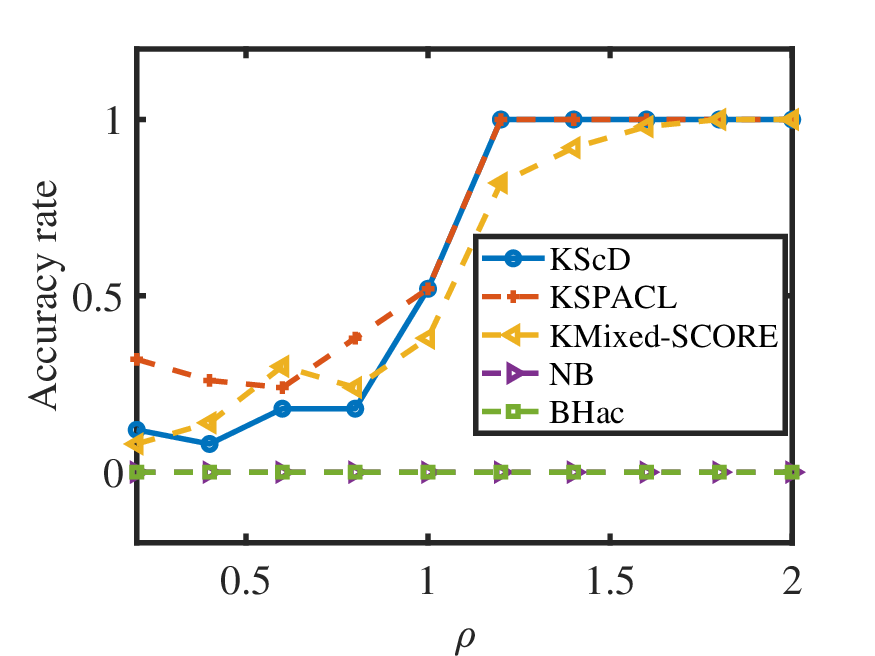}}
}
\resizebox{\columnwidth}{!}{
\subfigure[$\eta_{\mathcal{N}}=-1$]{\includegraphics[width=0.2\textwidth]{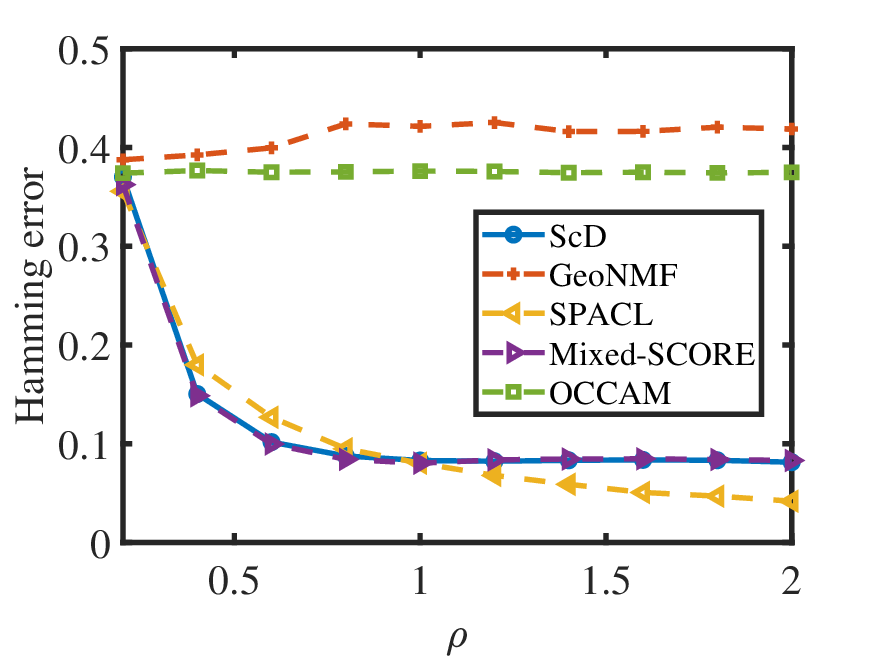}}
\subfigure[$\eta_{\mathcal{N}}=-1$]{\includegraphics[width=0.2\textwidth]{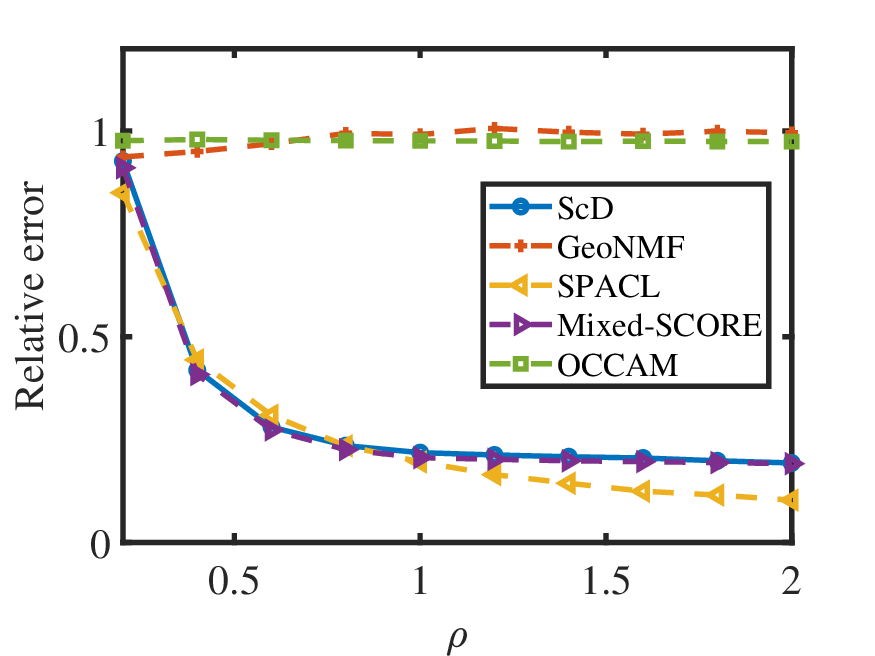}}
\subfigure[$\eta_{\mathcal{N}}=-1$]{\includegraphics[width=0.2\textwidth]{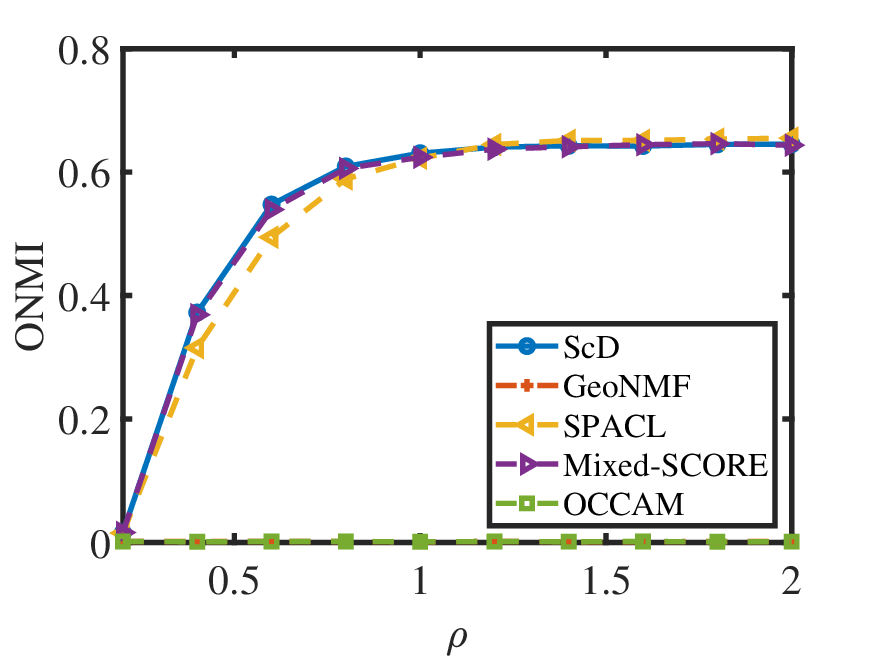}}
\subfigure[$\eta_{\mathcal{N}}=-1$]{\includegraphics[width=0.2\textwidth]{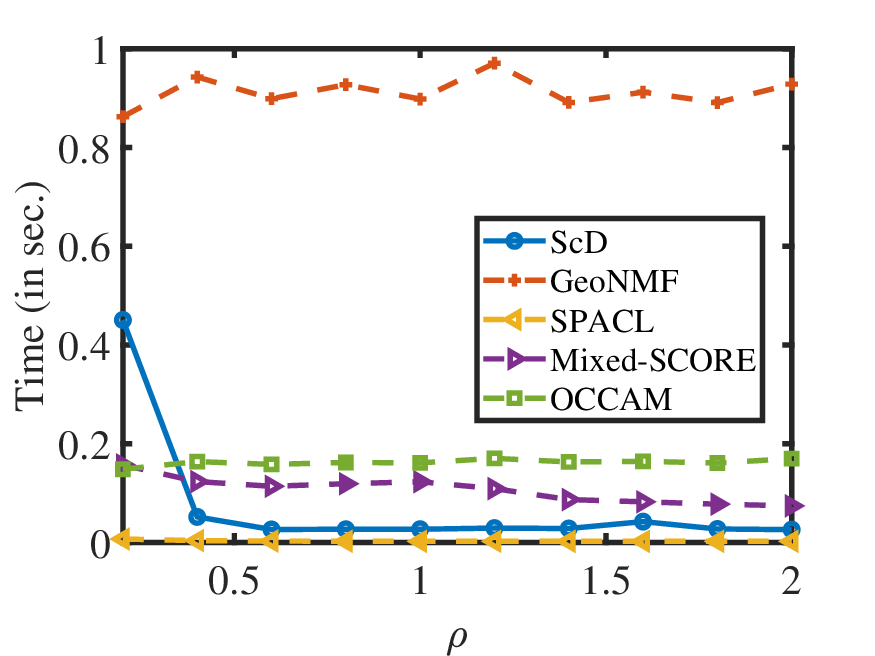}}
\subfigure[$\eta_{\mathcal{N}}=-1$]{\includegraphics[width=0.2\textwidth]{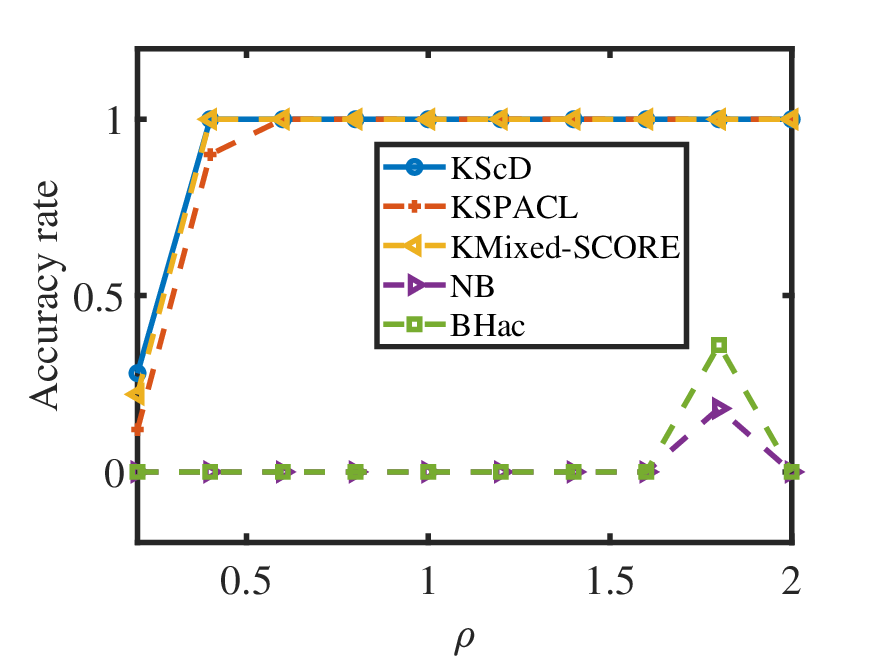}}
}
\caption{Numerical results of Experiment 1.}
\label{EX1} 
\end{figure}
For computer-generated networks, we set $n=400$ and $K=3$. Let each community have $n_{0}$ number of pure nodes. For pure node $i$, let $\Pi(i,k)=1$ if $1+(k-1)n_{0}\leq i\leq kn_{0}$ for $i\in[Kn_{0}], k\in[K]$. Set $n_{0}=40$.  Let all mixed nodes have four different memberships $(0.1, 0.1, 0.8), (0.1, 0.8, 0.1), (0.8, 0.1, 0.1)$, and $(1/3,1/3,1/3)$, each with $\frac{n-3n_{0}}{4}$ number of nodes. Let $P$'s diagonal elements be 1 and non-diagonal elements be $p$. For Normal distribution and signed networks, $p$ can be negative. For other distributions, $p$ should be positive. Let $\theta(i)=\sqrt{\rho}\times (\mathrm{rand}(1)/2+0.5)$, where $\mathrm{rand}(1)$ is a random value ranged in $(0,1)$. In this way, $n, K, P, \Pi$, and $\Theta$ have been set. Then following Steps (a)-(c), we can generate an adjacency matrix $A$ with true membership matrix $\Pi$ for any distribution $\mathcal{F}$ satisfying Equation (\ref{DefinOmega}) under our WDCMM model. For the task of community detection, applying ScD (and its competitors) to $A$ with $K$ communities obtains the estimated membership matrix $\hat{\Pi}$, and then we can obtain Hamming error and Relative error. For each parameter setting considered in this section, we report the averaged Hamming error (Relative error, and ONMI) over 50 repetitions for each method. We also report the average running time for each method over the 50 repetitions. For the task of determining the number of communities, applying KScD (and its competitors) to $A$ obtains the estimated number of communities, and then we can obtain the Accuracy rate over the 50 repetitions.
\begin{rem}
For our synthetic networks, $P, \Pi$, and $\Theta$ can be set arbitrarily as long as Equations (\ref{DefinePI})-(\ref{thetapositive}) hold and $P$'s non-diagonal elements should be positive or nonnegative or can be negative depending on distribution $\mathcal{F}$ just as analyzed in Examples \ref{NormalF}-\ref{SignedF}. Meanwhile, for simulations, $\mathcal{F}$ can be any distribution since our WDCMM is a distribution-free model and it has no constraint on a specific distribution $\mathcal{F}$ as long as Equation (\ref{DefinOmega}) holds. Furthermore, if we consider missing edges, ScD performs better as the number of 1s in $\mathcal{A}$ increases. Therefore, to save space, we do not consider missing edges in the simulations.
\end{rem}
\textbf{Experiment 1: Normal distribution.} Set $\mathcal{F}$ as Normal distribution such that $A(i,j)\sim\mathrm{Normal}(\Omega(i,j),\sigma^{2}_{A})$. Let $\sigma^{2}_{A}=2$ and $\rho$ range in $\{0.1,0.2,\ldots,2\}$. For assortative weighted networks (i.e., $\eta_{\mathcal{N}}=1$), we set $p=-0.2$. For dis-assortative weighted networks (i.e., $\eta_{\mathcal{N}}=-1$), we set $p=5$. The results are in Figure \ref{EX1}. We observe the following: (i) ScD performs better as $\rho$ increases, and this is consistent with our theoretical analysis in Example \ref{NormalF}. (ii) For assortative networks, all methods perform similarly and they all successfully estimate memberships for large $\rho$. For dis-assortative networks, GeoNMF and OCCAM fail to estimate mixed memberships while the other three methods behave similar and perform better when we increase $\rho$. Meanwhile, SPCAL runs slightly faster than ScD, Mixed-SCORE, and OCCAM while GeoNMF runs slowest. (iii) For the task of inferring the number of communities, our methods KScD, KSPACL, and KMixed-SCORE designed based on our overlapping weighted modularity enjoy similar and satisfactory performances because their accuracy rates increase as $\rho$ increases, which indicates the effectiveness of our overlapping weighted modularity for both assortative and dis-assortative networks. For comparison, NB and BHac fail to estimate $K$ for Normal distribution. (iv) We observe that $\mathrm{ONMI}<0.7$ for all methods (this observation is also found in Experiments 2-5). This is acceptable since ONMI works for binary overlapping memberships and we compute it by using $\Pi^{0}$ and $\hat{\Pi}^{0}$, where $\Pi^{0}$ (and $\hat{\Pi}^{0}$) differs from $\Pi$ (and $\hat{\Pi}$), i.e., using ONMI to measure the difference between $\Pi^{0}$ and $\hat{\Pi}^{0}$ can not fully capture the difference between $\Pi$ and $\hat{\Pi}$.
\begin{rem}
For the task of estimating memberships, in our experimental studies, we aim at using simulations to verify our theoretical analysis that ScD has different behaviors for different distributions when $\rho$ is changed as analyzed in Examples \ref{NormalF}-\ref{SignedF}. We also want to investigate the performances of GeoNMF, SPACL, Mixed-SCORE, and OCCAM when they are applied to estimate memberships for weighted networks generated from our model WDCMM instead of showing that ScD outperforms its competitors.
\end{rem}
\begin{figure}
\centering
\resizebox{\columnwidth}{!}{
\subfigure[$\eta_{\mathcal{N}}=1$]{\includegraphics[width=0.2\textwidth]{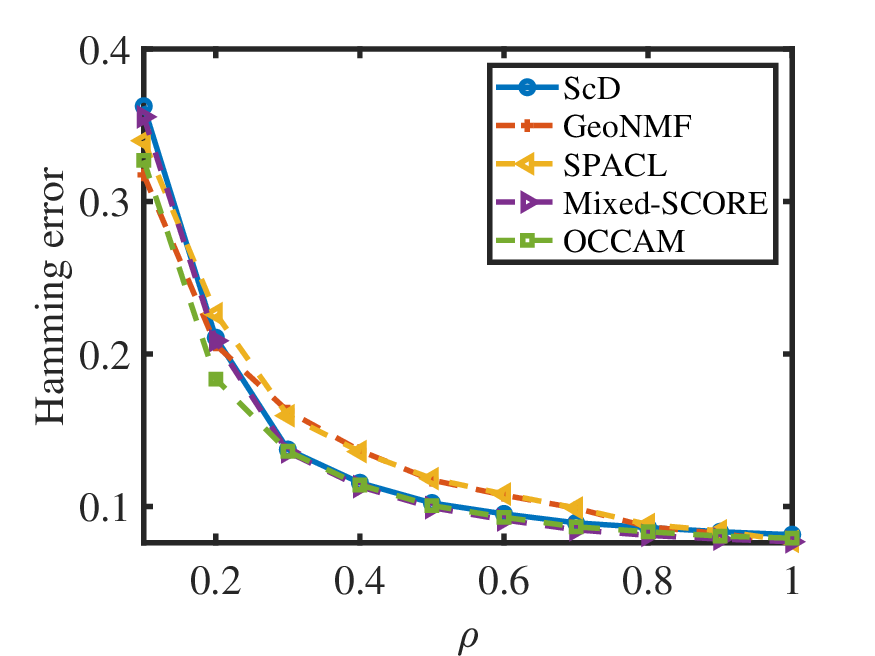}}
\subfigure[$\eta_{\mathcal{N}}=1$]{\includegraphics[width=0.2\textwidth]{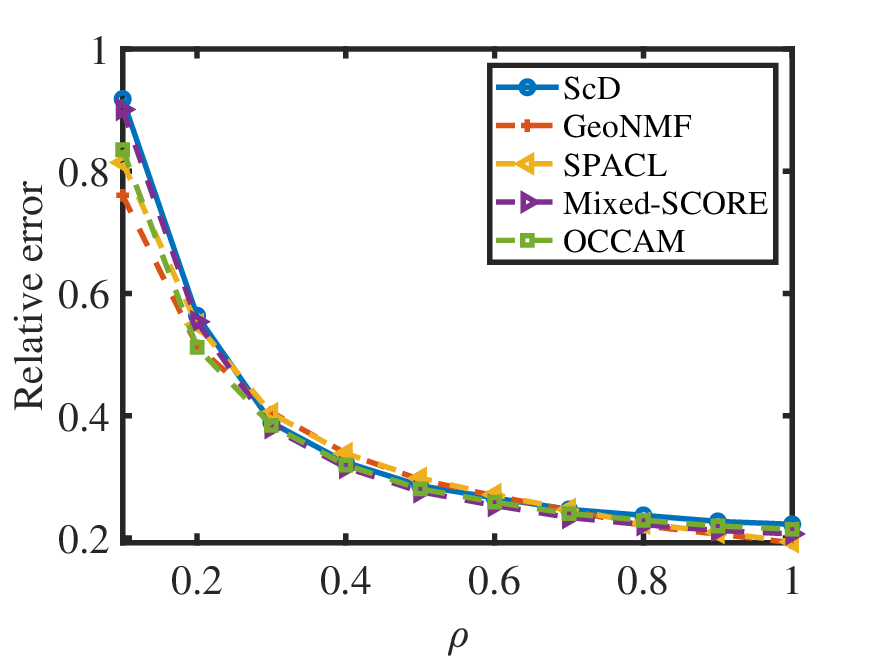}}
\subfigure[$\eta_{\mathcal{N}}=1$]{\includegraphics[width=0.2\textwidth]{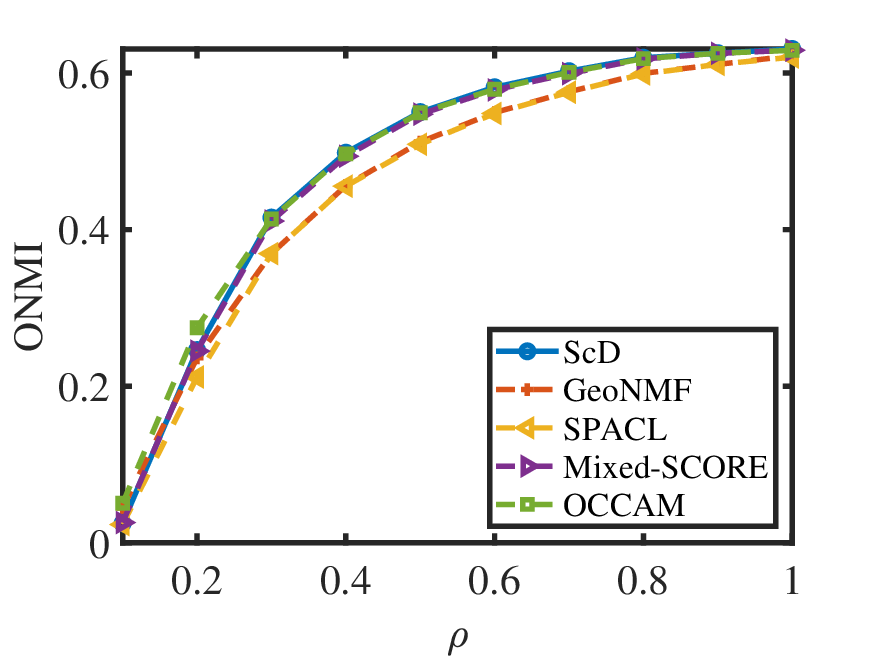}}
\subfigure[$\eta_{\mathcal{N}}=1$]{\includegraphics[width=0.2\textwidth]{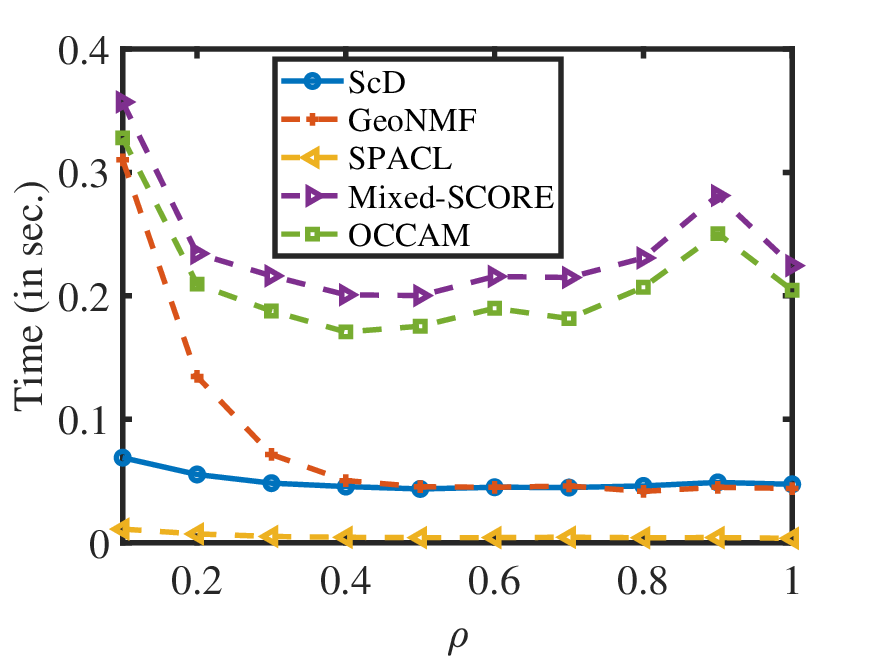}}
\subfigure[$\eta_{\mathcal{N}}=1$]{\includegraphics[width=0.2\textwidth]{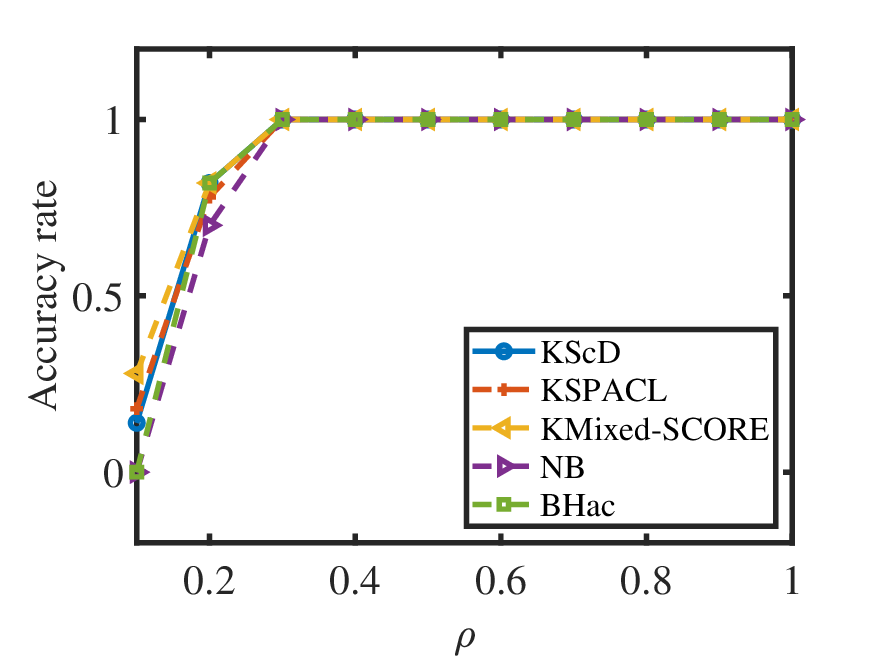}}
}
\resizebox{\columnwidth}{!}{
\subfigure[$\eta_{\mathcal{N}}=-1$]{\includegraphics[width=0.2\textwidth]{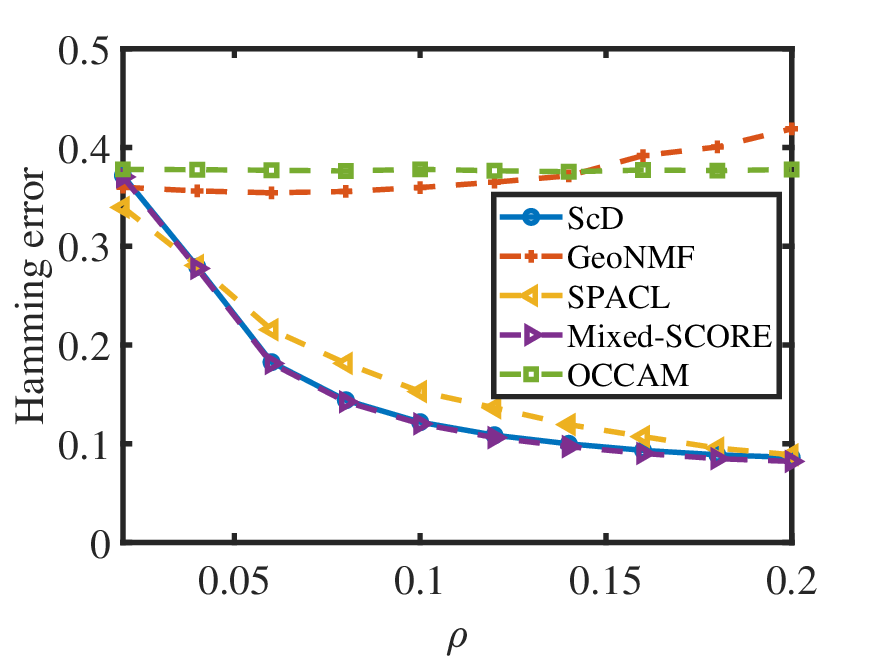}}
\subfigure[$\eta_{\mathcal{N}}=-1$]{\includegraphics[width=0.2\textwidth]{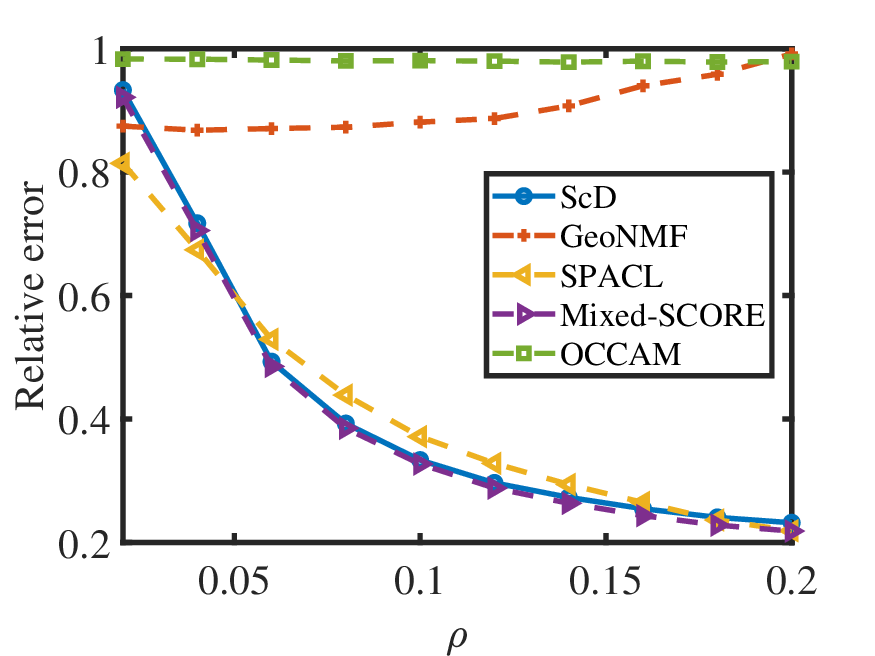}}
\subfigure[$\eta_{\mathcal{N}}=-1$]{\includegraphics[width=0.2\textwidth]{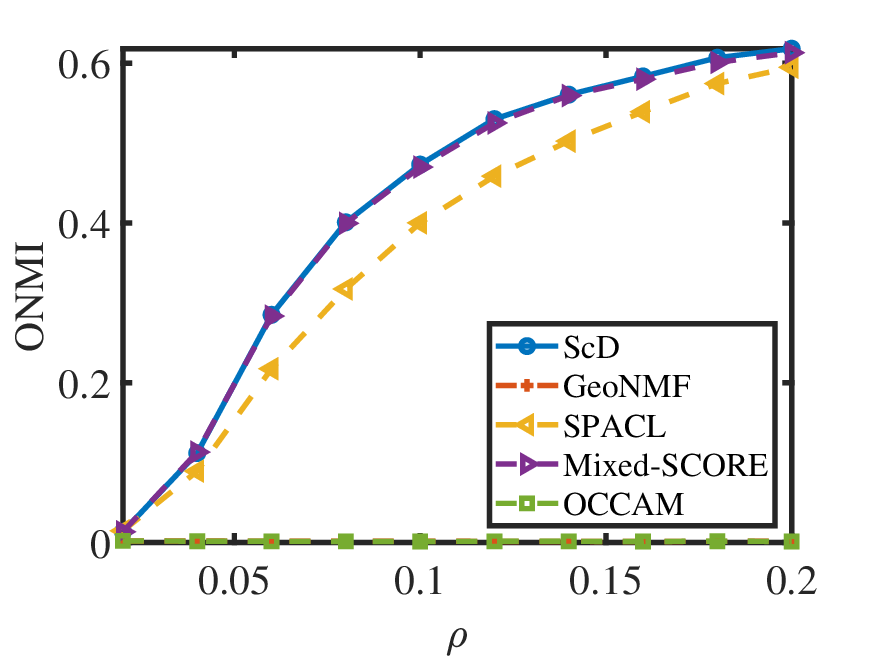}}
\subfigure[$\eta_{\mathcal{N}}=-1$]{\includegraphics[width=0.2\textwidth]{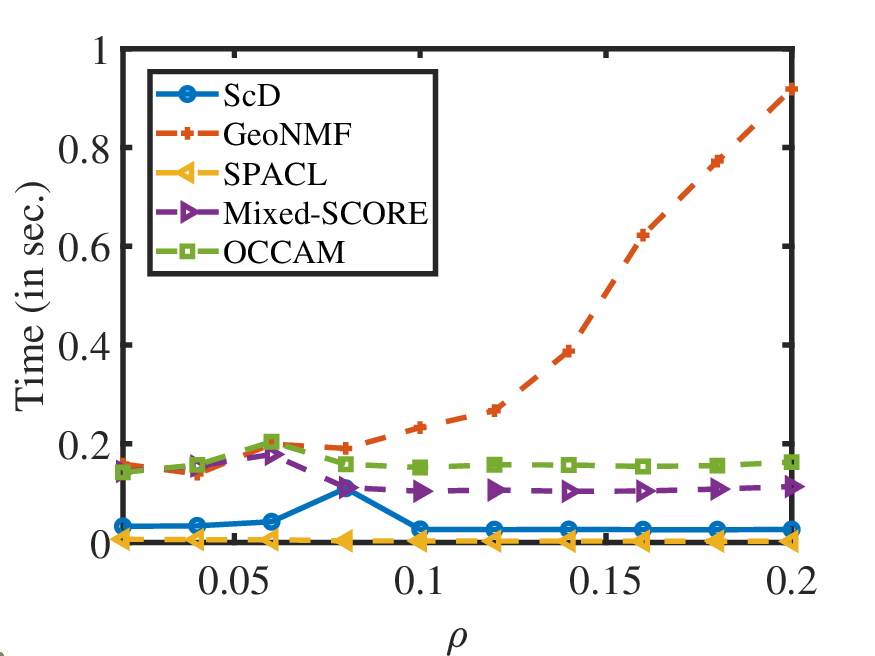}}
\subfigure[$\eta_{\mathcal{N}}=-1$]{\includegraphics[width=0.2\textwidth]{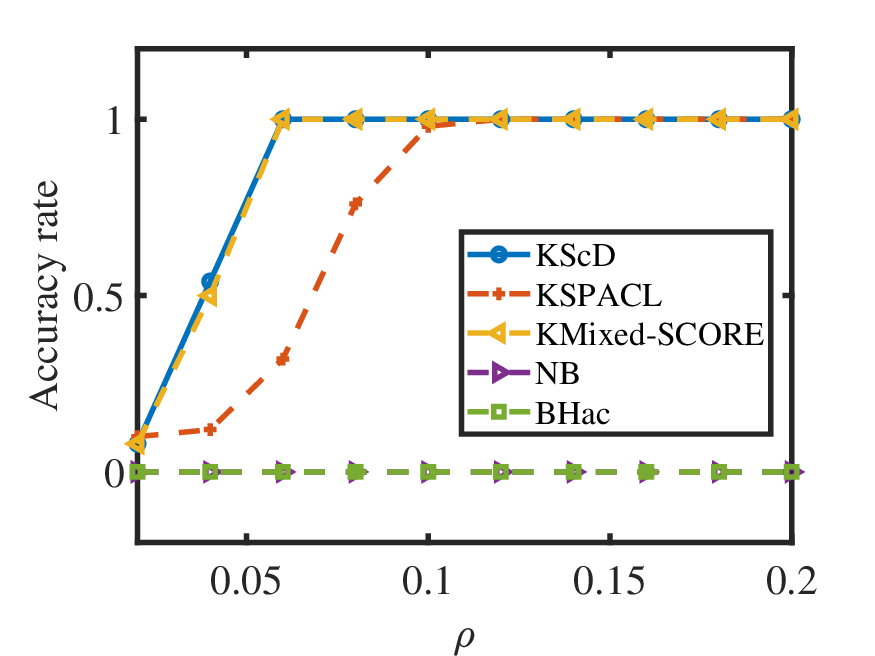}}
}
\caption{Numerical results of Experiment 2.}
\label{EX2} 
\end{figure}
\begin{figure}
\centering
\resizebox{\columnwidth}{!}{
\subfigure[$\eta_{\mathcal{N}}=1$]{\includegraphics[width=0.2\textwidth]{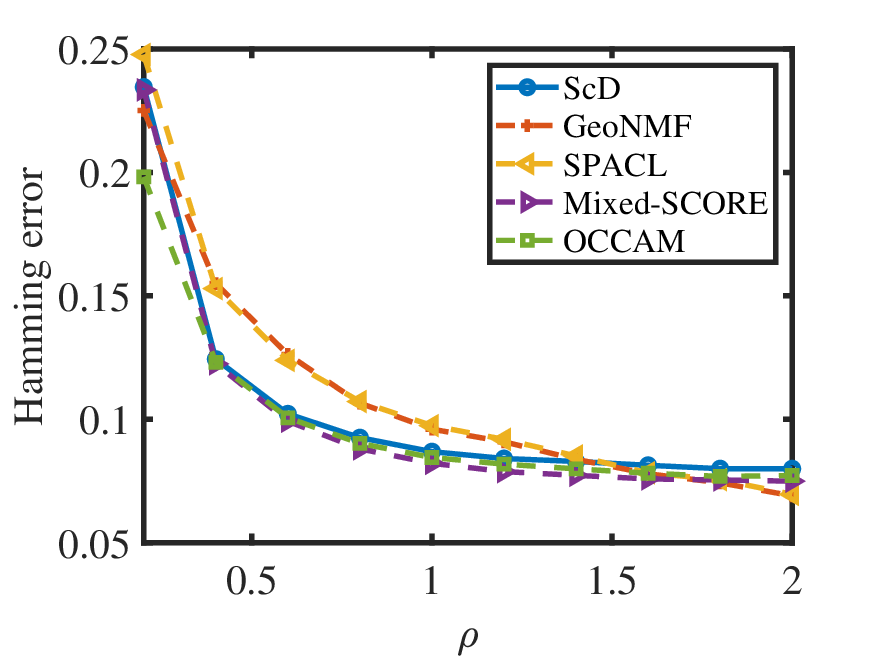}}
\subfigure[$\eta_{\mathcal{N}}=1$]{\includegraphics[width=0.2\textwidth]{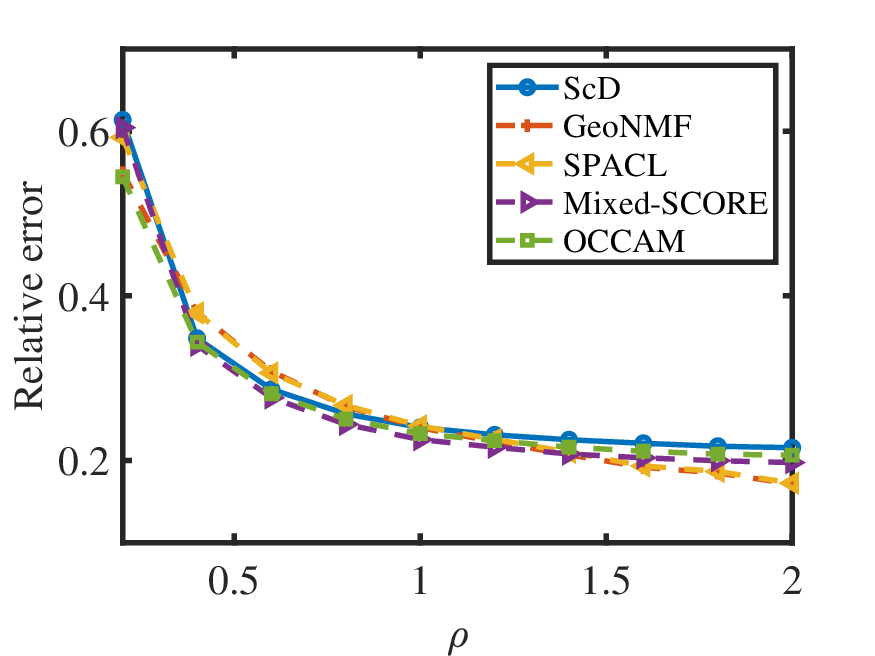}}
\subfigure[$\eta_{\mathcal{N}}=1$]{\includegraphics[width=0.2\textwidth]{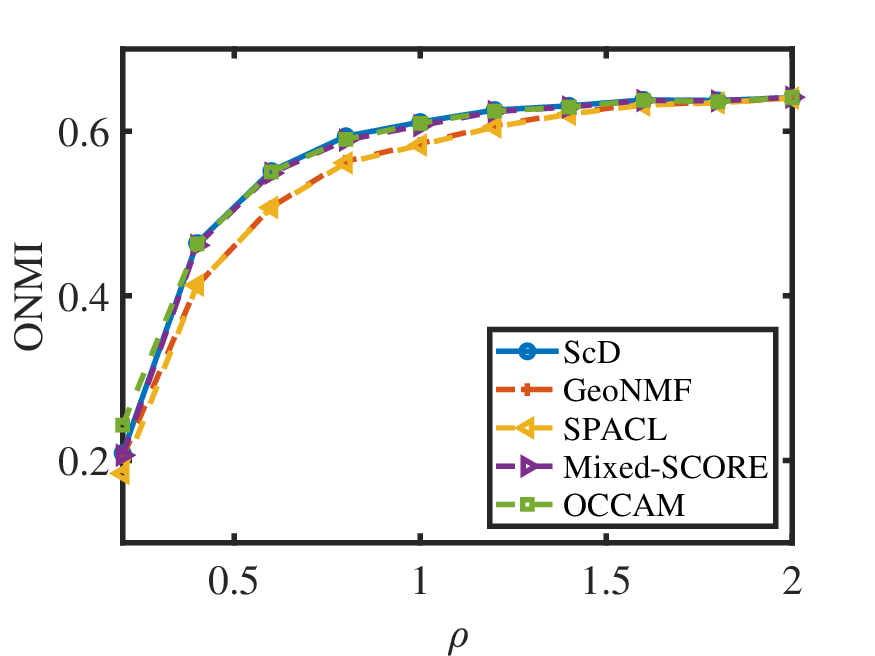}}
\subfigure[$\eta_{\mathcal{N}}=1$]{\includegraphics[width=0.2\textwidth]{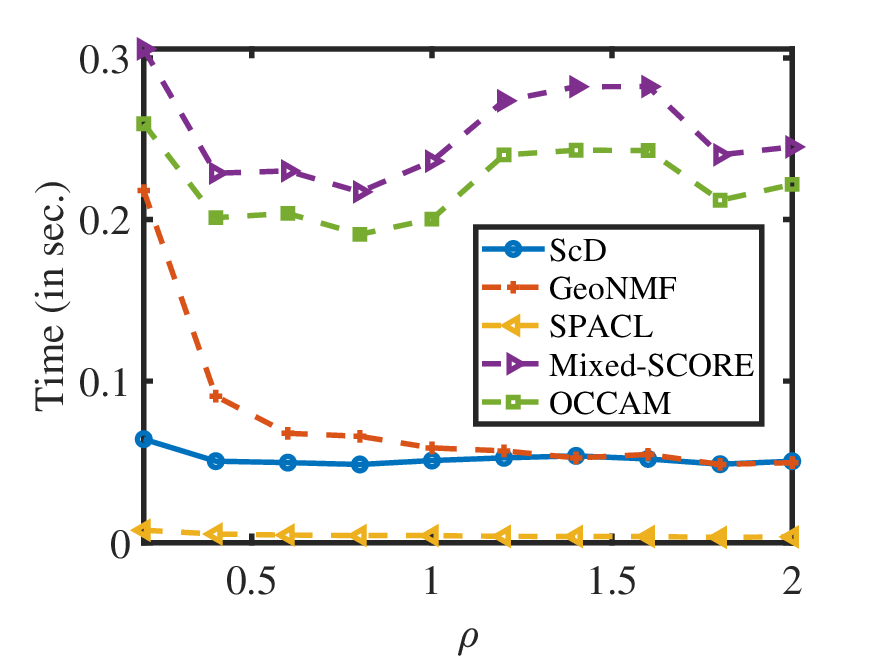}}
\subfigure[$\eta_{\mathcal{N}}=1$]{\includegraphics[width=0.2\textwidth]{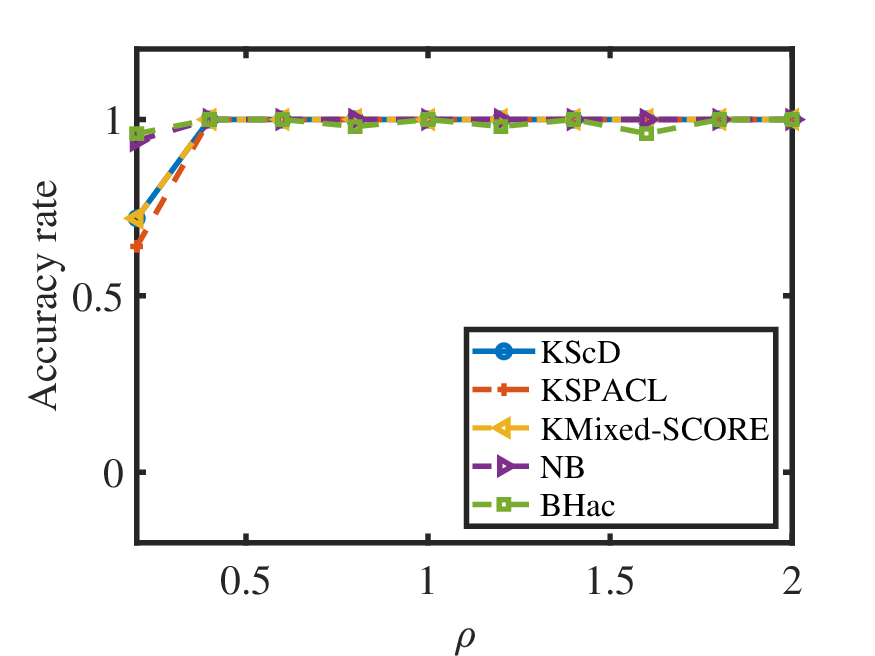}}
}
\resizebox{\columnwidth}{!}{
\subfigure[$\eta_{\mathcal{N}}=-1$]{\includegraphics[width=0.2\textwidth]{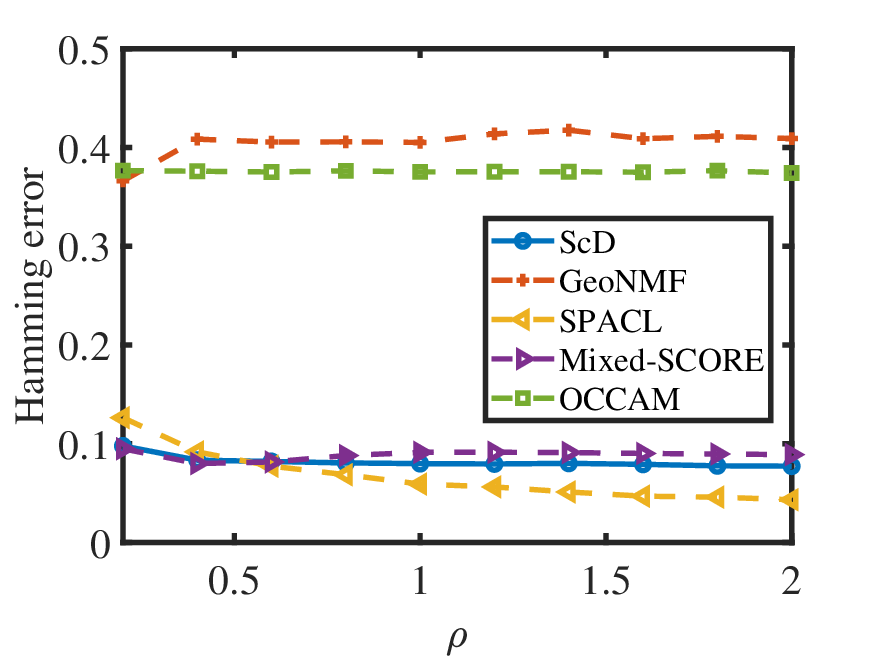}}
\subfigure[$\eta_{\mathcal{N}}=-1$]{\includegraphics[width=0.2\textwidth]{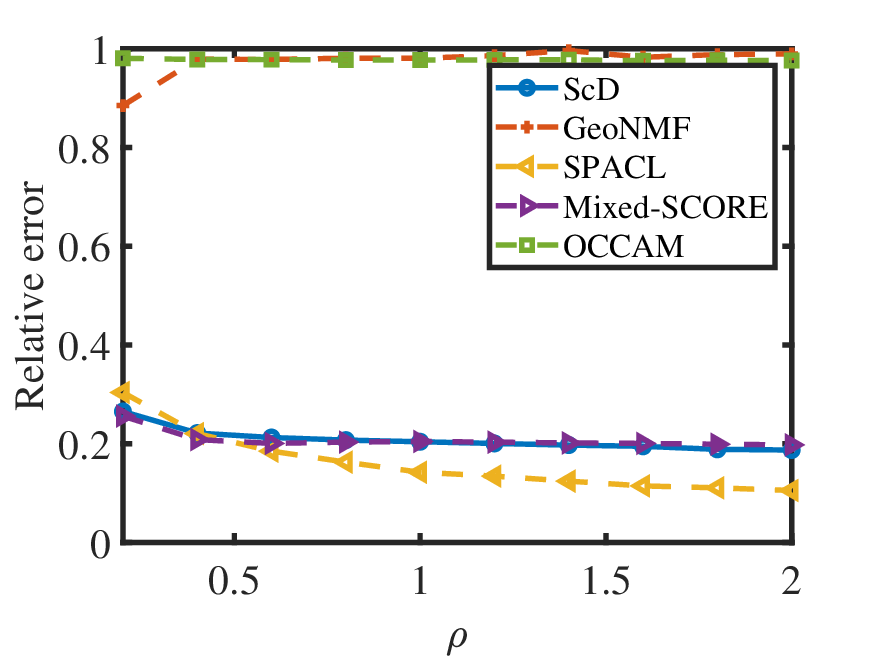}}
\subfigure[$\eta_{\mathcal{N}}=-1$]{\includegraphics[width=0.2\textwidth]{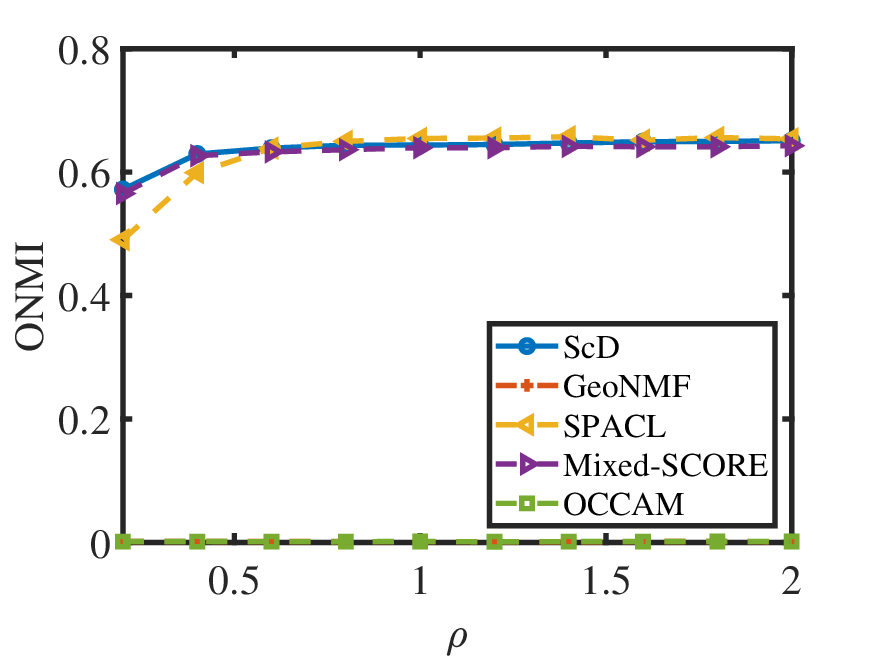}}
\subfigure[$\eta_{\mathcal{N}}=-1$]{\includegraphics[width=0.2\textwidth]{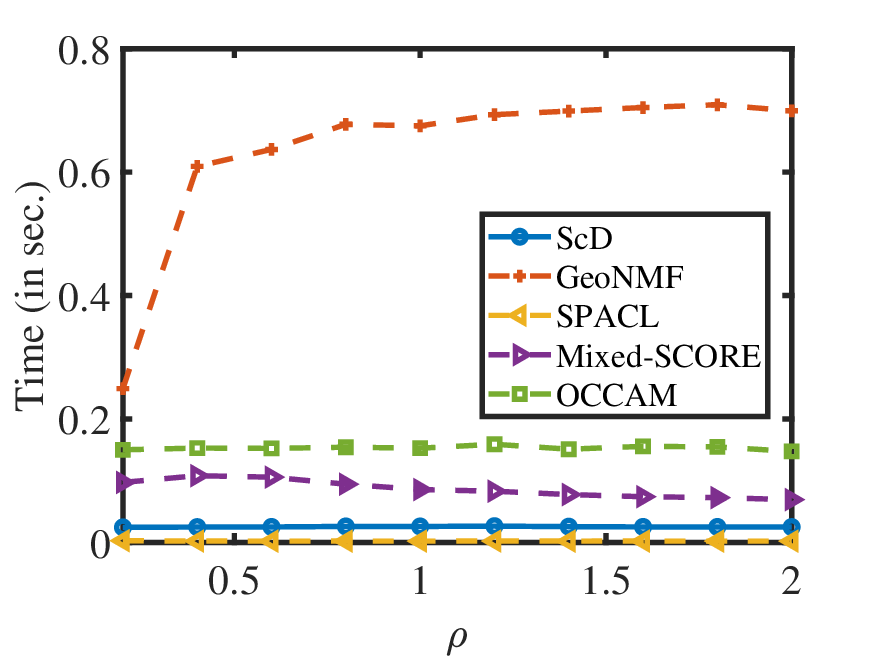}}
\subfigure[$\eta_{\mathcal{N}}=-1$]{\includegraphics[width=0.2\textwidth]{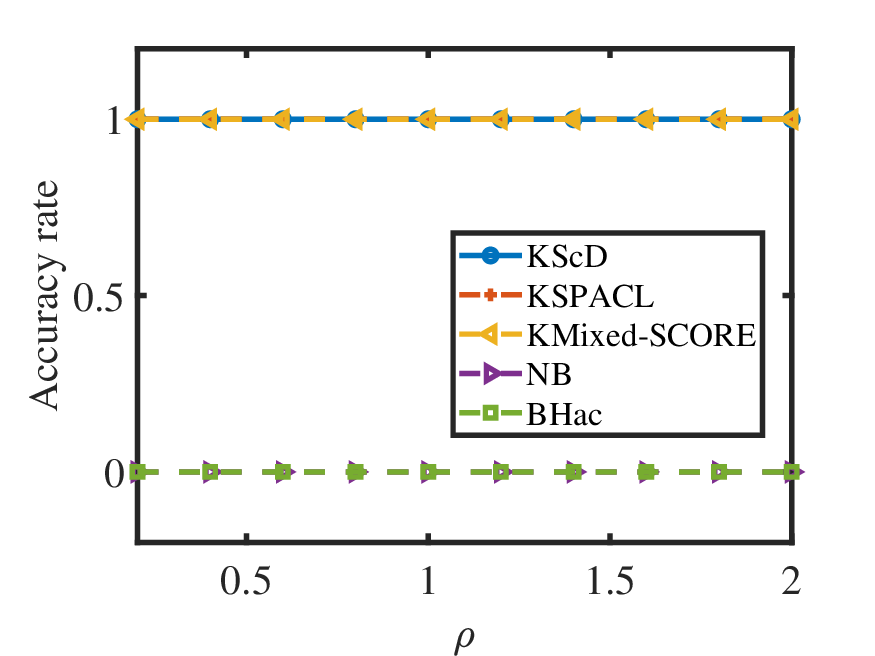}}
}
\caption{Numerical results of Experiment 3.}
\label{EX3} 
\end{figure}

\textbf{Experiment 2: Bernoulli distribution.} Set $\mathcal{F}$ as Bernoulli distribution such that $A(i,j)\sim\mathrm{Bernoulli}(\Omega(i,j))$. For assortative networks (i.e., $\eta_{\mathcal{N}}=1$), we let $p=0.2$ and $\rho$ range in $\{0.1,0.2,\ldots,1\}$. For dis-assortative networks (i.e., $\eta_{\mathcal{N}}=-1$), we let $p=5$ and $\rho$ range in $\{0.02,0.04,\ldots,0.2\}$. Figure \ref{EX2} shows the results. For the task of estimating memberships, with the exception of GeoNMF and OCCAM, which fail to work for dis-assortative networks, all methods enjoy better performances as $\rho$ increases which is consistent with our analysis in Example \ref{BernoulliF}. Meanwhile, we see that SPCAL and ScD run faster than the other three methods. For the task of determining $K$, all methods have larger accuracy rates when $\rho$ increases in assortative networks. Notably, our KScD, KSPACL, and KMixed-SCORE methods also exhibit high accuracy in estimating $K$, whereas NB and BHac fail to work for dis-assortative networks.

\begin{figure}
\centering
\resizebox{\columnwidth}{!}{
\subfigure[$\eta_{\mathcal{N}}=1$]{\includegraphics[width=0.2\textwidth]{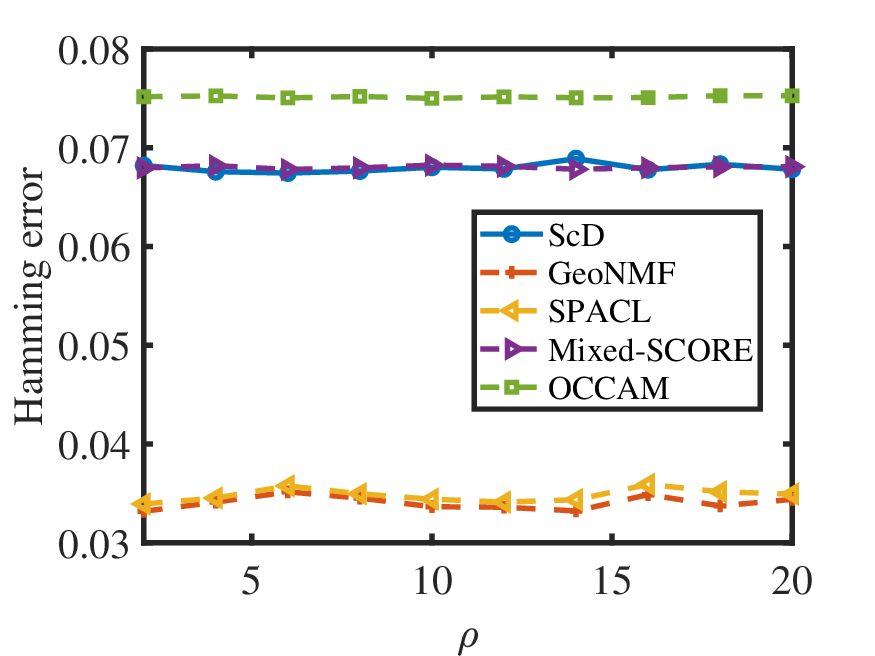}}
\subfigure[$\eta_{\mathcal{N}}=1$]{\includegraphics[width=0.2\textwidth]{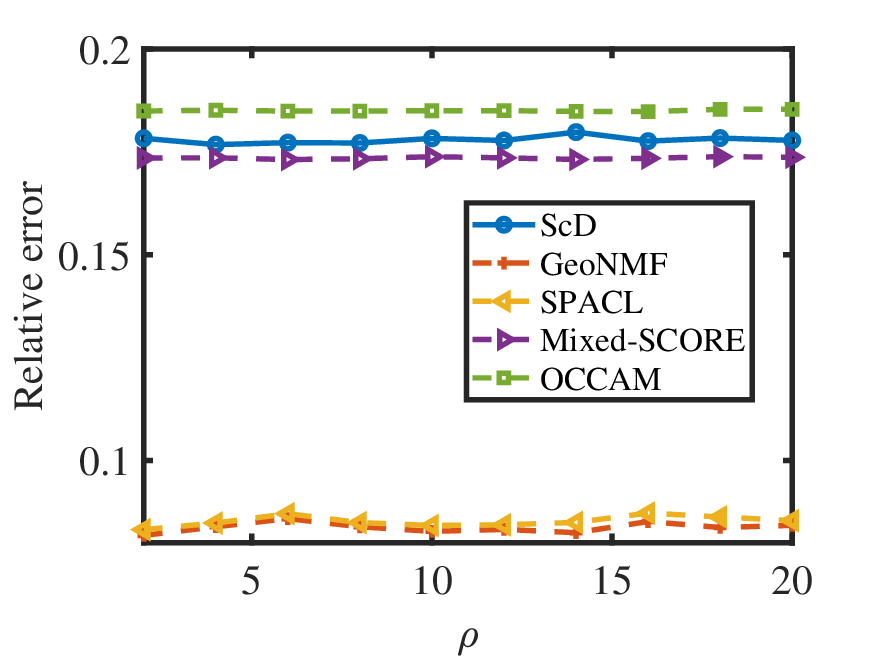}}
\subfigure[$\eta_{\mathcal{N}}=1$]{\includegraphics[width=0.2\textwidth]{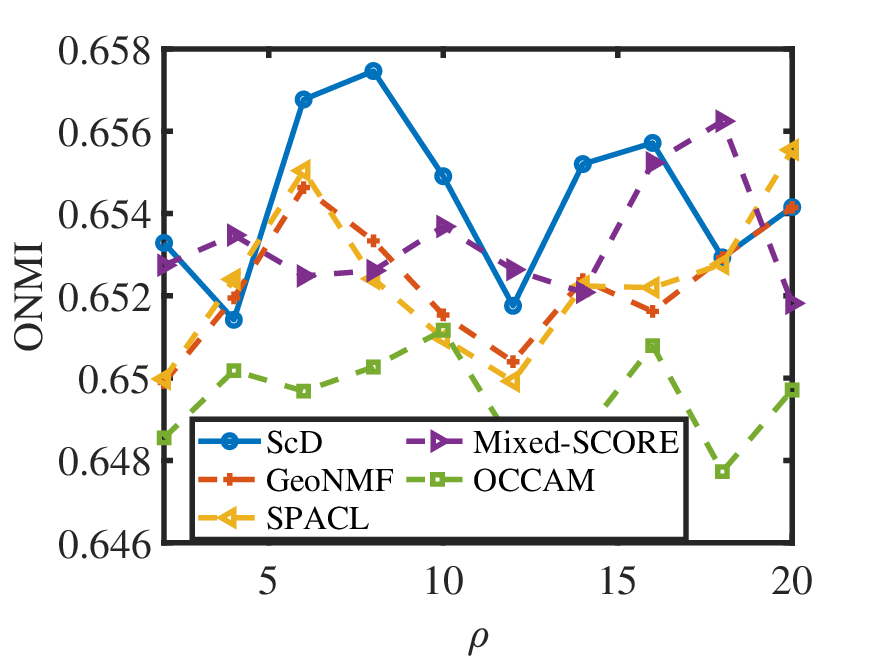}}
\subfigure[$\eta_{\mathcal{N}}=1$]{\includegraphics[width=0.2\textwidth]{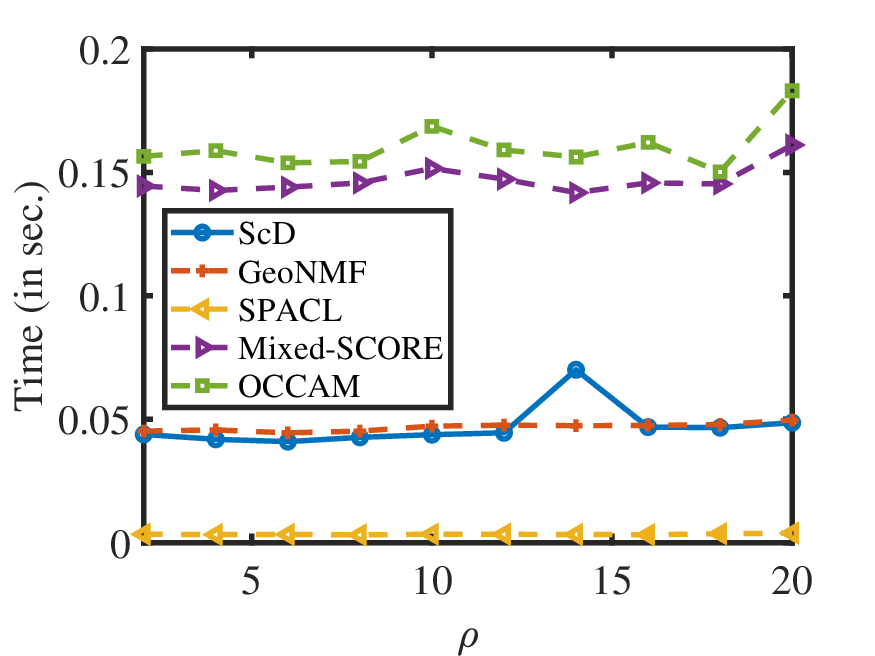}}
\subfigure[$\eta_{\mathcal{N}}=1$]{\includegraphics[width=0.2\textwidth]{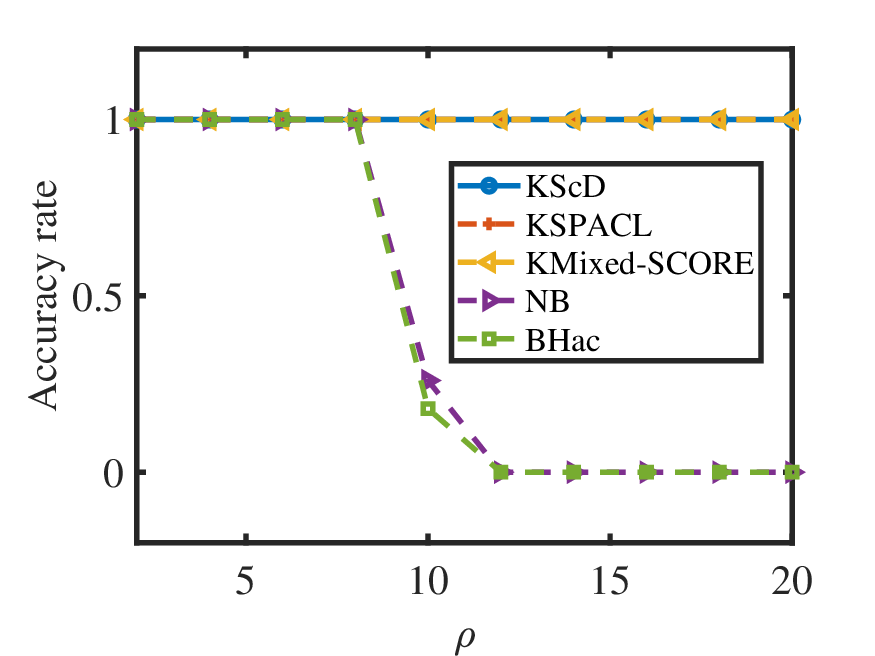}}
}
\resizebox{\columnwidth}{!}{
\subfigure[$\eta_{\mathcal{N}}=-1$]{\includegraphics[width=0.2\textwidth]{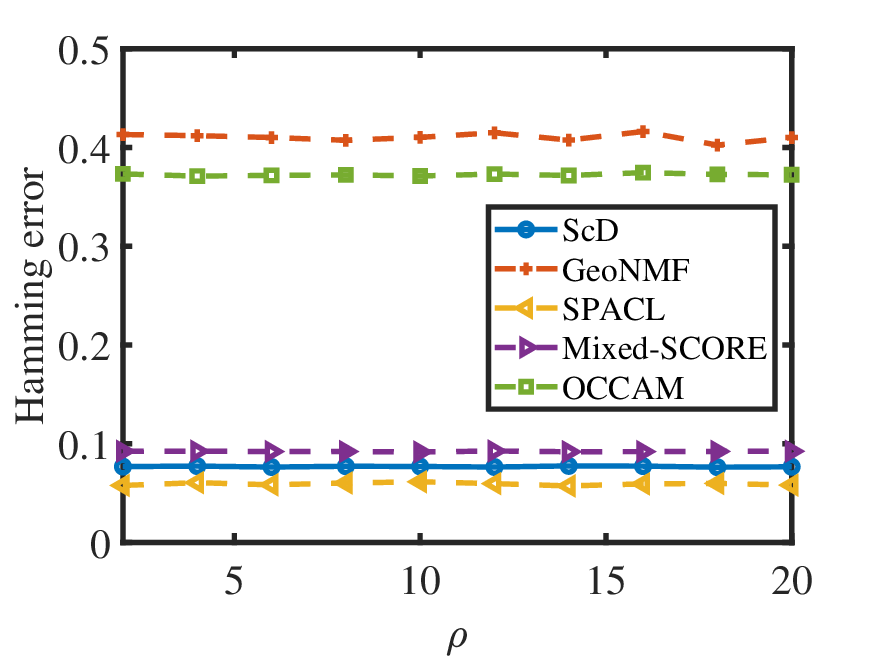}}
\subfigure[$\eta_{\mathcal{N}}=-1$]{\includegraphics[width=0.2\textwidth]{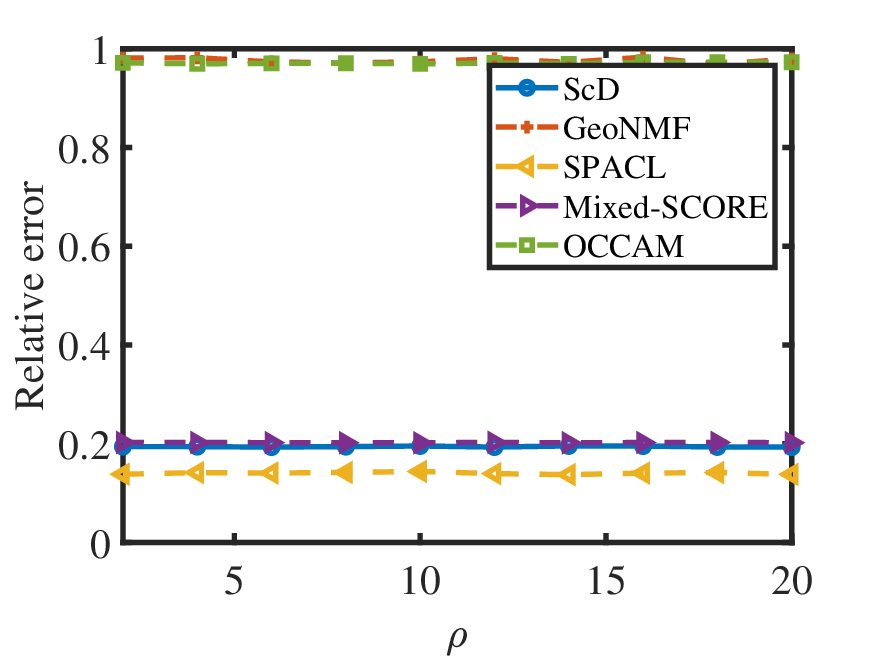}}
\subfigure[$\eta_{\mathcal{N}}=-1$]{\includegraphics[width=0.2\textwidth]{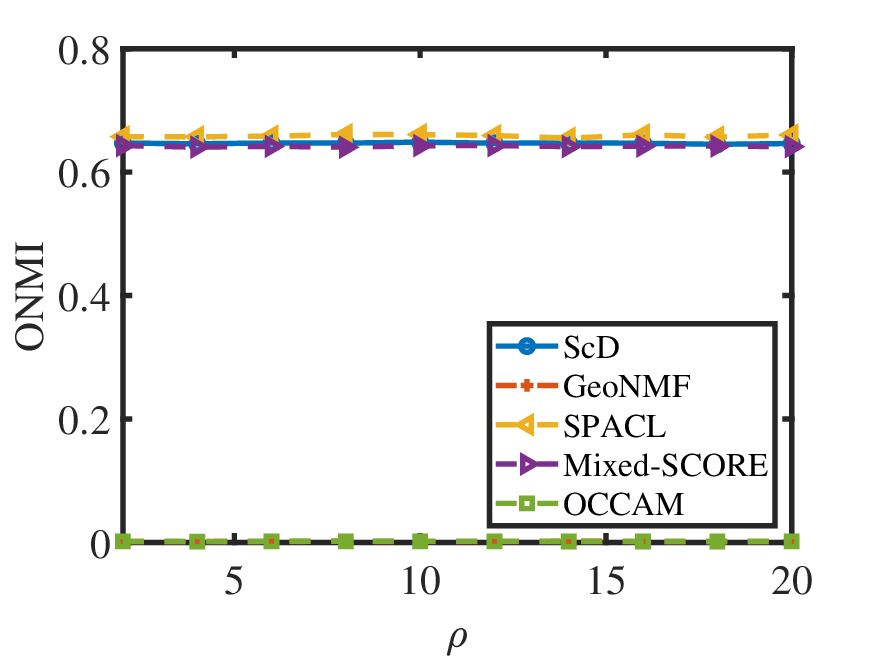}}
\subfigure[$\eta_{\mathcal{N}}=-1$]{\includegraphics[width=0.2\textwidth]{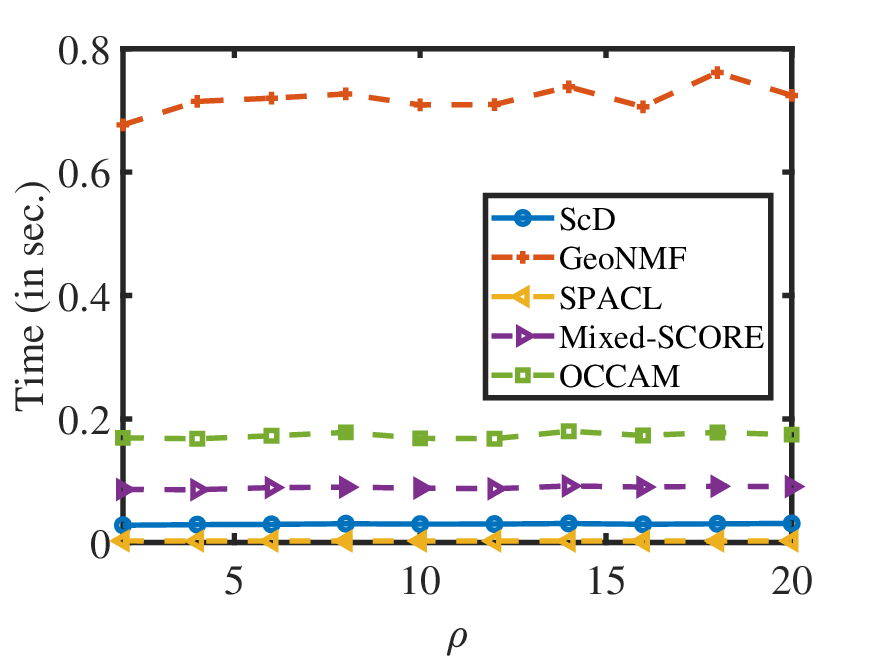}}
\subfigure[$\eta_{\mathcal{N}}=-1$]{\includegraphics[width=0.2\textwidth]{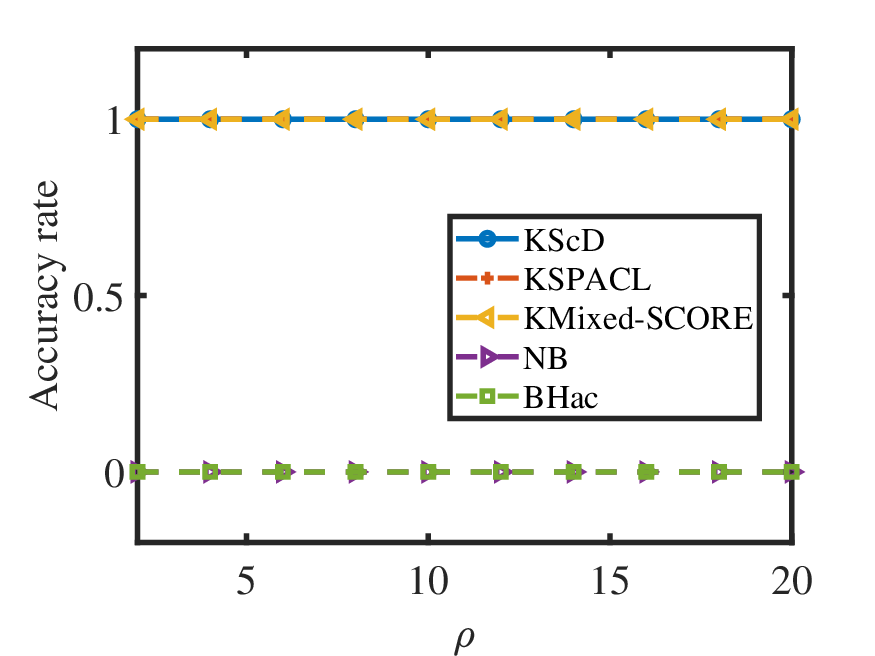}}
}
\caption{Numerical results of Experiment 4.}
\label{EX4} 
\end{figure}
\textbf{Experiment 3: Poisson distribution.} Set $\mathcal{F}$ as Poisson distribution such that $A(i,j)\sim\mathrm{Poisson}(\Omega(i,j))$. Let $p=0.2$ for assortative weighted networks (i.e., $\eta_{\mathcal{N}}=1$), $p=5$ for dis-assortative weighted networks (i.e., $\eta_{\mathcal{N}}=-1$), and $\rho$ range in $\{0.2,0.4,\ldots,2\}$. The results displayed in Figure \ref{EX3} are similar to that of Experiment 2 and we omit the analysis here.

\textbf{Experiment 4: Uniform distribution.} Set $\mathcal{F}$ as Uniform distribution such that $A(i,j)\sim\mathrm{Uniform}(0,2\Omega(i,j))$. Let $p=0.2$ for assortative weighted networks (i.e., $\eta_{\mathcal{N}}=1$), $p=5$ for dis-assortative weighted networks (i.e., $\eta_{\mathcal{N}}=-1$), and $\rho$ range in $\{2,4,\ldots,20\}$. Figure \ref{EX4} displays the results. We observe the following: (i) ScD's performance is almost unchanged when $\rho$ becomes larger, which is consistent with our analysis in Example \ref{UniformF}. (ii) for assortative networks, all methods perform satisfactorily and similarly because their Hamming errors are lesser than 0.07, their Relative errors are lesser than 0.18, and their ONMIs are larger than 0.65. For dis-assortative networks, ScD, SPACL, and Mixed-SCORE enjoy satisfactory performances while GeoNMF and OCCAM fail to work. (iii) ScD, SPACL, and GeoNMF run faster than Mixed-SCORE and OCCAM in assortative networks while  GeoNMF is the slowest in the dis-assortative case. (iv) Our KScD, KSPACL, and KMixed-SCORE correctly determine $K$ for this experiment in both assortative and dis-assortative networks. It is interesting to see that NB and BHac successfully estimate $K$ when $\rho<10$ while they fail to infer $K$ when $\rho\geq10$ in assortative networks and these two methods fail to determine $K$ for dis-assortative networks.

\begin{figure}
\centering
\resizebox{\columnwidth}{!}{
\subfigure[$\eta_{\mathcal{N}}=1$]{\includegraphics[width=0.2\textwidth]{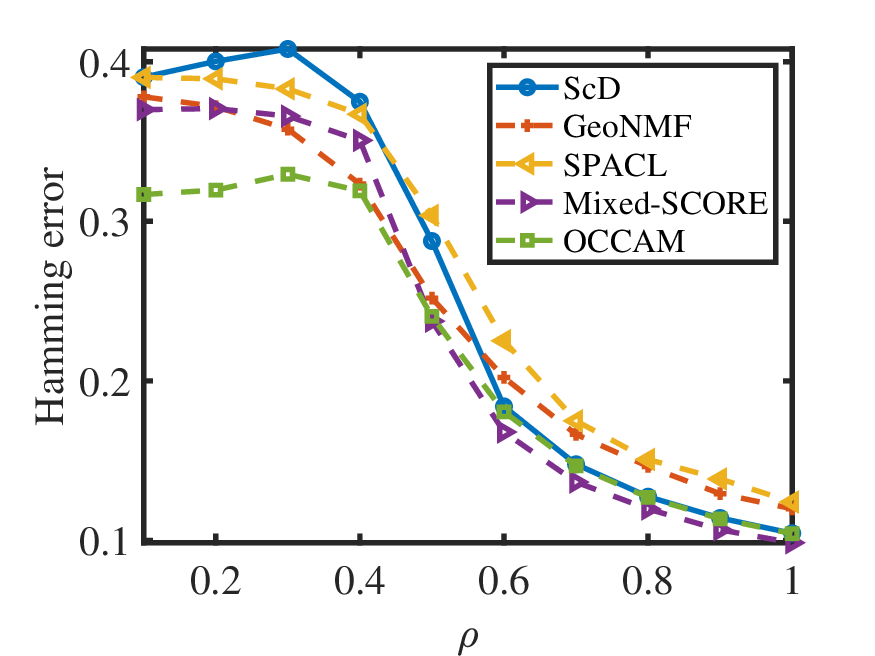}}
\subfigure[$\eta_{\mathcal{N}}=1$]{\includegraphics[width=0.2\textwidth]{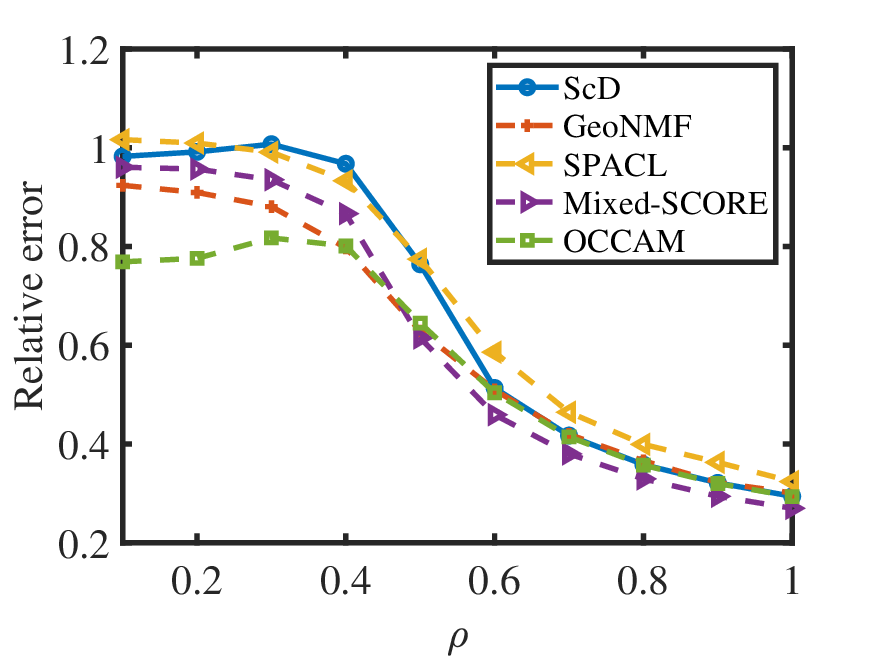}}
\subfigure[$\eta_{\mathcal{N}}=1$]{\includegraphics[width=0.2\textwidth]{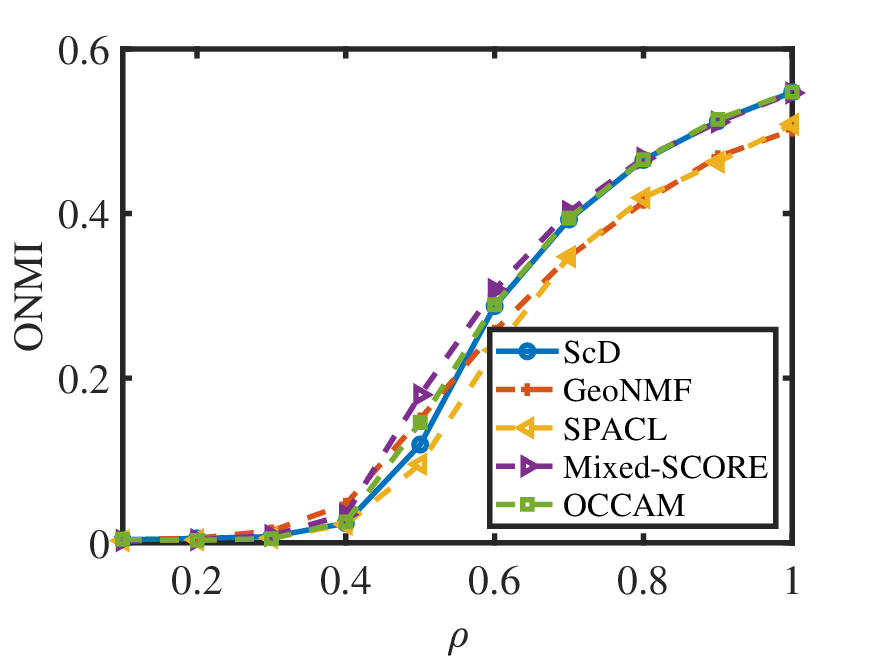}}
\subfigure[$\eta_{\mathcal{N}}=1$]{\includegraphics[width=0.2\textwidth]{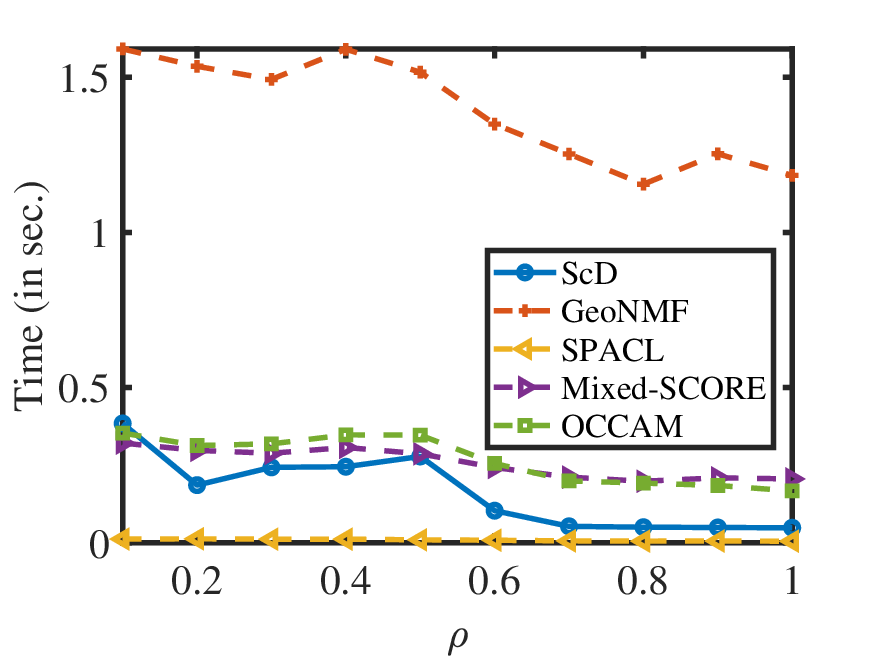}}
\subfigure[$\eta_{\mathcal{N}}=1$]{\includegraphics[width=0.2\textwidth]{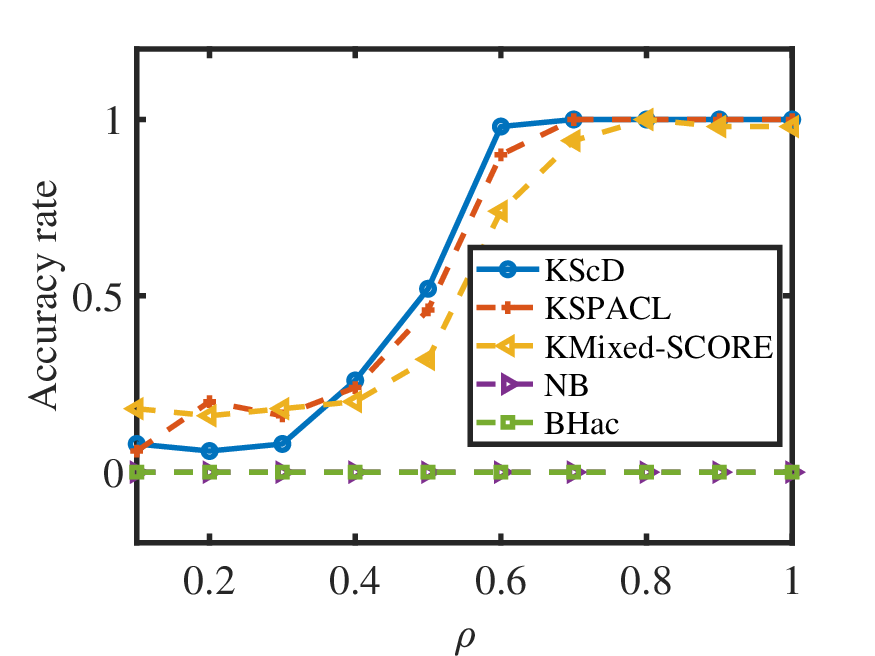}}
}
\resizebox{\columnwidth}{!}{
\subfigure[$\eta_{\mathcal{N}}=-1$]{\includegraphics[width=0.2\textwidth]{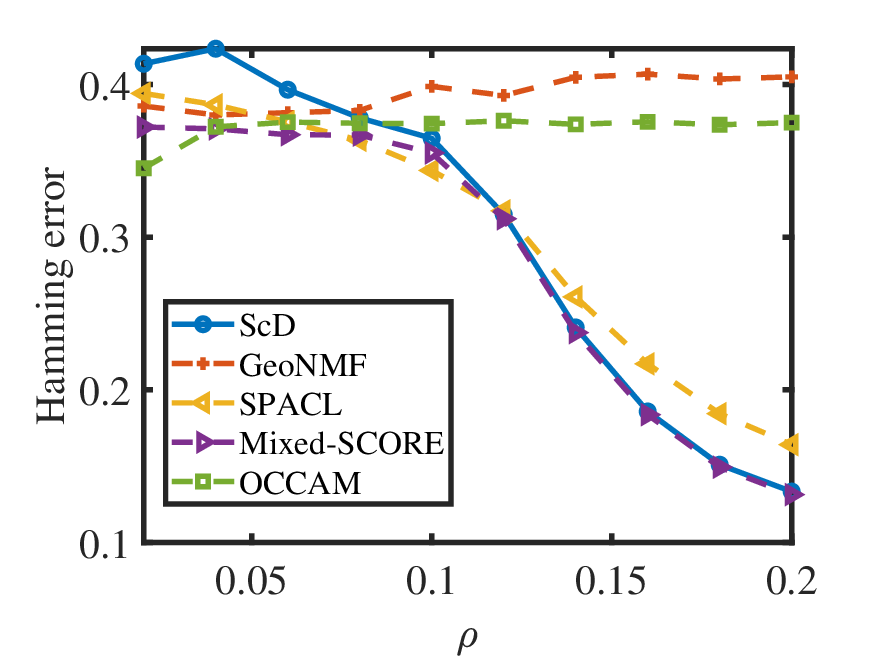}}
\subfigure[$\eta_{\mathcal{N}}=-1$]{\includegraphics[width=0.2\textwidth]{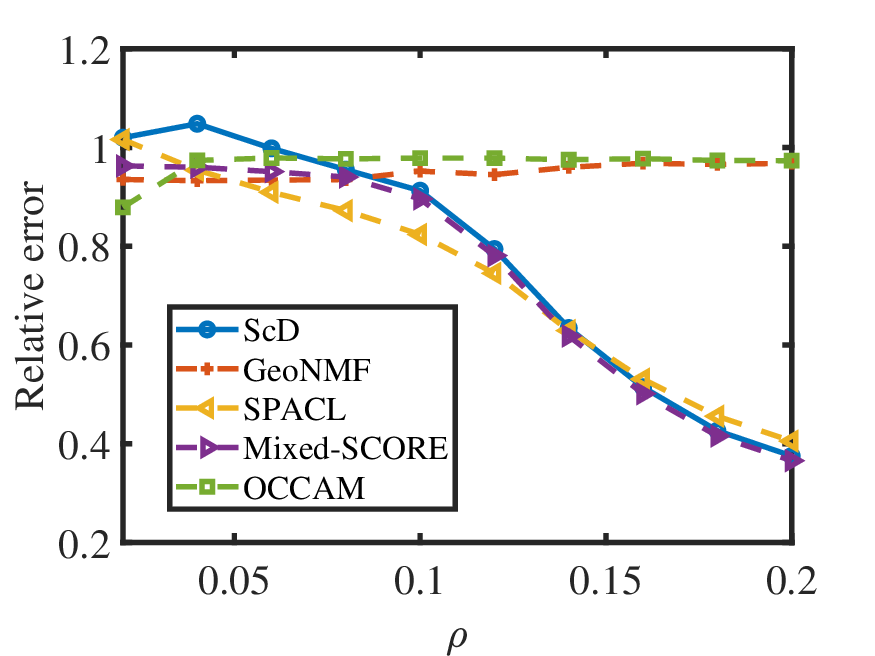}}
\subfigure[$\eta_{\mathcal{N}}=-1$]{\includegraphics[width=0.2\textwidth]{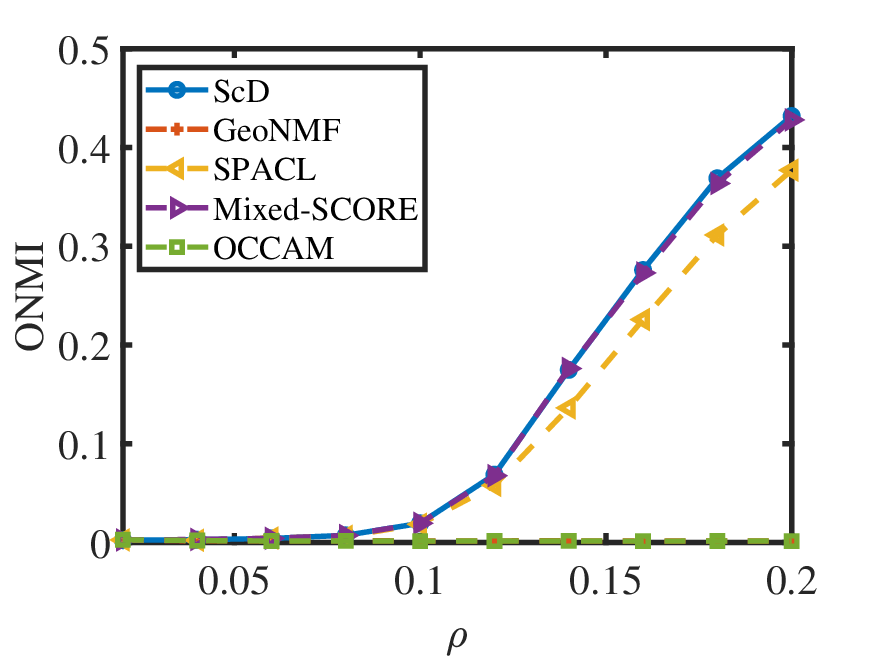}}
\subfigure[$\eta_{\mathcal{N}}=-1$]{\includegraphics[width=0.2\textwidth]{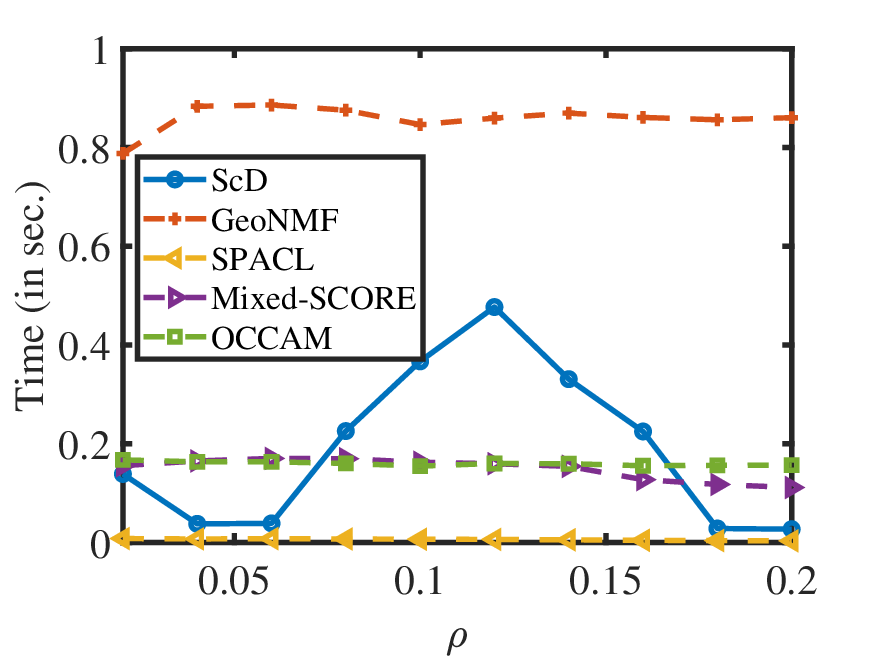}}
\subfigure[$\eta_{\mathcal{N}}=-1$]{\includegraphics[width=0.2\textwidth]{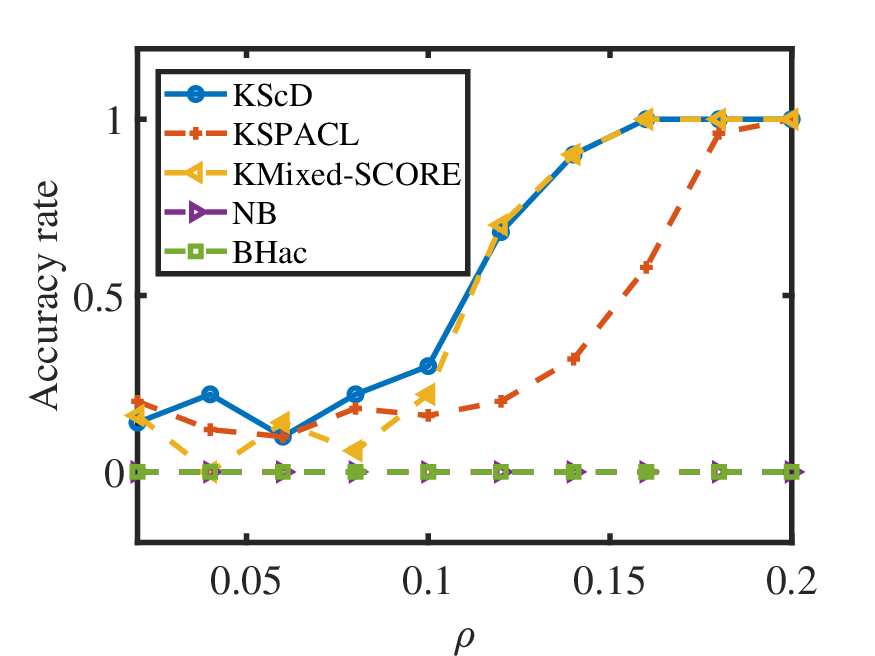}}
}
\caption{Numerical results of Experiment 5.}
\label{EX5} 
\end{figure}

\textbf{Experiment 5: Signed network.} For signed network when $\mathbb{P}(A(i,j)=1)=\frac{1+\Omega(i,j)}{2}$ and $\mathbb{P}(A(i,j)=-1)=\frac{1-\Omega(i,j)}{2}$. Let $p=-0.2$ and $\rho$ range in $\{0.1,0.2,\ldots,1\}$ for assortative signed networks (i.e., $\eta_{\mathcal{N}}=1$). Let $p=5$ and $\rho$ range in $\{0.02,0.04,\ldots,0.2\}$ for dis-assortative signed networks (i.e., $\eta_{\mathcal{N}}=-1$). Figure \ref{EX5} displays the results. We see that: (i) ScD exhibits improved performance in estimating memberships as $\rho$ increases, which supports our analysis in Example \ref{SignedF}. (ii) All methods, with the exception of GeoNMF and OCCAM, demonstrate similar behavior and improved performance as $\rho$ grows in both assortative and dis-assortative networks. For this experiment, SPACL runs faster than the other four methods while GeoNMF runs slowest. For determining $K$, the accuracy rates of our methods KScD, KSPACL, and KMixed-SCORE increase when increasing $\rho$ while their two competitors fail to find $K$ for signed networks.

Generally speaking, from the numerical results of Experiments 1-5, we have the following conclusions:
\begin{itemize}
  \item For the task of estimating memberships, the numerical results in Examples 1-5 support our analysis that ScD exhibits varying performance for different edge weight distributions as $\rho$ changes. GeoNMF, SPACL, Mixed-SCORE, and OCCAM show competitive performances with ScD and can estimate memberships for assortative networks generated from the WDCMM model. However, for dis-assortative networks, GeoNMF and OCCAM fail to work, while ScD, SPACL, and Mixed-SCORE demonstrate similar and improved performance as $\rho$ increases.
  \item Our methods enjoy satisfactory performance in determining the number of communities in both assortative and dis-assortative weighted networks, highlighting the effectiveness of our overlapping weighted modularity $Q_{ovw}$. Thus, a higher $Q_{ovw}$ value indeed indicates a better community partition. Conversely, NB and BHac perform well only in assortative networks generated from Bernoulli and Poisson distributions. They fail to estimate $K$ for assortative networks derived from Normal and Uniform distributions, as well as signed networks. Moreover, they entirely fail to infer $K$ for dis-assortative networks.
\end{itemize}
\begin{rem}
This discrepancy in performance of different methods  stems from a fundamental methodological difference. The superiority of ScD, SPACL, and Mixed-SCORE stems from their distribution-free spectral approach, which utilizes the eigenvectors corresponding to the top $K$ eigenvalues in magnitude of the adjacency matrix $A$. This strategy impartially captures the strongest structural signals, whether they manifest as large positive or large negative eigenvalues—the latter being a characteristic of dis-assortative networks. In fact, in the proof of Theorem \ref{Main}, we only require the connectivity matrix \(P\) to satisfy Equation (\ref{definP}) without limiting it to be specifically designed for assortative or dis-assortative networks. Conversely, GeoNMF, grounded in non-negative matrix factorization, is fundamentally constrained by its requirement for non-negative inputs and factor matrices. This constraint renders it incapable of representing the core dis-assortative structural pattern where connections between different communities are stronger than those within communities. Similarly, OCCAM relies solely on the largest positive eigenvalues, thereby disregarding the dominant negative spectral information that is critical in dis-assortative networks. The flexibility of the spectral approach employed by ScD and its peers is thus the key to their universal applicability across both assortative and dis-assortative networks.
\end{rem}
\subsection{Real Data}\label{Real11data}
\begin{table}[h!]
\footnotesize
	\centering
\resizebox{\columnwidth}{!}{
	\begin{tabular}{cccccccccccc}
\hline\hline
Dataset&Source&Node meaning&Edge meaning&Weighted?&True memberships&$n$&$K$\\
\hline
Karate club&\cite{zachary1977information}&Member&Tie&No&Known&34&2\\
Dolphins&\cite{dolphins0}&Dolphin&Association&No&Known&62&2\\
UKfaculty&\cite{nepusz2008fuzzy}&Faculty&Friendship&No&Known&79&3\\
Political books& Krebs (unpublished)&Book&Co-purchase&No&Known&92&2\\
Political blogs&\cite{Polblogs1}&Blog&Hyperlink&No&Known&1222&2\\
Slovene Parliamentary Party&\cite{ferligoj1996analysis}&Party&Political space distance&Yes&Unknown&10&2\\
Gahuku-Gama subtribes&\cite{read1954cultures}&Tribe&Friendship &Yes&Known&16&3\\
Karate-club-weighted&\cite{zachary1977information}&Member&Tie&Yes&Known&34&2\\
Train bombing&\cite{hayes2006connecting}&Terrorist&Contact&Yes&Unknown&64&Unknown\\
Les Mis\'erables&\cite{knuth1993stanford}&Character&Co-occurence&Yes&Unknown&77&Unknown\\
US Top-500 Airport Network&\cite{colizza2007reaction}&Airport&\#Seats&Yes&Unknown&500&Unknown\\
CE-CX&\citep{cho2014wormnet}&Gene&Line&Yes&Unknown&15063&Unknown\\
WormNet-v3&\citep{cho2014wormnet}&Gene&Line&Yes&Unknown&16258&Unknown\\
Human-gene1&\citep{bansal2007infer}&Gene&Interaction&Yes&Unknown&21853&Unknown\\
\hline\hline
\end{tabular}
}
\caption{Basic information and summarized statistics of real-world networks studied in this paper.} \label{realdata}
\end{table}

In addition to the computer-generated networks, we also consider some real-world datasets. In this study, we are interested in the largest connected component since real-world networks may have several disconnected components. Table \ref{realdata} displays basic information for real data considered in this article. For the five un-weighted networks in Table \ref{realdata}, their adjacency matrices and true memberships can be downloaded from \url{http://zke.fas.harvard.edu/software/SCOREplus/Matlab/datasets/}.  For the Slovene Parliamentary Party network, it can be downloaded from \url{http://vlado.fmf.uni-lj.si/pub/networks/data/soc/Samo/Stranke94.htm}. For Gahuku-Gama subtribes, its community information is shown in Figure 9 (b) \cite{yang2007community}. Karate-club-weighted is the weighted version of Karate club and can be downloaded from \url{http://vlado.fmf.uni-lj.si/pub/networks/data/ucinet/ucidata.htm#kazalo}. For US Top-500 Airport Network, it can be downloaded from \url{https://toreopsahl.com/datasets/#online_social_network}. The last three networks in Table \ref{realdata} are biological networks and they can be downloaded from \url{https://networkrepository.com/bio.php} \citep{rossi2015network}. To simplify our analysis, we treat these real networks as assortative, i.e., $\eta_{\mathcal{N}}=1$ for these networks.

\begin{figure}
\centering
\normalsize
\resizebox{\columnwidth}{!}{
\subfigure[Karate club]{\includegraphics[width=0.32\textwidth]{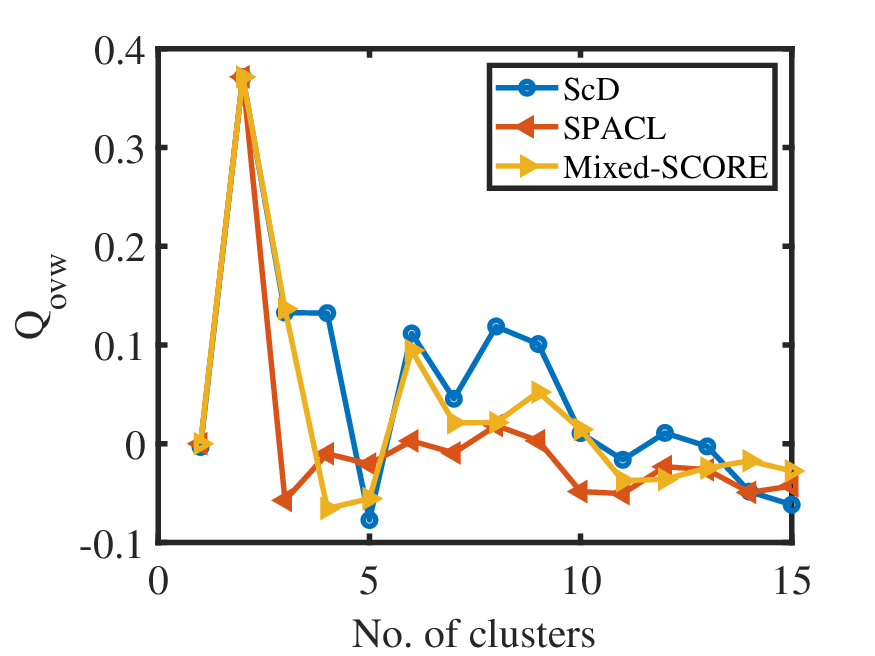}}
\subfigure[Dolphins]{\includegraphics[width=0.32\textwidth]{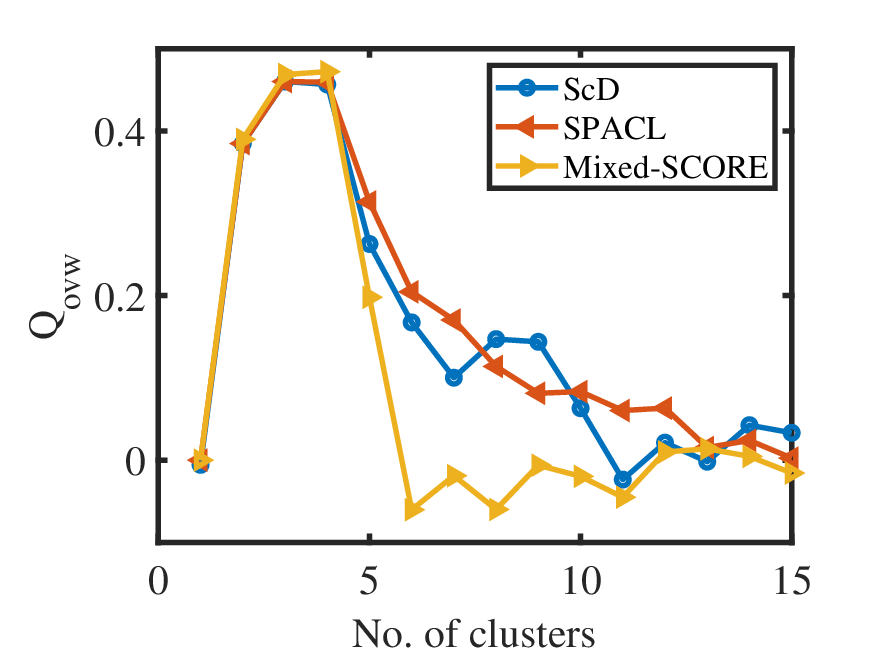}}
\subfigure[UKfaculty]{\includegraphics[width=0.32\textwidth]{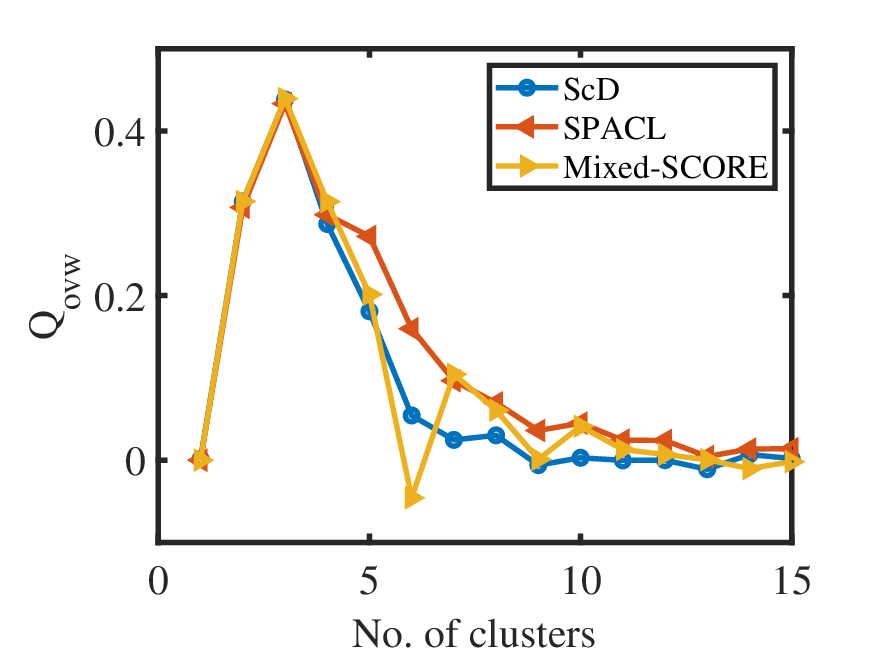}}
\subfigure[Political books]{\includegraphics[width=0.32\textwidth]{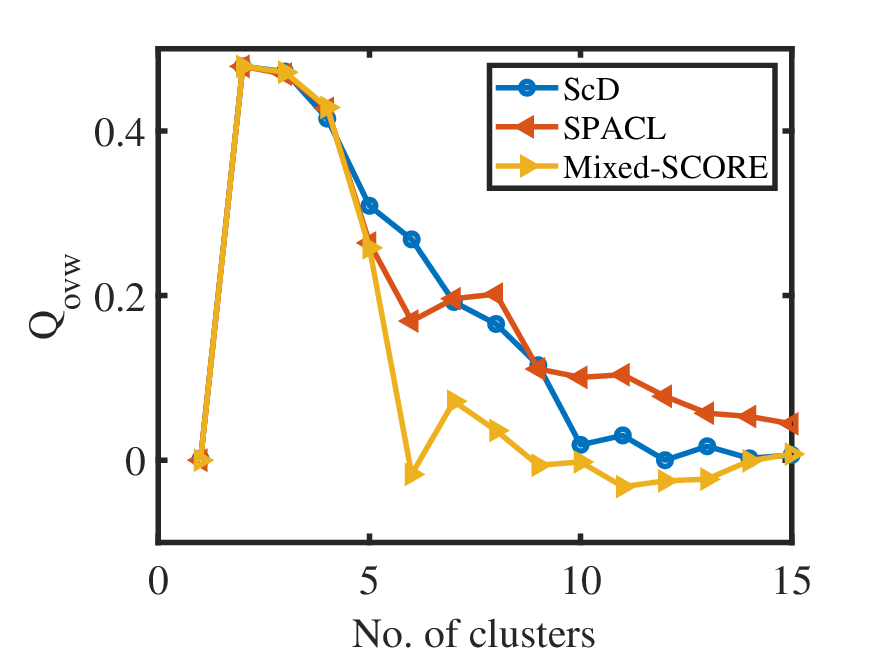}}
}
\resizebox{\columnwidth}{!}{
\subfigure[Political blogs]{\includegraphics[width=0.32\textwidth]{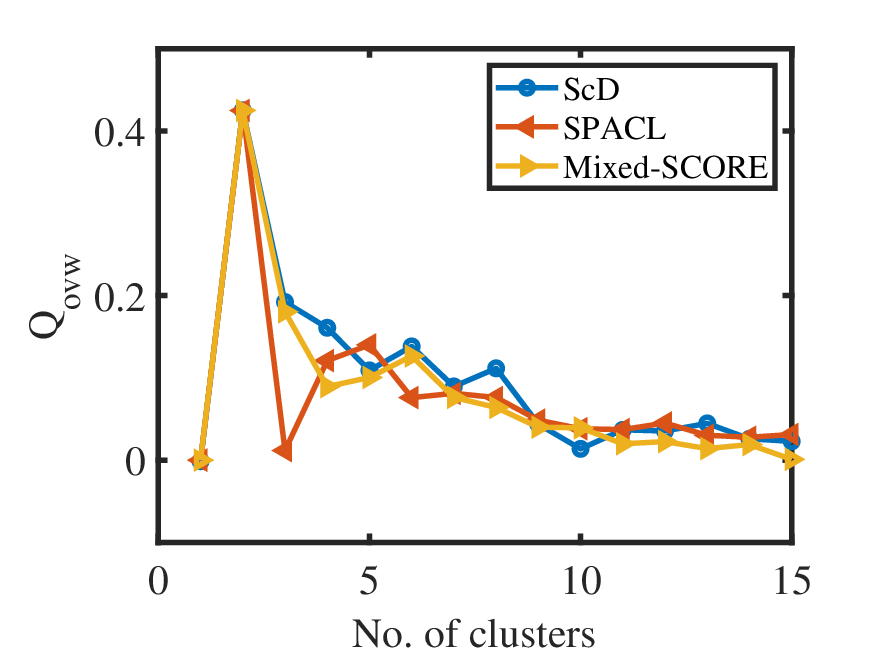}}
\subfigure[Slovene Parliamentary Party]{\includegraphics[width=0.32\textwidth]{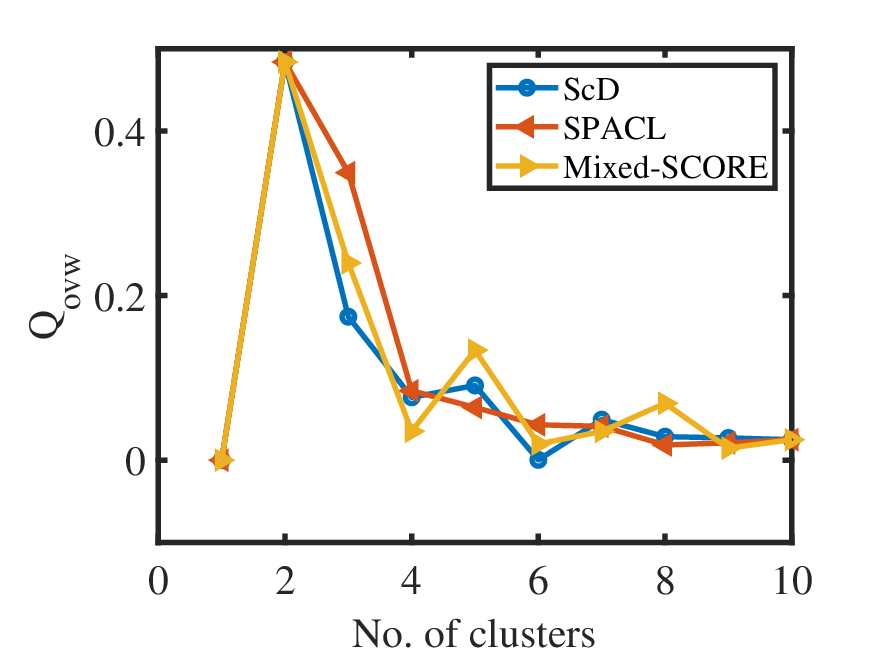}}
\subfigure[Gahuku-Gama subtribes]{\includegraphics[width=0.32\textwidth]{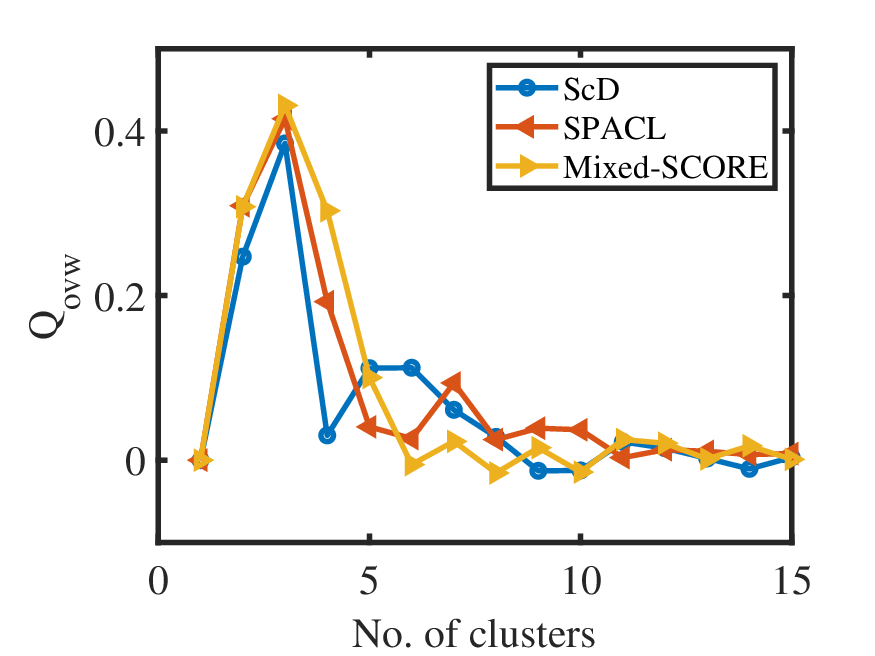}}
\subfigure[Karate-club-weighted]{\includegraphics[width=0.32\textwidth]{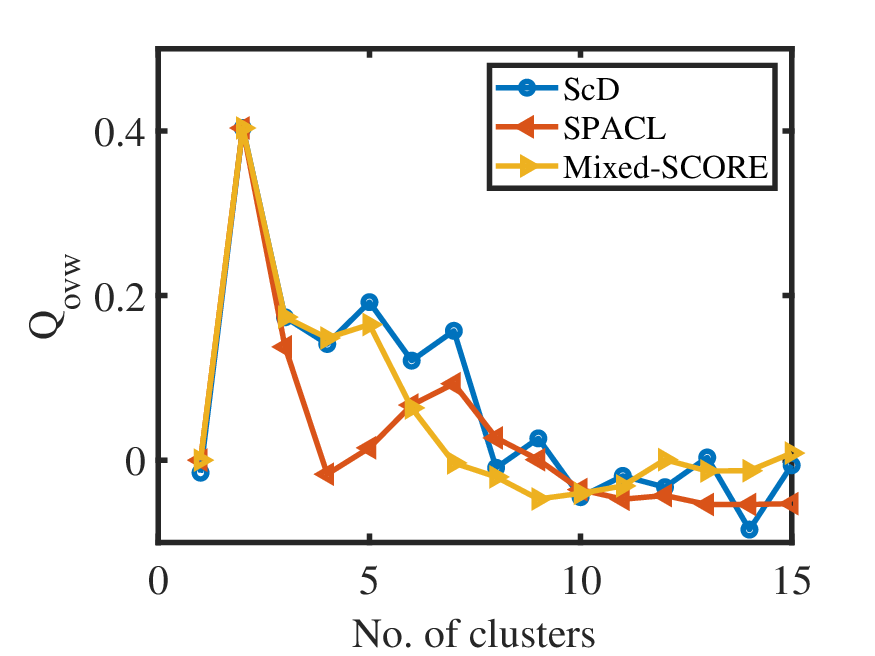}}
}
\resizebox{\columnwidth}{!}{
\subfigure[Train bombing]{\includegraphics[width=0.32\textwidth]{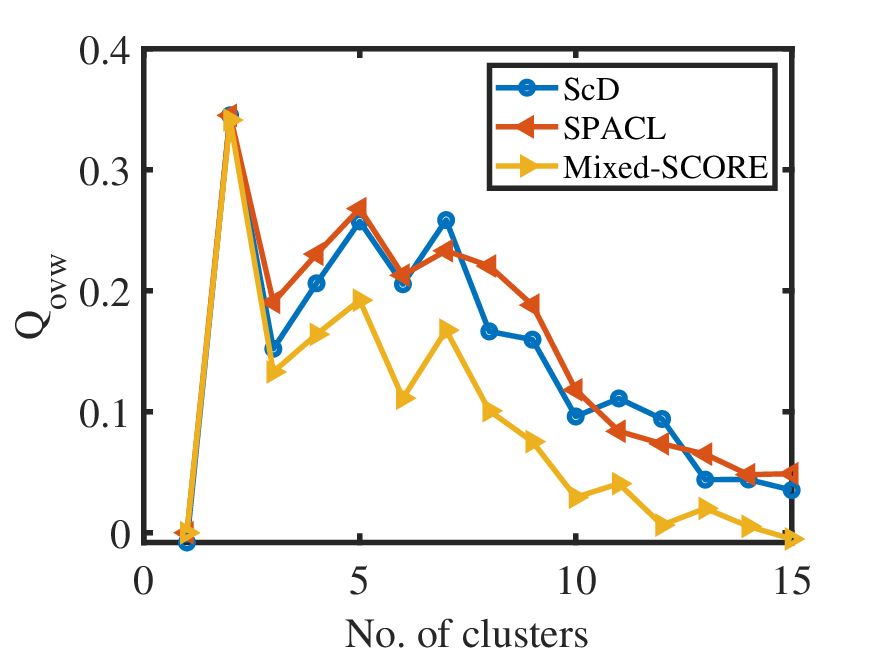}}
\subfigure[Les Mis\'erables]{\includegraphics[width=0.32\textwidth]{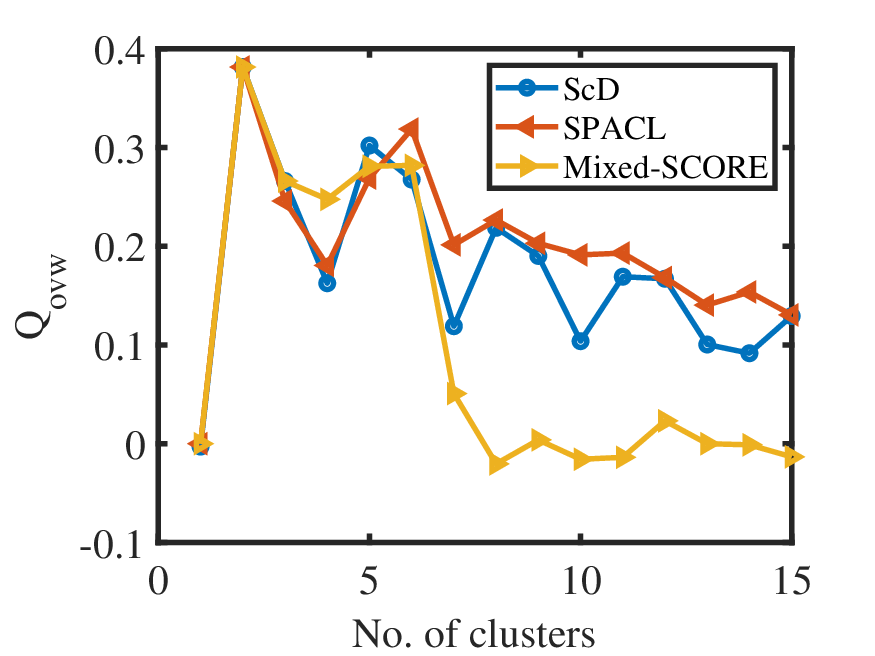}}
\subfigure[US Top-500 Airport Network]{\includegraphics[width=0.32\textwidth]{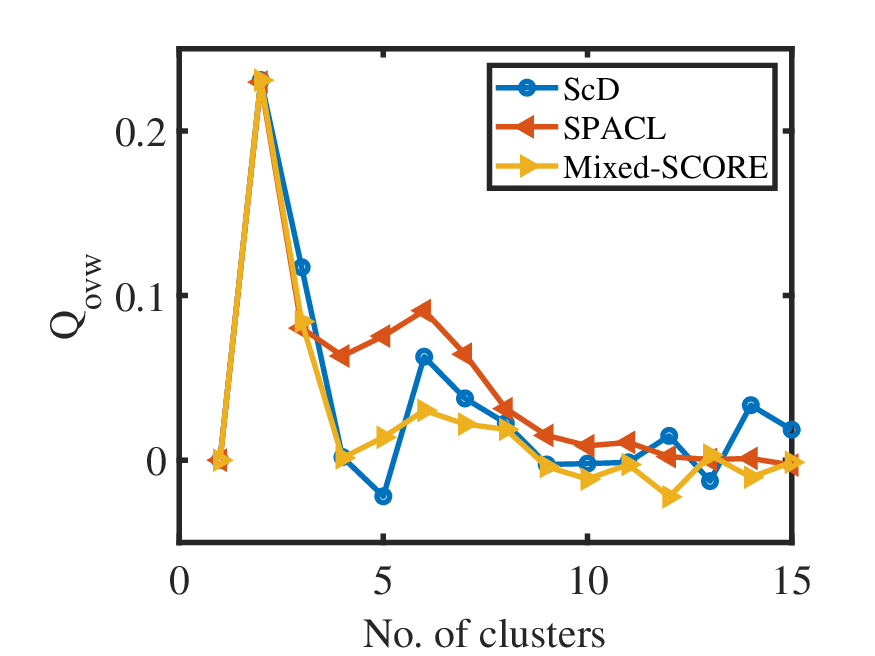}}
\subfigure[CE-CX]{\includegraphics[width=0.32\textwidth]{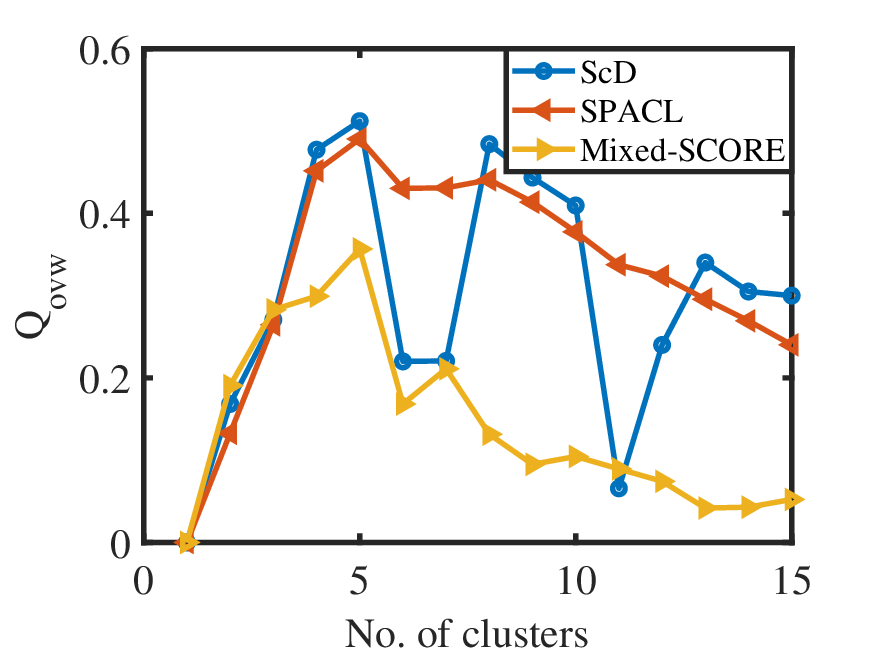}}
}
\subfigure[\scriptsize WormNet-v3]{\includegraphics[width=0.25\textwidth]{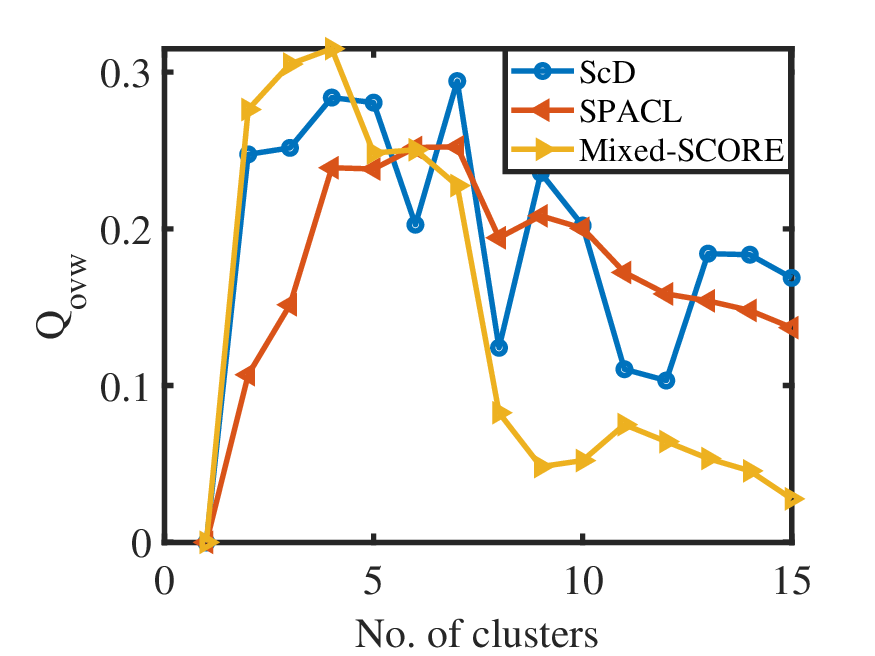}}
\subfigure[\scriptsize Human-gene1]{\includegraphics[width=0.25\textwidth]{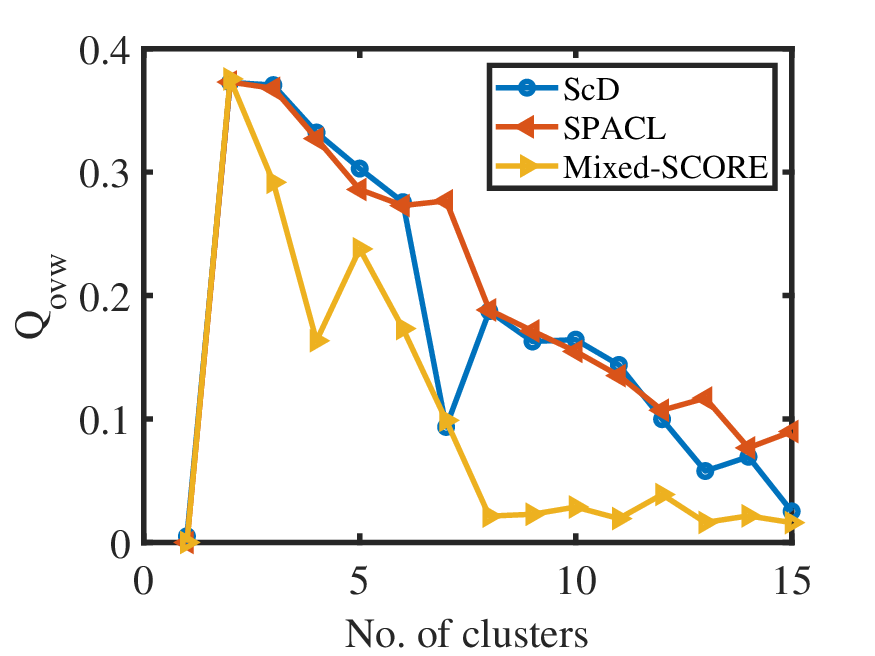}}
\caption{Overlapping weighted modularity $Q_{ovw}$ computed by Equation (\ref{Modularity}) against the number of clusters for real data used in this paper.}
\label{Qreal} 
\end{figure}

For the task of determining the number of communities, we apply KScD, KSPACL, KMixed-SCORE, and their competitors to real-world networks in Table \ref{realdata}. Here, for comparison, we also consider another three approaches BHa, BHm, and BHmc designed based on the Bethe Hessian matrix in \cite{le2022estimating}. Figure \ref{Qreal} displays the overlapping weighted modularity via Equation (\ref{Modularity}) using ScD (and SPACL and Mixed-SCORE) for different choices of the number of communities. From Figure \ref{Qreal}, we can directly find the $k$ maximizing $Q_{ovw}$ for each method. Meanwhile, results shown in Table \ref{Qreal} also indicate that the overlapping weighted modularity of ScD is comparable to its competitors as $K$ increases for all data considered in this article. Table \ref{realdataEstimatedK} shows the estimated number of communities of KScD, KSPACL, KMixed-SCORE, and their competitors for real networks considered in this paper. For networks with known $K$, KScD, KSPCAL, and KMixed-SCORE correctly infer $K$ for all networks except for the Dophins network while NB, BHac, BHa, BHm, and BHmc fail to determine $K$ for Political blogs and all weighted networks. Again, the fact that our methods significantly outperform NB, BHac, BHa, BHm, and BHmc for the task of inferring $K$ on both un-weighted and weighted real-world networks, suggests the effectiveness and advantage of our overlapping weighted modularity in measuring the quality of overlapping community detection. For Train bombing, Les Mis\'erables, and US Top-500 Airport Network, our methods designed based on overlapping weighted modularity infer their $K$ as 2 while $K$ estimated by NB, BHac, BHa, BHm, and BHmc is larger. For CE-CX and Human-gene1, our methods determine their $K$ as 5 and 2, respectively, whereas their competitors fail to output due to memory constraints. For WormNet-v3, KScD and KSPACL determine its $K$ as 7 which is different from that of KMixed-SCORE. As a result, for networks with unknown $K$, we tend to believe that $K$ estimated by our KScD, KSPACL, and KMixed-SCORE is more accurate than their competitors because our approaches outperform their competitors for both simulated weighted networks and real-world networks with known $K$.

For the task of estimating memberships, Table \ref{realdataQ} reports the overlapping weighted modularity $Q_{ovw}$ of ScD, GeoNMF, SPACL, and Mixed-SCORE. The results show that all methods have comparable or identical $Q_{ovw}$ values for all networks, except for the CE-CX network. In particular, from Table \ref{realdataQ}, we find that $Q_{ovw}$ of Karate-club-weighted is larger than that of Karate club, which supports the statement in \cite{newman2004analysis} that weights provide more useful information about community structure than edges. Meanwhile, all methods process the Human-gene1 network of up to 21853 nodes within 14 seconds when there are two communities.
\begin{table}[h!]
\footnotesize
	\centering
	\resizebox{\columnwidth}{!}{
	\begin{tabular}{cccccccccccc}
\hline\hline
Dataset&True $K$&KScD&KSPACL&KMixed-SCORE&NB&BHac&BHa&BHm&BHmc\\
\hline
Karate club&2&2&2&2&2&2&2&2&2\\
Dolphins&2&3&3&4&2&2&2&2&2\\
UKfaculty&3&3&3&3&3&3&3&3&3\\
Political books&2&2&2&2&2&3&3&2&3\\
Political blogs&2&2&2&2&7&8&7&7&8\\
Slovene Parliamentary Party&2&2&2&2&N/A&N/A&N/A&N/A&N/A\\
Gahuku-Gama subtribes&3&3&3&3&1&13&12&1&N/A\\
Karate-club-weighted&2&2&2&2&4&4&4&4&4\\
Train bombing&Unknown&2&2&2&3&4&4&3&3\\
Les Mis\'erables&Unknown&2&2&2&6&7&7&6&7\\
US Top-500 Airport Network&Unknown&2&2&2&147&158&158&147&147\\
CE-CX&Unknown&5&5&5&N/A&N/A&N/A&N/A&N/A\\
WormNet-v3&Unknown&7&7&4&N/A&N/A&N/A&N/A&N/A\\
Human-gene1&Unknown&2&2&2&N/A&N/A&N/A&N/A&N/A\\
\hline\hline
\end{tabular}
}
	\caption{Estimated number of communities by KScD, KSPACL, KMixed-SCORE, and their competitors for real-world networks in Table \ref{realdata}.}\label{realdataEstimatedK}
\end{table}

\begin{table}[h!]
\footnotesize
	\centering
\resizebox{\columnwidth}{!}{
	\begin{tabular}{cccccccccccc}
\hline\hline
Dataset&ScD&GeoNMF&SPACL&Mixed-SCORE\\
\hline
Karate club&\textbf{0.3715}&\textbf{0.3715}&\textbf{0.3715}&\textbf{0.3715}\\
Dolphins&0.3848&0.3848&0.3848&\textbf{0.3899}\\
UKfaculty&0.4384&0.4362&0.4333&\textbf{0.4393}\\
Political books&\textbf{0.4786}&\textbf{0.4786}&\textbf{0.4786}&\textbf{0.4786}\\
Political blogs&\textbf{0.4251}&0.4249&0.4249&0.4249\\
Slovene Parliamentary Party&\textbf{0.4839}&\textbf{0.4839}&\textbf{0.4839}&\textbf{0.4839}\\
Gahuku-Gama subtribes&0.3825&0.4149&0.4149&\textbf{0.4310}\\
Karate-club-weighted&\textbf{0.4036}&\textbf{0.4036}&\textbf{0.4036}&\textbf{0.4036}\\
Train bombing&\textbf{0.3450}&0.2792&\textbf{0.3450}&0.3411\\
Les Mis\'erables&\textbf{0.3814}&\textbf{0.3814}&\textbf{0.3814}&\textbf{0.3814}\\
US Top-500 Airport Network&0.2310&0.2297&0.2297&0.2309\\
CE-CX&\textbf{0.5120}&0.4900&0.4903&0.3566\\
WormNet-v3&\textbf{0.2943}&0.2552&0.2523&0.2276\\
Human-gene1&0.3727&0.3754&0.3730&\textbf{0.3755}\\
\hline\hline
\end{tabular}
}
\caption{Overlapping weighted modularity $Q_{ovw}$ (calculated by Equation (\ref{Modularity}) of ScD and its competitors for real-world networks in Table \ref{realdata}. For networks with known $K$, all four methods use the true $K$ to compute $Q_{ovw}$. For networks with unknown $K$, to compute $Q_{ovw}$, all methods use $K$ estimated by KScD in Table \ref{realdataEstimatedK}. For each data, the largest $Q_{ovw}$ is marked in bold.}\label{realdataQ}
\end{table}

\begin{figure}
\centering
\resizebox{\columnwidth}{!}{
\subfigure[Karate club]{\includegraphics[width=0.32\textwidth]{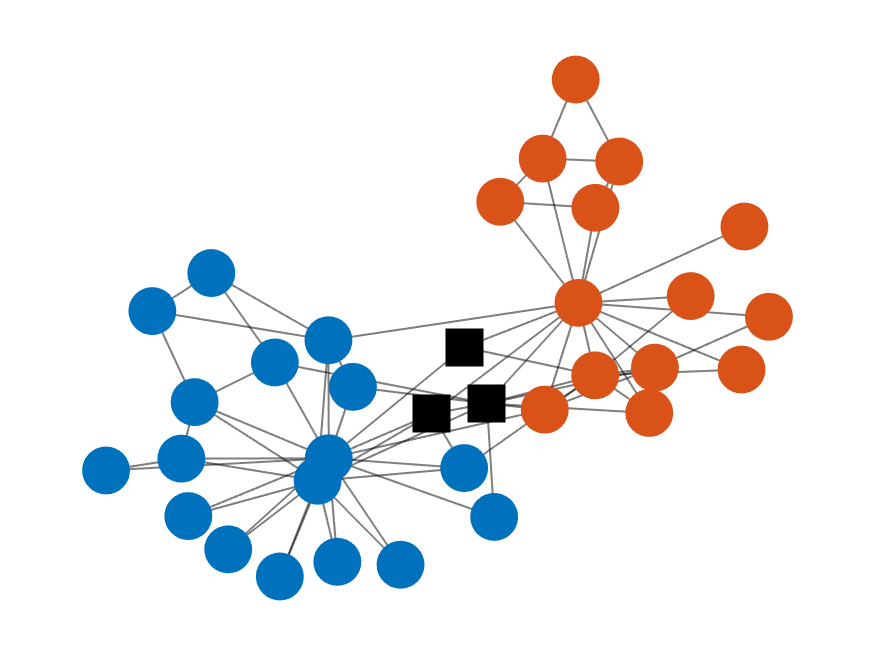}}
\subfigure[Dolphins]{\includegraphics[width=0.32\textwidth]{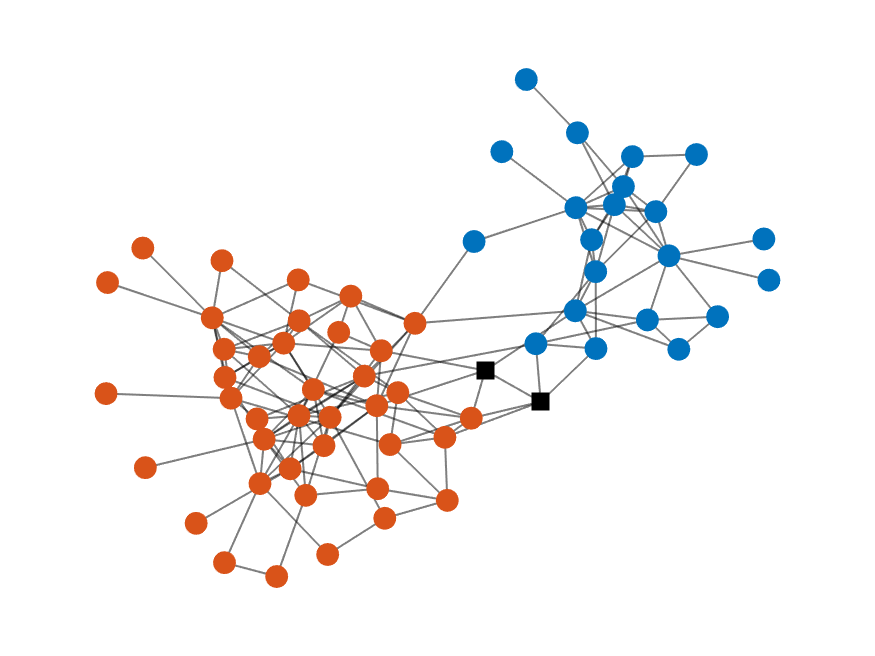}}
\subfigure[UKfaculty]{\includegraphics[width=0.32\textwidth]{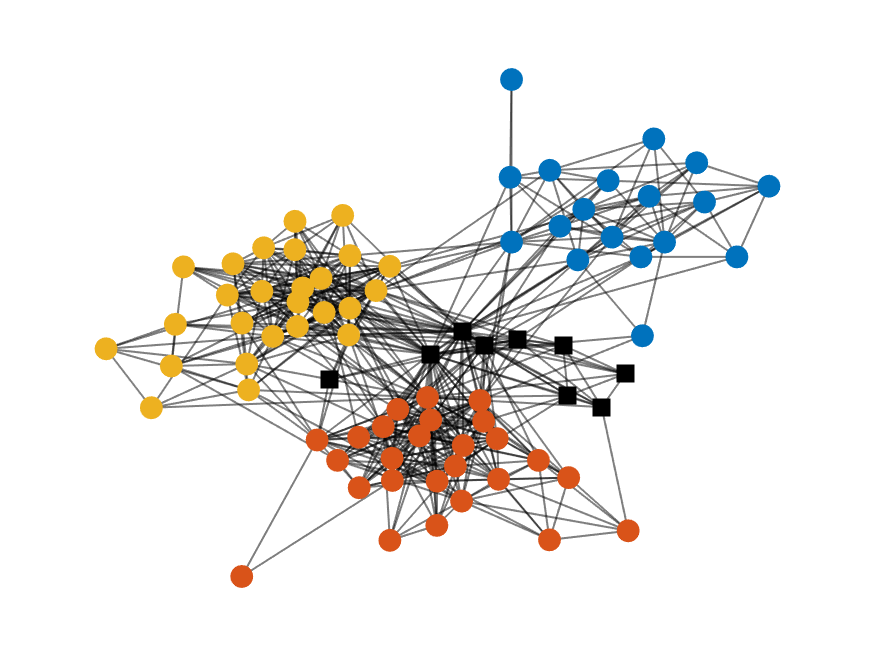}}
\subfigure[Political books]{\includegraphics[width=0.32\textwidth]{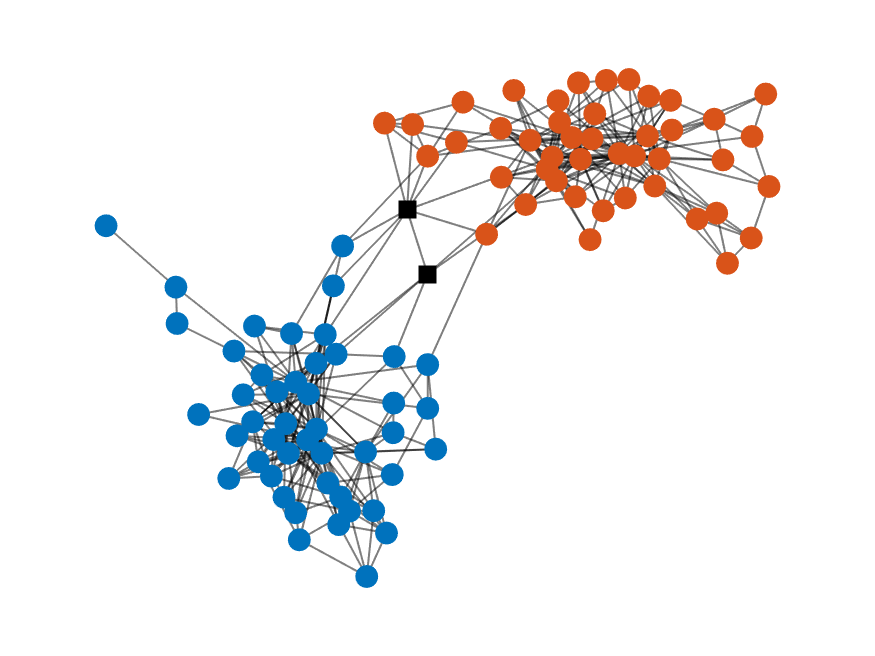}}
}
\resizebox{\columnwidth}{!}{
\subfigure[Political blogs]{\includegraphics[width=0.32\textwidth]{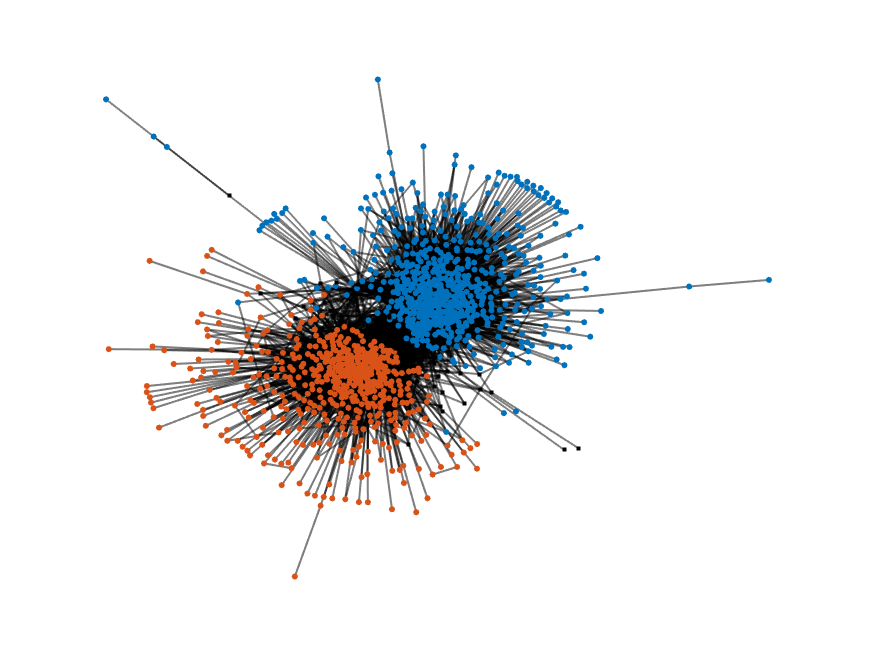}}
\subfigure[Slovene Parliamentary Party]{\includegraphics[width=0.32\textwidth]{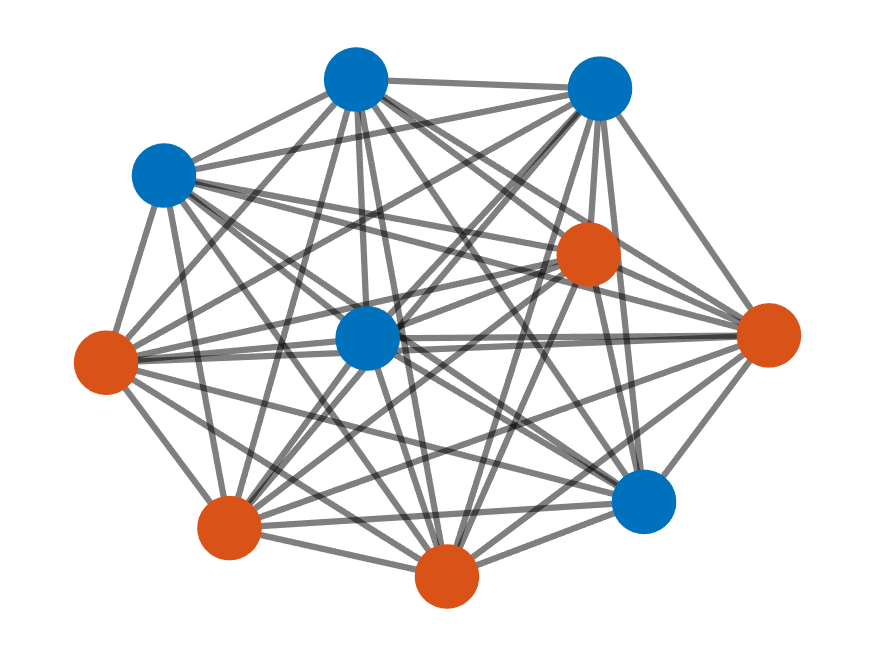}}
\subfigure[Gahuku-Gama subtribes]{\includegraphics[width=0.32\textwidth]{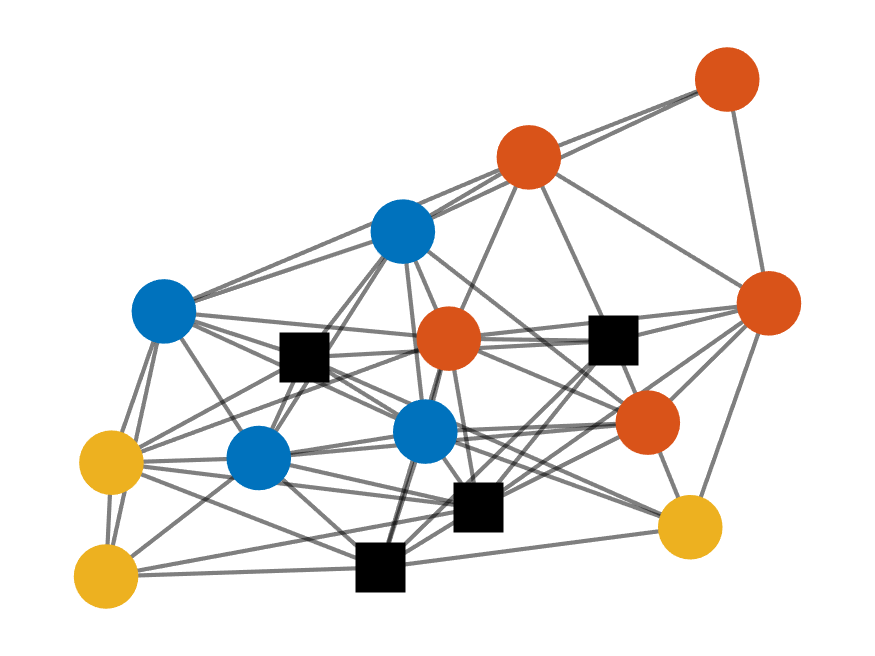}}
\subfigure[Karate-club-weighted]{\includegraphics[width=0.32\textwidth]{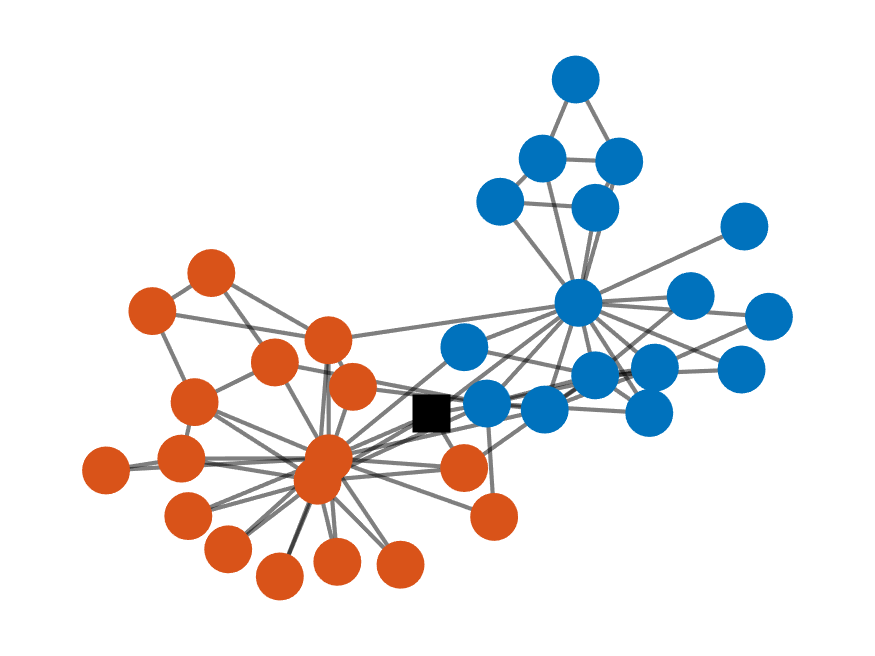}}
}
\resizebox{\columnwidth}{!}{
\subfigure[Train bombing]{\includegraphics[width=0.32\textwidth]{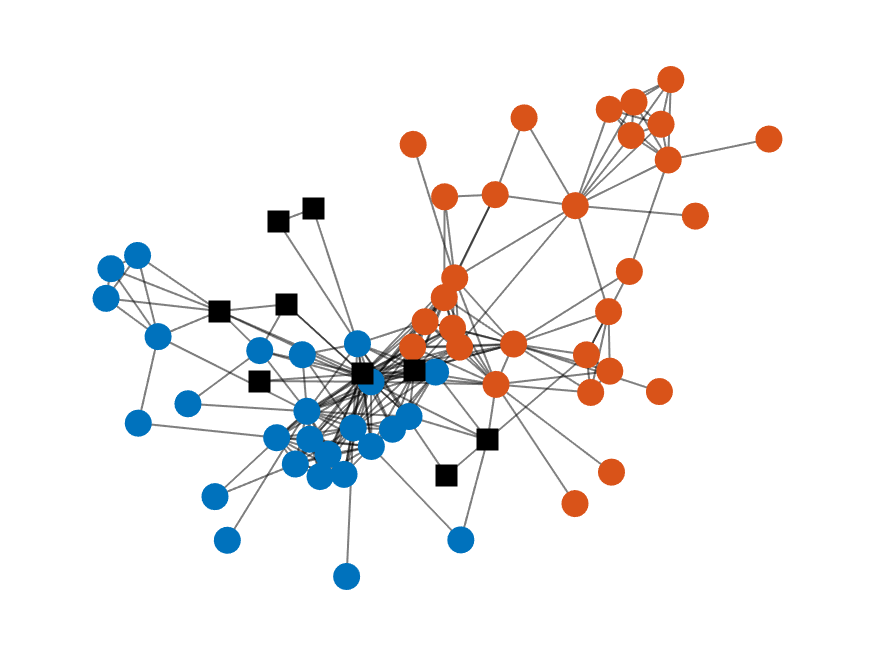}}
\subfigure[Les Mis\'erables]{\includegraphics[width=0.32\textwidth]{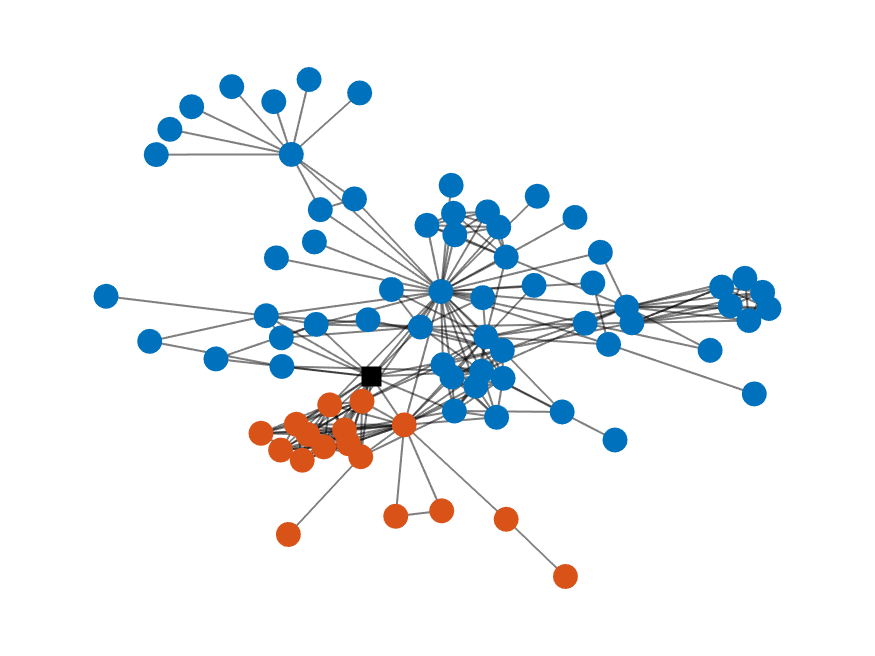}}
\subfigure[US Top-500 Airport Network]{\includegraphics[width=0.32\textwidth]{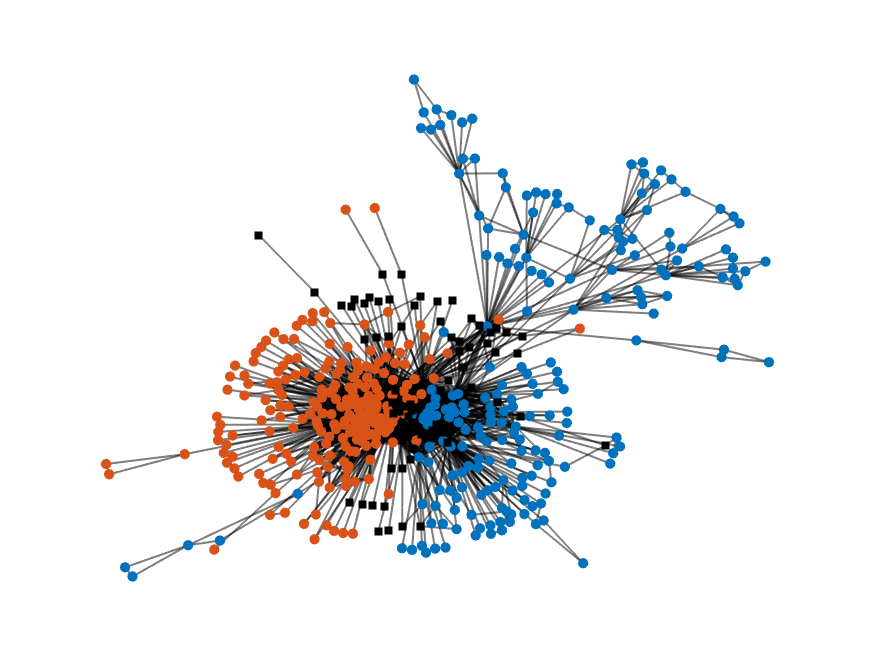}}
\subfigure[CE-CX]{\includegraphics[width=0.32\textwidth]{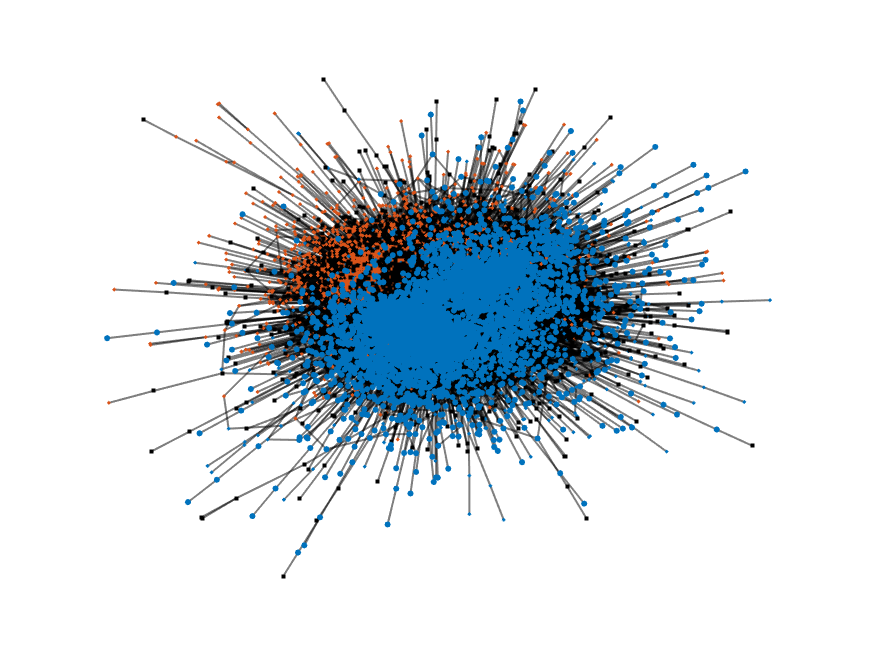}}
}
\subfigure[\scriptsize WormNet-v3]{\includegraphics[width=0.25\textwidth]{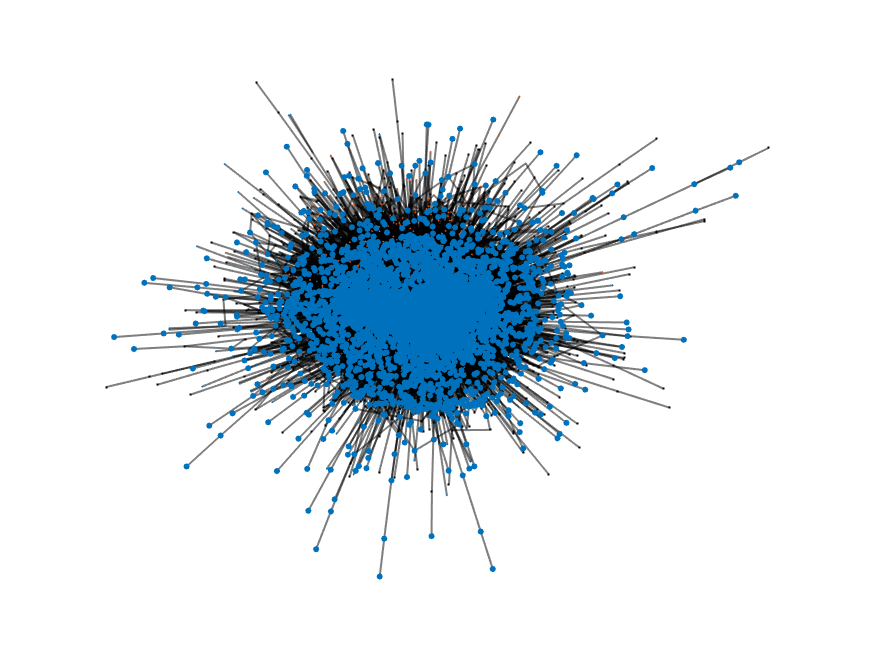}}
\subfigure[\scriptsize Human-gene1]{\includegraphics[width=0.25\textwidth]{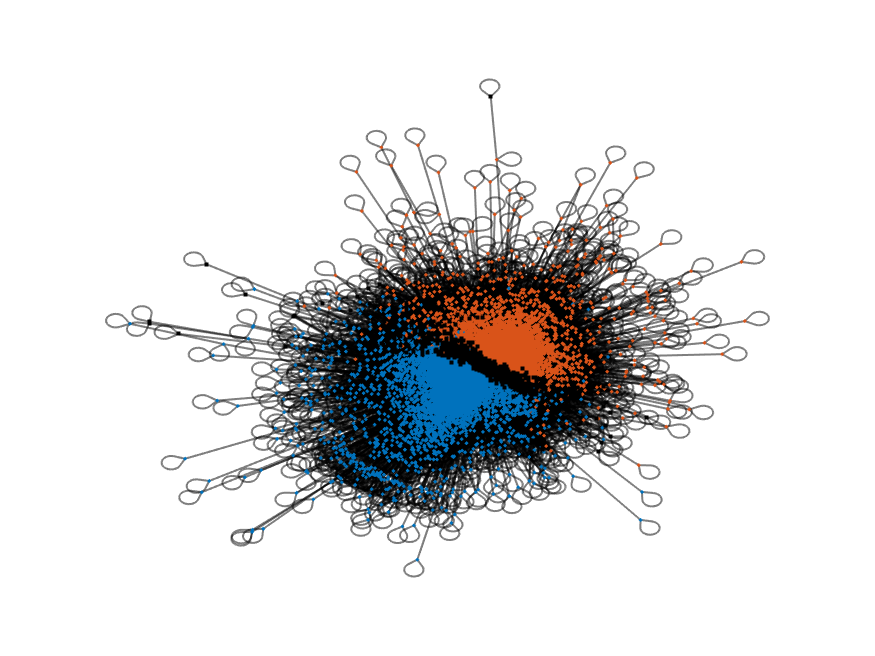}}
\caption{Communities detected by ScD, where $K$ used here is the same as that of Table \ref{realdata3indices} for each network. Colors indicate home base communities obtained from $\ell$ and the black square indicates highly mixed nodes, where we let $\ell(i)=\mathrm{arg~max}_{k\in[K]}\hat{\Pi}(i,k)$ be the home base community for node $i\in[n]$. For visualization, we do not show node labels and edge weights.}
\label{NetReal} 
\end{figure}

More than our overlapping weighted modularity, to have a better understanding of community structure for real data, we define three indices $\zeta_{\mathrm{mixed}}, \zeta_{\mathrm{pure}}$, and $\varsigma$ to measure the mixedness, the purity, and the equilibrium of a real network. Call node $i$ a highly mixed node if $\mathrm{max}_{k\in[K]}\hat{\Pi}(i,k)\leq0.7$ and a highly pure node if $\mathrm{max}_{k\in[K]}\hat{\Pi}(i,k)\geq0.9$ for $i\in[n]$. $\zeta_{\mathrm{mixed}}$ and $\zeta_{\mathrm{pure}}$ are the proportions of highly mixed nodes and highly pure nodes in a network, respectively. Larger $\zeta_{\mathrm{mixed}}$ (and $\zeta_{\mathrm{pure}}$) means a higher proportion of highly mixed (and pure) nodes in a network. $\varsigma$ is defined as $\varsigma=\frac{\mathrm{min}_{k\in[K]}\|\hat{\Pi}(:,k)\|_{1}}{\mathrm{max}_{k\in[K]}\|\hat{\Pi}(:,k)\|_{1}}$. $\varsigma$ ranges in $(0,1]$ and a larger $\varsigma$ indicates a more balanced network. Table \ref{realdata3indices} shows the three indices detected by ScD for real data considered in this paper and we have the following conclusions:
\begin{itemize}
  \item For Karate club, it has $34\times0.882\approx3$ highly mixed nodes and $34\times0.7353\approx25$ highly pure nodes. This data is quite balanced for its large equilibrium index $\varsigma$.
  \item For Dolphins, it has $62\times0.0323\approx2$ highly mixed nodes and $62\times0.8871\approx55$ highly pure nodes. This data is quite unbalanced for its small $\varsigma$.
  \item For UKfaculty, it has $79\times 0.1139\approx9$ highly mixed nodes and $79\times0.6582\approx52$ highly pure nodes. It is more balanced than the Dolphins and more unbalanced than the Karate club.
  \item For Political books, it has $92\times0.0217\approx2$ highly mixed nodes and $92\times0.9457\approx87$ highly pure nodes. It is the most balanced network among all networks because its equilibrium index $\varsigma$ is 0.9771, a value close to 1.
  \item For Political blogs, it has $1222\times0.0426\approx52$ highly mixed nodes and $1222\times0.8331\approx1018$ highly pure nodes.
  \item For Slovene Parliamentary Party, it has zero highly mixed nodes and 7 highly pure nodes. It is quite balanced because 0.9199 is a value close to 1.
  \item For Gahuku-Gama subtribes, it has 4 highly mixed nodes and 8 highly pure nodes. Its equilibrium index is 0.7159, so it is not that balanced.
  \item For Karate-club-weighted, the results are similar to that of Karate club, and we omit the analysis here.
  \item For Train bombing and US Top-500 Airport Network, their mixedness, purity, and equilibrium are similar to that of UKfaculty, and we omit the analysis here.
  \item For Les Mis\'erables, it has $77\times 0.0130\approx1$ highly mixed node and $77\times0.8052\approx62$ highly pure nodes. It is quite unbalanced for its small equilibrium index.
  \item For CE-CX and WormNet-v3, both of them have a large proportion of highly mixed nodes and they are even more unbalanced than Les Mis\'erables.
  \item For Human-gene1, most of its nodes are highly pure and it is much more balanced than the other two biological networks CE-CX and WormNet-v3.
\end{itemize}
For visibility, Figure \ref{NetReal} displays communities detected by ScD and we also highlight highly mixed nodes for each data.
\begin{table}[h!]
\footnotesize
\centering
\resizebox{\columnwidth}{!}{
	\begin{tabular}{cccccccccccc}
\hline\hline
Dataset&$\zeta_{\mathrm{mixed}}$&$\zeta_{\mathrm{pure}}$&$\varsigma$\\
\hline
Karate club&0.0882&0.7353&0.9227\\
Dolphins&0.0323&0.8871&0.5709\\
UKfaculty&0.1139&0.6582&0.8041\\
Political books&0.0217&0.9457&0.9771\\
Political blogs&0.0426&0.8331&0.8288\\
Slovene Parliamentary Party&0&0.7000&0.9199\\
Gahuku-Gama subtribes&0.2500&0.5000&0.7159\\
Karate-club-weighted&0.0294&0.7941&0.9375\\
Train bombing&0.1406&0.6250&0.8937\\
Les Mis\'erables&0.0130&0.8052&0.3335\\
US Top-500 Airport Network&0.1480&0.6700&0.8981\\
CE-CX&0.2551&0.5355&0.1043\\
WormNet-v3&0.4752&0.2497&0.2578\\
Human-gene1&0.0460&0.8902&0.7838\\
\hline\hline
\end{tabular}
}
\caption{$\zeta_{\mathrm{mixed}},\zeta_{\mathrm{pure}},\varsigma$ detected by ScD for real data in Table \ref{realdata}. For networks with known $K$, ScD uses the true $K$ to estimate mixed memberships. For networks with unknown $K$, ScD uses $K$ estimated by KScD in Table \ref{realdataEstimatedK} to estimate mixed memberships.}\label{realdata3indices}
\end{table}

\section{Conclusion}\label{SecConclusion}
Overlapping community detection plays an important role in understanding latent community memberships of complex networks. In this article, we present the Weighted Degree-Corrected Mixed Membership (WDCMM) model for overlapping weighted networks. WDCMM releases
DCMM’s distribution restriction by allowing the elements of the adjacency matrix to be generated from distributions more than Bernoulli. In this way, WDCMM can model weighted networks in which nodes belong to multiple communities and edge weights can be generated from any distribution as long as the expectation adjacency matrix $\Omega$ has a block structure related to the community memberships.

We use an efficient spectral algorithm to estimate the community memberships of weighted networks generated from WDCMM. We show that the algorithm is asymptotically consistent under mild conditions for any distribution and we can obtain a theoretical guarantee of consistency for a specific distribution from our main theoretical results immediately by analyzing the variance parameter related to the distribution. We present some examples of how to analyze the algorithm's performance when the adjacency matrix is generated from different distributions under our WDCMM.

We also present overlapping weighted modularity to evaluate the quality of overlapping community partition for both assortative and dis-assortative weighted networks with positive and negative edge weights. We propose a method to determine the number of communities based on the overlapping weighted modularity. Experiments support our theoretical analysis and show that some classical spectral algorithms also successfully estimate community memberships for weighted networks generated from the proposed model. The advantage and effectiveness of our overlapping weighted modularity are supported by substantial experimental results that our method successfully estimates the number of communities for weighted networks and it outperforms its competitors.

Our work has several implications: First, the ground truth
community memberships of overlapping weighted networks generated from WDCMM provide a reliable way for evaluating mixed membership community detection approaches. Second, the algorithms studied in this paper can broaden our understanding of the community structure of weighted networks. Third, our overlapping weighted modularity sheds light on evaluating the quality of overlapping community partitions for both assortative and dis-assortative weighted networks with positive and negative edge weights. And last, our KScD provides a baseline for determining the number of communities for weighted networks.

Future works will be studied from five aspects. First,a theoretical guarantee of Mixed-SCORE under WDCMM should be developed. Second, given a weighted network, analyzing which distribution the edge weights are coming from is an interesting and challenging task. Third, extending WDCMM and overlapping weighted modularity from un-directed weighted networks to directed weighted networks is appealing. Fourth, the ScD algorithm and its competitors can be accelerated by some random-projection techniques \citep{zhang2022randomized} to detect mixed memberships of large-scale weighted networks. Fifth, extending the higher-order spectral clustering developed in \citep{guo2024efficacy} from weighted SBMs to the WDCMM model introduced in this work is appealing. And last, in this paper, we estimate $K$ by maximizing our overlapping weighted modularity for both assortative and dis-assortative networks. Nevertheless, this method fails to infer $K$ in cases where the elements of $P$ are not restricted. Consequently, the development of a more general method, equipped with theoretical guarantees, for estimating $K$ in weighted networks generated from models such as WSBMs and our WDCMM, presents a challenging yet appealing avenue for future research.

\bmhead{Acknowledgements} \textbf{Huan Qing} was sponsored by the Scientific Research Foundation of Chongqing University of Technology (Grant No. 2024ZDR003), and the Science and Technology Research Program of Chongqing Municipal Education Commission (Grant No. KJQN202401168).
\bmhead{Author Contributions} \textbf{Huan Qing:} Conceptualization; Data curation; Formal analysis; Funding acquisition; Methodology; Project administration; Resources; Software; Validation; Visualization; Writing-original draft; Writing-review $\&$ editing.
\section*{Declarations}
\textbf{Conflict of interest} The author declares no conflict of interest.
\begin{appendices}
\section{Proof of Theorem \ref{Main}}\label{AppendixMain}
\begin{proof}
Let $H_{\hat{U}}=\hat{U}'U$. Let $H_{\hat{U}}=U_{H_{\hat{U}}}\Sigma_{H_{\hat{U}}}V'_{H_{\hat{U}}}$ be the top $K$ singular value decomposition of $H_{\hat{U}}$. Let $\mathrm{sgn}(H_{\hat{U}})=U_{H_{\hat{U}}}V'_{H_{\hat{U}}}$. Under the WDCMM, the following results are useful for our theoretical analysis.
\begin{itemize}
  \item $\mathbb{E}[A(i,j)-\Omega(i,j)]=0$ under the WDCMM.
  \item $\mathbb{E}[(A(i,j)-\Omega(i,j))^{2}]=\mathrm{Var}(A(i,j))\leq \gamma\theta(i)\theta(j)\leq \gamma\theta^{2}_{\mathrm{max}}$.
  \item $|A(i,j)-\Omega(i,j)|\leq \tau$ for $i,j\in[n]$.
  \item Let $\mu=\frac{n\|U\|^{2}_{2\rightarrow\infty}}{K}$ be the incoherence parameter of $\Omega$ defined in Equation (4.23) \cite{chen2021spectral}. By Lemma H.1 of \cite{MaoSVM} and Lemma A1 of \cite{Criterion}, we have $\frac{\theta^{2}_{\mathrm{min}}}{\theta^{2}_{\mathrm{max}}K\lambda_{1}(\Pi'\Pi)}\leq \|U\|^{2}_{2\rightarrow\infty}\leq\frac{\theta^{2}_{\mathrm{max}}}{\theta^{2}_{\mathrm{min}}\lambda_{K}(\Pi'\Pi)}$, which gives $\frac{\theta^{2}_{\mathrm{min}}n}{\theta^{2}_{\mathrm{max}}K^{2}\lambda_{1}(\Pi'\Pi)}\leq\mu\leq\frac{\theta^{2}_{\mathrm{max}}n}{\theta^{2}_{\mathrm{min}}K\lambda_{K}(\Pi'\Pi)}$. By Condition \ref{condition}, we have $\mu=O(1)$.
  \item Let $c_{b}=\frac{\tau}{\theta_{\mathrm{max}}\sqrt{\gamma n/(\mu \mathrm{log}(n))}}$. By Assumption \ref{assumesparsity}, Condition \ref{condition}, and the fact that $\mu=O(1)$, we have $c_{b}=O(\frac{\tau\sqrt{\mathrm{log}(n)}}{\theta_{\mathrm{max}}\sqrt{\gamma n}})=O(\sqrt{\frac{\tau^{2}\mathrm{log}(n)}{\gamma\theta^{2}_{\mathrm{max}}n}})=O(\sqrt{\frac{\tau^{2}\mathrm{log}(n)}{\gamma\theta_{\mathrm{max}}\|\theta\|_{1}}})\leq O(1)$.
\end{itemize}

The above five bullets guarantee that Equations (4.28) and (4.29) of Assumption 4.1. in \cite{chen2021spectral} hold. Therefore, by Theorem 4.2.1 \cite{chen2021spectral}, with probability exceeding $1-O(\frac{1}{n^{5}})$, we have
\begin{align*}
\|\hat{U}\mathrm{sgn}(H_{\hat{U}})-U\|_{2\rightarrow\infty}\leq C\frac{\theta_{\mathrm{max}}\kappa(\Omega)\sqrt{\mu\gamma K}+\theta_{\mathrm{max}}\sqrt{\gamma K\mathrm{log}(n)}}{\sigma_{K}(\Omega)},
\end{align*}
provided that $\sigma_{K}(\Omega)\gg \theta_{\mathrm{max}}\sqrt{\gamma n\mathrm{log}(n)}$. Since $\mu=O(1)$ and $K=O(1)$, we have
\begin{align*}
\|\hat{U}\mathrm{sgn}(H_{\hat{U}})-U\|_{2\rightarrow\infty}=O(\frac{\theta_{\mathrm{max}}\kappa(\Omega)\sqrt{\gamma}+\theta_{\mathrm{max}}\sqrt{\gamma\mathrm{log}(n)}}{\sigma_{K}(\Omega)}).
\end{align*}

For convenience, set $\varpi=\|\hat{U}\hat{U}'-UU'\|_{2\rightarrow\infty}$ as the row-wise eigenspace error. Since $\|\hat{U}\hat{U}'-UU'\|_{2\rightarrow\infty}\leq2\|U-\hat{U}\mathrm{sgn}(H_{\hat{U}})\|_{2\rightarrow\infty}$,
we have
\begin{align*}	
\varpi=O(\frac{\theta_{\mathrm{max}}\kappa(\Omega)\sqrt{\gamma}+\theta_{\mathrm{max}}\sqrt{\gamma\mathrm{log}(n)}}{\sigma_{K}(\Omega)}).
\end{align*}

Since $\sigma_{K}(\Omega)=\sigma_{K}(\Theta \Pi P\Pi'\Theta)\geq\theta^{2}_{\mathrm{min}}\sigma_{K}(\Pi P\Pi')\geq\theta^{2}_{\mathrm{min}}\sigma_{K}(P)\lambda_{K}(\Pi'\Pi)$ and $\sigma_{1}(\Omega)=\sigma_{1}(\Theta \Pi P\Pi'\Theta)\leq\theta^{2}_{\mathrm{max}}\sigma_{1}(\Pi P\Pi')\leq\theta^{2}_{\mathrm{max}}\sigma_{1}(P)\lambda_{1}(\Pi'\Pi)$ under $WDCMM_{n}(K, P,\Pi, \Theta, \mathcal{F})$, by Condition \ref{condition}, we have $\kappa(\Omega)=O(1)$, which gives
\begin{align}\label{rowwise}	
\varpi&=O(\frac{\theta_{\mathrm{max}}\kappa(\Omega)\sqrt{\gamma}+\theta_{\mathrm{max}}\sqrt{\gamma\mathrm{log}(n)}}{\theta^{2}_{\mathrm{min}}\sigma_{K}(P)\lambda_{K}(\Pi'\Pi)})
=O(\frac{\kappa(\Omega)\sqrt{\gamma}+\sqrt{\gamma\mathrm{log}(n)}}{\theta_{\mathrm{min}}\sigma_{K}(P)n})=O(\frac{\sqrt{\gamma\mathrm{log}(n)}}{\theta_{\mathrm{min}}\sigma_{K}(P)n})\notag\\
&=O(\frac{\sqrt{\gamma\mathrm{log}(n)}}{\sigma_{K}(P)n\sqrt{\rho}}).
\end{align}

Corollary 2 of \cite{Criterion} is a distribution-free theoretical result, it says that there exists a permutation matrix $\mathcal{P}\in\mathbb{R}^{K\times K}$ such that for $i\in[n]$,
\begin{align}\label{FinalBound}	
\|e'_{i}(\hat{\Pi}-\Pi\mathcal{P})\|_{1}=O(\varpi\sqrt{n})=O(\frac{\sqrt{\gamma\mathrm{log}(n)}}{\sigma_{K}(P)\sqrt{\rho n}}).
\end{align}

Finally, by Condition \ref{condition}, since $\sigma_{K}(\Omega)\geq\theta^{2}_{\mathrm{min}}\sigma_{K}(P)\lambda_{K}(\Pi'\Pi)=O(\sigma_{K}(P)\rho n)$. To make the requirement $\sigma_{K}(\Omega)\gg \theta_{\mathrm{max}}\sqrt{\gamma n\mathrm{log}(n)}=O(\sqrt{\gamma\rho n\mathrm{log}(n)})$ hold, we only need $\sigma_{K}(P)\rho n\gg\sqrt{\gamma \rho n\mathrm{log}(n)}\Leftrightarrow\sigma_{K}(P)\gg\sqrt{\frac{\gamma\mathrm{log}(n)}{\rho n}}$, and this holds naturally because we need the theoretical upper bound of ScD's error rate provided in Equation (\ref{FinalBound}) go to zero.
\end{proof}
\section{Proof of $\tau$'s FWHP property}\label{ExtraSec}
\begin{lem}\label{NormalTau}
When $\mathcal{F}$ is Normal distribution such that $A(i,j)\sim \mathrm{Normal}(\Omega(i,j),\sigma^{2}_{A})$ for $i,j\in[n]$, $\tau$ is finite with high probability. In detail, with probability at least $1-\alpha$, we have
\begin{align*}
\tau\leq\sigma_{A}\sqrt{-2\mathrm{log}(\frac{\alpha}{n(n+1)})}.
\end{align*}
\end{lem}

\begin{proof}
For any random variable $x$ following a standard normal distribution $\mathrm{Normal}(0,1)$, by Theorem 4.1.1 (Matrix Gaussian $\&$ Rademacher Series) of \citep{tropp2015introduction}, we know that $\mathbb{P}(|x|\geq t)\leq2\mathrm{exp}(-\frac{t^{2}}{2})$ for all $t\geq0$. Thus, when $A(i,j)\sim \mathrm{Normal}(\Omega(i,j),\sigma^{2}_{A})$, we have $\mathbb{P}(|\frac{A(i,j)-\Omega(i,j)}{\sigma_{A}}|\geq t)\leq2\mathrm{exp}(-\frac{t^{2}}{2})$ for $i,j\in[n]$. Since $A$ is symmetric, there are $\frac{n(n+1)}{2}$ independent entries in $A$. Using the union bound, the probability that any entry deviates by more than $t\sigma_{A}$ is
\begin{align*}
\mathbb{P}(\mathrm{max}_{i,j\in[n]}|A(i,j)-\Omega(i,j)|\geq t\sigma_{A})=\mathbb{P}(\tau\geq t\sigma_{A})\leq n(n+1)\mathrm{exp}(-\frac{t^{2}}{2}).
\end{align*}

To ensure that $\tau\leq t\sigma_{A}$ with high probability, we set the right hand side to a small probability $\alpha$:
\begin{align*}
n(n+1)\mathrm{exp}(-\frac{t^{2}}{2})\leq\alpha,
\end{align*}
which gives that
\begin{align*}
t\geq\sqrt{-2\mathrm{log}(\frac{\alpha}{n(n+1)})}.
\end{align*}

Therefore, with probability at least $1-\alpha$, we have
\begin{align*}
\tau=\mathrm{max}_{i,j\in[n]}|A(i,j)-\Omega(i,j)|\leq\sigma_{A}\sqrt{-2\mathrm{log}(\frac{\alpha}{n(n+1)})}.
\end{align*}

For example, if we let $\alpha=\frac{1}{n^{2}}$, then with probability at least $1-\frac{1}{n^{2}}$, we have $\tau\leq\sigma_{A}\sqrt{2\mathrm{log}(n^{3}(n+1))}$. Suppose that $n=10000$ and $\sigma_{A}=1$, we see that with probability at least $1-\frac{1}{10000^{2}}$, we have $\tau\leq\sqrt{2\mathrm{log}(10000^{3}(10000+1))}\approx8.5839$.
\end{proof}

\begin{lem}\label{PoissonTau}
When $\mathcal{F}$ is Poisson distribution such that $A(i,j)\sim \mathrm{Poisson}(\Omega(i,j))$ for $i,j\in[n]$, $\tau$ is finite with high probability.
\end{lem}

\begin{proof}
First, we focus on the upper bound of $\tau_{i,j}\equiv|A(i,j)-\Omega(i,j)|$ for $i,j\in[n]$. If $A(i,j)\leq\Omega(i,j)$, the upper bound of $\tau_{i,j}$ is $\Omega(i,j)$, a value smaller than $\Omega_{\mathrm{max}}\equiv\mathrm{max}_{i,j\in[n]}\Omega(i,j)$. Otherwise, by Bennet inequality given in Equation (26) of \citep{zhang2020non}, for any $t>0$, we have
\begin{align*}
\mathbb{P}(\tau_{i,j}\geq t)&\leq\mathrm{exp}(-\frac{t^{2}}{2\Omega(i,j)}\frac{(1+t/\Omega(i,j))\mathrm{log}(1+t/\Omega(i,j))-t/\Omega(i,j)}{(t^{2}/\Omega^{2}(i,j))/2})\\
&=\mathrm{exp}(t-(\Omega(i,j)+t)\mathrm{log}(1+\frac{t}{\Omega(i,j)}))\\
&=\mathrm{exp}(t+(t+\Omega(i,j))\mathrm{log}(\Omega(i,j))-(t+\Omega(i,j))\mathrm{log}(t+\Omega(i,j))).
\end{align*}

Let $g(\Omega(i,j))=t+(t+\Omega(i,j))\mathrm{log}(\Omega(i,j))-(t+\Omega(i,j))\mathrm{log}(t+\Omega(i,j))$ be a function of $\Omega(i,j)$. We get
\begin{align*}
g'(\Omega(i,j))&=\mathrm{log}(\Omega(i,j))+\frac{t}{\Omega(i,j)}-\mathrm{log}(t+\Omega(i,j)),\\
g''(\Omega(i,j))&=\frac{1}{\Omega(i,j)}-\frac{t}{\Omega^{2}(i,j)}-\frac{1}{t+\Omega(i,j)}=-\frac{t^{2}}{\Omega^{2}(i,j)(t+\Omega(i,j))}<0.
\end{align*}

Thus, $g'(\Omega(i,j))$ is monotonically decreasing with respect to $\Omega(i,j)$. Since $\mathrm{lim}_{\Omega(i,j)\rightarrow+\infty}g'(\Omega(i,j))=0$, we see that $g'(\Omega(i,j))>0$ for all $\Omega(i,j)>0$. Thus, we get that $g(\Omega(i,j))$ is monotonically decreasing with respect to $\Omega(i,j)$. We get
\begin{align*}
\mathbb{P}(\tau_{i,j}&\geq t)\leq\mathrm{exp}(t+(t+\Omega(i,j))\mathrm{log}(\Omega(i,j))-(t+\Omega(i,j))\mathrm{log}(t+\Omega(i,j)))\\
&\leq\mathrm{exp}(t+(t+\Omega_{\mathrm{max}})\mathrm{log}(\Omega_{\mathrm{max}})-(t+\Omega_{\mathrm{max}})\mathrm{log}(t+\Omega_{\mathrm{max}})).
\end{align*}

Then, using the union bound, we get
\begin{align*}
\mathbb{P}(\mathrm{max}_{i,j\in[n]}|A(i,j)-\Omega(i,j)|\geq t)\leq\frac{n(n+1)}{2}\mathrm{exp}(t+(t+\Omega_{\mathrm{max}})\mathrm{log}(\Omega_{\mathrm{max}})-(t+\Omega_{\mathrm{max}})\mathrm{log}(t+\Omega_{\mathrm{max}})).
\end{align*}

To ensure that $\tau\leq t$ with high probability, let $\alpha$ be a small probability such that
\begin{align*}
\frac{n(n+1)}{2}\mathrm{exp}(t+(t+\Omega_{\mathrm{max}})\mathrm{log}(\Omega_{\mathrm{max}})-(t+\Omega_{\mathrm{max}})\mathrm{log}(t+\Omega_{\mathrm{max}}))\leq\alpha,
\end{align*}
which gives
\begin{align*}
t+(t+\Omega_{\mathrm{max}})\mathrm{log}(\Omega_{\mathrm{max}})-(t+\Omega_{\mathrm{max}})\mathrm{log}(t+\Omega_{\mathrm{max}})\leq\mathrm{log}(\frac{2\alpha}{n(n+1)}).
\end{align*}

Set $f(t)=t+(t+\Omega_{\mathrm{max}})\mathrm{log}(\Omega_{\mathrm{max}})-(t+\Omega_{\mathrm{max}})\mathrm{log}(t+\Omega_{\mathrm{max}})-\mathrm{log}(\frac{2\alpha}{n(n+1)})$ as a function of $t$. Note that simply setting $f(t)=0$ can not yield an explicit solution for $t$. For further analysis, we get
\begin{align*}
f'(t)=\mathrm{log}(\Omega_{\mathrm{max}})-log(t+\Omega_{\mathrm{max}})<0,
\end{align*}
which implies that $f(t)$ is monotonically decreasing with respect to $t$. We observe that
\begin{align*}
\mathrm{lim}_{t\rightarrow0}f(t)&=\mathrm{log}(n(n+1))-\mathrm{log}(2\alpha)>0,\\
\mathrm{lim}_{t\rightarrow+\infty}f(t)&=\mathrm{lim}_{t\rightarrow+\infty}[t-(t+\Omega_{\mathrm{max}})\mathrm{log}(1+\frac{t}{\Omega_{\mathrm{max}}})+\mathrm{log}(n(n+1))-\mathrm{log}(2\alpha)]=-\infty,
\end{align*}
which implies that there must exist a $t_{\alpha}$ such that $f(t_{\alpha})=0$ based on $f(t)$'s monotonicity. Thus, with probability at least $1-\alpha$, we get
\begin{align*}
\tau\leq t_{\alpha},
\end{align*}
which implies that $\tau$ is finite with high probability.
\end{proof}

\begin{lem}\label{TauRoughLemma}
For any distribution $\mathcal{F}$ that satisfies Equation (\ref{DefinOmega}), assuming that $\mathcal{F}$'s variance exists, $\tau$ is finite with high probability.
\end{lem}

\begin{proof}
Given that the variance of $\mathcal{F}$ exists, by the definition of $\gamma$, we have $\mathrm{Var}(A(i,j)) \leq \gamma \theta_{\mathrm{max}}^2$ for all $i,j \in [n]$. Applying Chebyshev’s inequality, for any $t > 0$, we obtain
\begin{align*}
\mathbb{P}(|A(i,j) - \Omega(i,j)| \geq t) \leq \frac{\mathrm{Var}(A(i,j))}{t^2} \leq \frac{\gamma \theta_{\mathrm{max}}^2}{t^2}.
\end{align*}

Using the union bound, we get
\begin{align*}
\mathbb{P}(\tau \geq t) \leq \frac{n(n+1) \gamma \theta_{\mathrm{max}}^2}{2t^2}.
\end{align*}

Let $\alpha > 0$ be a small probability such that
\begin{align*}
\frac{n(n+1) \gamma \theta_{\mathrm{max}}^2}{2t^2} \leq \alpha,
\end{align*}
which implies
\begin{align*}
t \geq \sqrt{\frac{n(n+1) \gamma \theta_{\mathrm{max}}^2}{2\alpha}}.
\end{align*}

Thus, with probability at least $1 - \alpha$, we have
\begin{align}\label{taurough}
\tau \leq \sqrt{\frac{n(n+1) \gamma \theta_{\mathrm{max}}^2}{2\alpha}}.
\end{align}

According to the 5th column of Table \ref{table-bounds}, $\gamma$ always has a finite upper bound. Consequently, the last inequality indicates that $\tau$ is finite with high probability. Note that the upper bound for $\tau$ in Equation (\ref{taurough}) is quite coarse, as it is derived using Chebyshev’s inequality, which does not account for the specific type of distribution. For instance, in Example \ref{NormalF}, when $\mathcal{F}$ is a Normal distribution, we know that $\gamma \leq \sigma_A^2 / \theta_{\mathrm{min}}^2$. If we set $\alpha = \frac{1}{n^2}$, $\sigma_A = 1$, and $n = 10000$, Equation (\ref{taurough}) yields $\tau \leq \sqrt{\frac{n^3 (n+1) \theta_{\mathrm{max}}^2}{2 \theta_{\mathrm{min}}^2}} = \sqrt{\frac{10000^3 \times 10001}{2}} \frac{\theta_{\mathrm{max}}}{\theta_{\mathrm{min}}}$. Although $\sqrt{\frac{10000^3 \times 10001}{2}} \frac{\theta_{\mathrm{max}}}{\theta_{\mathrm{min}}}$ is finite, it is significantly larger than the value of 8.5839 obtained in the proof of Lemma \ref{NormalTau}. Therefore, while Equation (\ref{taurough}) ensures that $\tau$ is finite with high probability, the bound is quite loose. To obtain a tighter bound, the specific characteristics of the distribution $\mathcal{F}$ should be used, as demonstrated in the analysis of Lemmas \ref{NormalTau} and \ref{PoissonTau}.
\end{proof}
\end{appendices}


\bibliography{refwDCMM}

\end{document}